\documentclass{jfm}
\usepackage{natbib}
\usepackage{graphics}
\usepackage{graphicx}
\usepackage{epstopdf, epsfig}
\usepackage{upgreek}
\usepackage[normalem]{ulem}
\usepackage{cancel}
\usepackage{subcaption}
\usepackage{amssymb}
\usepackage{amsbsy}
\usepackage{amsmath}
\newcommand*\diff{\mathop{} \mathrm{d}}

\usepackage{color}
\definecolor{cyan}{cmyk}{1,0,0,0}
\definecolor{black}{cmyk}{0,0,0,1}
\definecolor{magenta}{cmyk}{0,1,0,0}
\definecolor{yellow}{cmyk}{0,0,1,0}
\definecolor{blue}{cmyk}{1,1,0,0}
\definecolor{red}{cmyk}{0,1,1,0}
\definecolor{green}{cmyk}{1,0,1,0}
\def\bea{\begin{eqnarray}}
\def\eea{\end{eqnarray}}
\def\be{\begin{equation}}
\def\ee{\end{equation}}

\def\bn{\mathbf{n}}
\def\bD{\mathbf{D}}
\def\bQ{\mathbf{Q}}

\def\bv{\mathbf{v}}
\def\bx{\mathbf{x}}
\def\bg{\mathbf{g}}
\def\bT{\mathbf{T}}
\def\bE{\mathbf{E}}
\def\bA{\mathbf{A}}
\def\bQ{\mathbf{Q}}
\def\bN{\mathbf{N}}
\def\bup{ \hat{\boldsymbol{\upsilon}}  }
\def\bL{\mathbf{L}}
\def\bI{\mathbf{I}}
\def\bk{\mathbf{k}}

\def\bkap{\boldsymbol{\kappa}}

\def\DD{|\mathbf{D}|}
\def\DDD{ |\mathbf{D}^{(0)}|}

\newcommand{\ul}{\underline}
\def\dmu{\stackrel{\:\prime}{\mu}}
\def\ddmu{\stackrel{\:\prime\prime}{\mu}}

\newcommand{\ccdot}{\! \cdot \!}
\newcommand{\ML}{{\small MATLAB}\textsuperscript{\textregistered}}

\shorttitle{On the stability of $\mu(I)$}
\shortauthor{J. D. Goddard and Jaesung Lee}
\title{On the stability of the    $\mu(I)$-rheology for granular flow}
\author{J. D. Goddard \aff{1}\corresp{\email{jgoddard@ucsd.edu}}\and 
Jaesung Lee\aff{2}}
\affiliation{\aff{1} Department of Mechanical and Aerospace Engineering, 
University of California, San Diego, 9500 Gilman Drive, La Jolla, CA 92093-0411, USA
\aff{2} Department of Chemical and Environmental Technology, Inha Technical College, 100 Inha-ro, Nam-gu, 22212 Incheon, Republic of Korea}

\begin{document}

\maketitle

\begin{abstract} This article deals with the Hadamard instability of the so-called $\mu(I)$ model of dense rapidly-sheared granular flow, as reported recently  by Barker et al. (2015,
this journal, {\bf 779}, 794-818).  The present paper presents a more comprehensive 
study of the linear stability of   planar simple shearing and pure shearing flows, with account taken of    convective    Kelvin wave-vector
stretching by the base flow.   We provide a closed form solution for the linear stability problem and show that wave-vector stretching    leads to asymptotic stabilization of the non-convective  
instability found by Barker et al.   We also explore the  stabilizing effects of higher velocity gradients achieved by an enhanced-continuum model based on a dissipative analog of the van der Waals-Cahn-Hilliard  equation of equilibrium thermodynamics. This model involves 
a dissipative hyper-stress, as the analog of a special Korteweg stress,  with  surface viscosity representing the counterpart of elastic surface tension.     Based on the enhanced continuum model,     we also present a model of   steady    shear bands   and their non-linear stability against parallel shearing.       Finally, we propose a theoretical connection between  the
    non-convective    instability of Barker et al.  and the loss
of generalized ellipticity in the quasi-static field equations.  
Apart from the theoretical interest, the present work may suggest stratagems for the numerical simulation of continuum field equations involving the $\mu(I)$ rheology and variants thereof.   

\end{abstract}

%\begin{keywords}
%Authors should not enter keywords on the manuscript, as these must be chosen by the author during the online submission process and will then be added during the typesetting process (see http://journals.cambridge.org/data/\linebreak[3]relatedlink/jfm-\linebreak[3]keywords.pdf for the full list)
%\end{keywords}

\section{ Introduction}\label{sect:intro}
Granular flows are ubiquitous in nature and technology, a fact which accounts for a large body of research devoted to the development of continuum models for flows on length scales much larger than the typical grain diameter.  Of particular interest here is the phenomenological  ``$\mu(I)$" model proposed by  Jop, Forterre, Pouliquen and coworkers    \citep {GDR04,Jop05,JFP06}, which has proven useful for dense rapidly sheared 
flows in chutes and avalanching granular layers. However,  \cite {BSBG15}, hereinafter referred to as Ref.\! 1,  conclude that this model is generally ill-posed in the sense of Hadamard, exhibiting the classical linear instability against short wavelength perturbations.
We recall that such phenomena are part and parcel of {\it material instability}, long recognized in solid mechanics and more recently  in the mechanics of complex fluids  \citep[See e.g.][ and references therein ]{JG03}.\\

 While  the  work of Ref.\! 1 is highly relevant to the modeling of granular flow, particularly as it pertains to  numerical simulations, we do not share the authors' assessment of Hadamard instability as physically  ``unrealistic". On the contrary, we assert that this type of instability signals the emergence of spatiotemporal discontinuities,
as ``weak solutions" of the underlying field equations for numerous  physical phenomena, including aerodynamic shocks, 
hydraulic jumps, thermodynamic phase transitions, and, most relevant to the present work,  shear bands or other forms of localized deformation in complex solids and fluids.\\

  In the examples  cited above one should distinguish those involving dynamic or ``geometric" instability from those representing material instability \citep{JG03}, a matter discussed further in the following. In either case,
numerical simulation   generally  requires  advanced  techniques  such as  shock capturing, adaptive mesh-refinement, or mesh-less methods \citep[cf.][]{Bel94}, to represent certain discontinuous solutions that  find widespread
applicability in various fields of  mechanics and thermodynamics .\\

 Moreover,  as  particularly appreciated  in the field of plasticity \citep[See e.g.][]{Forest10,Hen14}, weak solutions may be regularized by means of enhanced continuum models that involve non-local or weakly non-local ``gradient"  effects. From a physical point of view, such effects represent   the emergence of microscopic or mesoscopic length scales or,  loosely, ``Knudsen effects". As one of the   benefits for numerical simulation, regularization   stabilizes against short-wavelength disturbances and imparts a diffuse structure to  otherwise sharp discontinuities. \\

The above considerations provide much of the  motivation for the present work whose  principal objectives are to:\\

\begin{enumerate}
\renewcommand{\theenumi}{1}
 \item explore a simple gradient regularization of the $\mu(I)$ model,  and
\renewcommand{\theenumi}{2}
  \item provide a more complete linear-stability analysis of the regularized model.
\end{enumerate}
\vspace{3mm}
For the purposes of  Item (1), we shall adopt  a visco-plastic tensorial analog of the scalar  van der Waals-Cahn-Hilliard (vdW-CH)  model of equilibrium thermodynamics, under isothermal conditions, with dissipation potential replacing 
Helmholtz free energy and with velocity gradient replacing density. We make no claim for the physical validity of this largely phenomenological model of gradient effects, merely noting that it  is one of the simplest models imaginable and that it embodies  the dissipative analog of a special form of Korteweg stress  \citep{And98}, with  surface viscosity 
arising as the counterpart of equilibrium surface tension. We recall that appeals have been made to the vdW-CH model  in the treatment of  other dissipative phenomena \citep[e.g. by][]{Forest10}, and  the thermodynamic version has been connected to microscopic forces by \cite{G96}. For our purposes,  it conveniently elucidates several theoretical issues  that have not been sufficiently emphasized in past works.  \\

In the case of Item (2), we shall show that the phenomenon of  wave-vector stretching, identified in several previous works  \citep{JG03} but neglected in the 
analysis of   Ref.\! 1,  results in the asymptotic stability of the $\mu(I)$ model, 
irrespective of the vdW-CH regularization. In effect,  initially unstable wave 
vectors are rotated by   simple   shearing     into an ultimately stable orientation, as  foreseen in the path breaking study  of hydrodynamic stability by W. Thomson  \citep[later Lord Kelvin]{Thomson1887}. To lend a certain plausibility to this scenario, we shall also explore an approximate model of a stable shear band with 
characteristic thickness derived from the vdW-CH model\footnote{   Following the original submission of the present work,   three  papers have  appeared \citep{Hey16,Barker17,Barker17b} the first two indicating that compressibility effects can also regularize the $\mu(I)$ model. The third paper  confirms our surmise that ill-posedness is associated with marginal convexity of the dissipation potential in the regimes of constant $\mu$ while offering a modified form of $\mu(I)$ to alleviate this difficulty.  However, whatever their other merits,  scale-independent models of the  kind provided in these works cannot describe the diffuse  shear  bands that   may   eventually emerge from material instability.   }. \\

    While our stability analysis  applies to any homogeneous shear as 
base state, we shall focus attention here    on the two important special cases of   planar flows, the simple shear treated in Ref.\! 1 and pure shear . \\

As a word on notation, we note that, when needed occasionally for clarity, we employ cartesian tensor notation, with sums over repeated indices, with commas denoting partial derivatives,  with tensor components indicated  in brackets $[\:\:]$,   and   with occasional listing of vector components in parentheses $(\:\:)$. In  coordinate-free notation, the linear transformation of vectors  $\bv$ into vectors by second-rank tensors $\bA$  is denoted by $\bA\bv=[A_{ij}v_j]$, 
and  the special case  of the  Euclidean scalar product (i.e. ordered contraction) of tensors having equal rank by the standard mathematical ``dot" product. We occasionally employ the
 brackets $[\:\:]$ as standard notation for skew-symmetrization of tensor components and for the arguments of functionals defined on fields.

 \section{ Gradient  regularization of $\mu(I)$ }\label{sect:vdwch}   As the method we employ applies to general models of viscoplasticity, we offer a derivation of   constitutive equation and momentum balance  based on the notion of a dissipation potential and the associated hyper-stresses and  variational principle. Readers not interested in these details can skip immediately to the 
 momentum balance presented below in (\ref{dg3}).\\  
 
    Consider a  strictly dissipative material endowed with 
  frame-indifferent   dissipation potential    \citep{ Edelen72, Edelen05,JG14,Saramito16}    depending on the first two spatial gradients of 
the material velocity field $\bv(t, \bx)$:

\be\label{strong}\begin{split} &\psi(\bD, \nabla \nabla \bv)=\psi(\bQ\bD\bQ^T,  \bQ\nabla \bQ\nabla \bv\bQ^T),
\\&\mbox{with}\:\:\bD=\frac{1}{2} \left(  \nabla \bv  + ( \nabla \bv )^{T} \right), 
\end{split}\ee
where $\bD$ is the deformation rate  and $\bQ=\bQ(t)$ is  an arbitrary time-dependent but spatially independent orthogonal tensor. \\

A simple special case is represented by the vdW-CH\footnote{The terminology ``van der Waals-Cahn-Hilliard" is  more accurate historically \citep{R79} than the   oft-used  ``Cahn-Hilliard". We recall that Korteweg  proposed a more general stress arising from  a  Helmholtz free energy contribution of 
form $\rho_{,i }\rho_{,j} \rho_{,ij}$.}
\bea\label{vdWCH}
	\psi (\bD, \nabla \nabla \bv) = \psi_0 (\bD) + \chi \left| \nabla \nabla \bv \right|^2 ,
\eea
where we take $\chi$ to be a positive constant,  and the term in higher-order velocity gradient $\vert  \nabla \nabla \bv \vert^2=(\nabla\nabla\bv)\ccdot(\nabla\nabla\bv)=v_{i,jk}v_{i, jk}$ serves to provide
a local regularization whenever $\psi_0$ becomes locally non-convex. 
We recall that the gradient term in the standard scalar   version of the vdW-CH equation involves fluid density in place of  $\nabla\bv$    \citep {CH58, G96}, providing an energy penalty on large gradients.     This yields 
  well- posed  field equations in the case of non-convex free-energy with phase transition  and serves to define equilibrium surface tension in the limit $\chi\rightarrow 0$.\\

  In the following, we deal with {\it strongly dissipative} materials, i.e.   strictly dissipative materials  devoid of gyroscopic or ``powerless" stress   \citep {JG14}, for which the partial derivatives of {\it dissipation potential}  $\psi(\bD, \nabla \nabla \bv)$  yield 
 work-conjugate (symmetric   Cauchy)   stress $ \bT^{(1) }$ and hyperstress  $\bT^{(2)}$, respectively, according to:
\bea\label{stress}
	\bT^{(1) } = \frac{\partial \psi}{\partial \bD} =  \frac{\partial \psi_0}{\partial \bD}  \:\: \mbox{and} \:\: 
	\bT^{(2) } = \frac{\partial \psi}{\partial ( \nabla \nabla \bv )} =  2\chi  \nabla \nabla \bv.
\eea 
and the volumetric rate of dissipation is given by  
\bea
	\mathfrak{D} = \bT^{(1)}\!\ccdot\bD +  \bT^{(2)}\!\ccdot\nabla \nabla \bv 
	      =  T_{ij}^{(1)} D_{ij} +  T_{ijk}^{(2)} v_{k,ij}.
\eea
The  hyperstress $\bT^{(2)}=[T^{(2)}_{ijk}]$ represents a generalized ``pinch" $\bn \ccdot \bT^{(2)}=[n_iT^{(2)}_{ijk}]$   acting on a material plane with unit normal $\bn$ which can be represented by a force dipole consisting of equal and opposite forces. The symmetric part $[n_iT^{(2)}_{i(jk)}]$  involves forces acting along the line of centers of their points of application, representing a normal pinch or symmetric ``stresslet", whereas the antisymmetric part $[n_iT^{(2)}_{i[jk]}]$ involves  forces acting perpendicular to their line of centers,  representing a ``torque" or ``rotlet".  Note that the  gradient of vorticity enters into the mechanical power  and that most  of the concepts of  kinematic and stress carry over from the 
various works on the elasticity of solids, where strain energy (Helmholtz free energy) rather than dissipation potential is involved\footnote{ Whereas the theory for elastic solid or fluids leads to elastic surface tension, the hyper-dissipative model can apparently yield a surface viscosity, which as far as we know would constitute a novel continuum approach to the subject.}. \\

 Thus, upon enslaving  micro-structural to continuum  kinematics in the classic work
of   \cite{M64}, achieved by  taking his relative displacement gradient $\gamma_{ij}=0$, one obtains the linear-elastic analog of the present work with his strains $\epsilon_{ij}$ replacing our strain rates $D_{ij}$.  Our  quasi-static
equation of equilibrium (\ref{qs})  follows from that of  Mindlin upon relaxing the assumption of incompressibility and combining his Eqs.  (4.1), effectively  eliminating his ``relative stress"  $\sigma_{ij}$. Mindlin's work also shows that,  within a linear isotropic gradient model,   one may anticipate further quadratic  terms in $\nabla\nabla\bv$ beyond that adopted in the simpler one-parameter model (\ref{vdWCH}) of the present study.\\

The Hamiltonian  momentum balances for the hyper-elastic system of Mindlin do not apply to hyper-dissipative systems,  except in the limit of quasi-static  (i.e.   inertialess  ) motion.   Hence,   further analysis is required to obtain the relevant balances for the latter, and we   begin with the variational principle leading to the  quasi-static  balance. We note that similar methods have been adopted in past works of   \cite{Hill56} and  \cite{Leonov88}, methods which are made rigorous mathematically by  the later works of Edelen,     e.g.  \cite{Edelen72,Edelen05},     that are highlighted   and simplified     in the survey by  \cite{JG14}. \\

Note that the  variational derivative of the functional of $\bv(\bx,t)$ representing global dissipation potential: 
\be\label{global}  \Psi[\bv] = \int_V[ \psi(\bD, \nabla\nabla\bv) -p \nabla\ccdot\bv]\diff V, \ee
 subject to incompressibility $\nabla\ccdot\bv=0$  in spatial domain V, is given by 
\be\label{variation}
\begin{split}
	\delta \Psi[\bv]&=
	\int_V \left[  
	\frac{\partial \psi}{\partial \bD}  \ccdot  \delta \nabla \bv +  \frac{\partial \psi}{\partial ( \nabla \nabla \bv )}  \ccdot  \delta \nabla \nabla \bv-p\nabla\ccdot\delta\bv
	\right] \diff V \\
	&= 
	\int_V \left[  
	T_{ij}^{(1)} \delta v_{i,j} +  T_{ijk}^{(2)} \delta v_{i,jk}- p\delta v_{j,j}
	\right] \diff V  \\
	&= 
	\int_V   \left[   
	( T_{ij}^{(1)} \delta v_{i} )_{,j}  -  T_{ij,j}^{(1)} \delta v_{i}  +   ( T_{ijk}^{(2)} \delta v_{i,j} )_{,k} \right.\\&-   \left. ( T_{ijk,k}^{(2)} \delta v_{i} )_{,j}  +   T_{ijk,jk}^{(2)} \delta v_{i} 
	 -  ( p \delta v_{j})_{,j}  +  p_{,j} \delta v_{j}  
	\right]  \diff V  \\
	&= 
	\int_{\partial V}  \bn \ccdot  \left[  ( \bT^{(1)}    -  \nabla  \ccdot   \bT^{(2)}  -  p\bI )   \ccdot  \delta \bv   +    \bT^{(2)}  \ccdot  \nabla\delta \bv \right] dS  \\
	&\quad\quad 	 - 
	\int_V   \left[   \nabla  \ccdot  \bT^{(1)}  -  \nabla p  -  (\nabla \nabla )  \ccdot  \bT^{(2)}\right]  \ccdot   \delta \bv \diff V,
\end{split}
\ee
where  pressure $p$ plays its usual role as Lagrange multiplier and  use has been made of the divergence theorem for  integration by parts. 
Noting that $\delta\bv=0$ on $\partial V$ implies that the surface-tangential gradient of  $\delta\bv$ vanishes on $\partial V$, 
this establishes the following variational theorem:
\begin{quote}  
{\it Stationarity of the  global dissipation potential  for all variations $\delta \bv (\bx )$ 
subject to incompressibility  $\nabla\ccdot\bv(\bx)=0$ in $V$ and to  fixed $\bv$ and $\bn\ccdot\nabla \bv$ on $\partial V$, yields  
the quasi-static  equation of equilibrium.} 
\end{quote}Ä
The latter is given by (\ref{variation}) as
\bea
	\label{qs}\begin{split}
	 & \nabla  \ccdot  \bT^{(1)}-\nabla p  - ( \nabla \nabla )  \ccdot  \bT^{(2)}= 0, \:\: \mbox{i.e. }\:\:   T_{ij,j}^{(1)}-p_{,i}-T_{jki, jk}^{(2)} =0, \\& \mbox{or} \:\:\nabla\ccdot\bT -\nabla p=0, \:\: \mbox{where} \:\:\bT:=\bT^{(1)} -  \nabla  \ccdot  \bT^{(2)}, \:\: \mbox{i.e.}\:\: T_{ij}=T^{(1)}_{ij}-T_{jki, k}^{(2)}, \end{split}
\eea where  $\bT$ obviously serves as effective stress tensor.   Despite the ostensible reduction  to a single stress tensor, we should re-emphasize that the  hyper-stress $\bT^{(2)}$ can  give rise to 
 singular surface stresses balancing discontinuities in  the Cauchy stress  $\bT^{(1)}$, as mentioned below in our analysis of shear bands. We further note that the presence of such effects at the nominal free surface 
of thin avalanching layers could invalidate theories based on variants of the $\mu(I)$ model, as already  suggested by certain strongly non-local models \citep{Hen14}.     \\

Now, one can  extend (\ref{qs}) to include gravitation or other fixed body forces $\bg$ by replacing $\psi$ with $\psi - \phi\rho_s\bg\ccdot\bv$ in the first term of (\ref{variation}).  Then, by a further appeal to {\it d'Alembert's principle}, one can then replace $\bg$ with the total acceleration $ \bg\!-\!d_t\bv$ in the resulting equation of equilibrium to obtain the complete
linear momentum balance:
\bea
	\label{dg2}
	\rho_s \phi \diff_t \bv
	=
	 - \nabla p 
	 + \nabla  \ccdot   \bT
	+ \rho_s \phi \bg \:\:\mbox{and}\:\:\nabla\ccdot\bv=0, \:\:\mbox{with}\:\: \diff_t=\partial_t+\bv\ccdot\nabla,
\eea
a relation which no longer follows  directly from the above extremum principle.
Here $\rho_s$ is the constant solid density of the grains, $p$  the pressure, 
$\bg$  the gravitational acceleration, and
$\diff_t$  the material (or ``substantial") time derivative. Thus, with re-interpretation of the stress tensor $\bT$, the linear momentum balance retains the same form as for a simple (``nonpolar") material. 

\subsection{ Application to the $\mu(I)$  model}\label{subsect:muI}
In the  model of     \cite{JFP06} (cf. Ref.\! 1) for granular rheology,   the Cauchy stress $\bT^{(1)}$  is given by a rate-dependent version of   Drucker-Prager plasticity with 
rate-dependent friction coefficient $\mu(I)$:
\bea\label{derivpot}\begin{split}
	&\bT^{(1)} =\partial_{\bD}\psi_0=\mu (I) p \bE,\:\:\mbox{with}\:\:\mu (I) = \mu_0 + \frac{ (\mu_{\infty} - \mu_0 )}{(I+ I_* ) } I\\&
	\mbox{and}\:\: \psi_0  =  \frac{p}{\theta}\left[\mu_\infty I  + (\mu_0 -\mu_\infty)I_* \ln
\left(\frac{I+I_*}{I_*}\right) \right].
%\\&\mbox{and} \:\:\mu (I) = \mu_0 + \frac{ (\mu_{\infty} - \mu_0 )}{(I+ I_* ) } I,	
\end{split}\eea 
Here, the normalized form or ``director" $\bE$, the Euclidean norm of the strain rate $  \vert \bD \vert $, and the {\it inertial number}  are defined, respectively, by
\be\label{eqsDI}
	\bE := \frac{\bD}{ \vert \bD \vert }, \:\:  \vert \bD \vert  := \sqrt{\mathrm{tr} (\bD^2) },\:\:
	\mbox{and}\:\:
	I  = \theta  \DD,\:\:\mbox{with}\:\: \theta = d\sqrt{2\rho_s/p},
\ee
where $I_*$ is an empirical constant (denoted by $I_0$ in several previous works),  $d$ is representative grain diameter, $p$ is local pressure, $\theta$ plays the role of  an inertial relaxation time, and the quantities $\mu_\infty \ge\mu_0$ represent limiting 
 friction coefficients. One obtains the standard version adapted to simple shear upon replacing the Euclidean norm by the  norm $\Vert\bD\Vert$=$|\bD|/\sqrt{2}$ employed in \cite{JFP06} and  Ref.\! 1. \\
 
  We note that 
$\psi_0$ is marginally convex in the special case of constant $\mu$  \citep{JG14}, which 
we  believe accounts for the general tendency to material instability in  perfectly-plastic models. The potential multiplicity of solutions will be made more evident by the model of shear-banding presented below in Section \ref{sect:shearband}.\\

The momentum balance (\ref{dg2}) can now be recast as 
\be
	\label{dg3}
  \begin{split}
	\rho_s \phi \diff_t \bv
	 =
	- \nabla p 
	+ \nabla  \ccdot  (  \mu (I) p \bE ) - 2 \chi \nabla^4 \bv 	+ \rho_s \phi \bg .	
	\end{split}
\ee
The third term on the r.h.s.  of (\ref{dg3})  arises from the hyperstress $\bT^{(2)}$, and the momentum balance of  Ref.\! 1     is obtained by taking $\chi\equiv 0$.  The preceding term
 represents the standard Cauchy stress and, confirming the analysis of  Ref.\! 1, can be expanded to yield
\begin{equation}
	\begin{split}
	\nabla   \ccdot  \left( \mu (I) p \bE \right) 
	 =   \frac{\mu (I) ( 2 -   \dmu)}{2} \bE \nabla p                
	   +  ( \dmu -1) \frac{\mu (I) p }{ \DD }  [(\bE\nabla) \nabla\bv]\ccdot\bE       
	   +   \frac{\mu (I) p }{2 \vert \bD \vert}   \nabla^2 \bv.
	\end{split}
\end{equation}
Then, the momentum balance in (\ref{dg2}) becomes 
\be
	\label{dg5a}
	\rho_s \phi  \diff_t \bv =
	 - \bN \nabla p 
	  +    \frac{(\dmu -1) \mu p }{ \vert \bD \vert} 
	[(\bE\nabla) \nabla\bv]\ccdot\bE 	  +   \frac{\mu  p }{2 \vert \bD \vert}   \nabla^2 \bv 
	 -  2 \chi \nabla^4 \bv  
	 +  \rho_s \phi \bg,	
\ee
where we employ the notation 
\be 
\bN = \bI  -  \frac{ (2 - \dmu) \mu}{2} \bE, \:\:
	\dmu 
	=  \frac{I}{\mu (I)}  \frac{\diff  \mu (I)}{\diff I} 
	=  \frac{\diff \log \mu (I)}{\diff \log I}    \:\:\mbox{and}\:\:
	\ddmu = \frac{\diff^2 \mu (I)}{\diff I^2} \frac{I^2}{\mu (I)}
\ee
here and below.

\section{ Linear Stability Analysis}\label{sect:linstab}

In the usual way, the perturbed velocity $\bv$ and pressure are written 
\be
	\label{lsa1}
	\bv = \bv^{(0)} + \bv^{(1)} \:\:\mbox{and}\:\:  p = p^{(0)} + p^{(1)},
\ee
where  superscripts $(0)$ and $(1)$ denote the base state and the perturbation, respectively.  
The perturbed friction coefficients are given by
\be
	\label{lsa2}
	\begin{split}
	&\mu  =  \mu^{(0)} 
		 +  \left( \frac{\partial \mu}{\partial p}  \right)^{(0)}     p^{(1)}
		 +  \left(  \frac{\partial \mu}{\partial \bD}  \right)^{(0)}    \ccdot \nabla  \bv^{(1)} \:\:\mbox{and}\:\: \\ 
	&\dmu \: = \:\dmu^{(0)} 
		 +  \left(  \frac{\partial \dmu}{\partial p}  \right)^{(0)}     p^{(1)}
		 +  \left(  \frac{\partial \dmu}{\partial \bD}  \right)^{(0)}    \ccdot \nabla \bv^{(1)}. 		
	\end{split}
\ee	
Substituting (\ref{lsa1}) and (\ref{lsa2}) into (\ref{dg5a}),
the linearized equations of perturbed motion become
\be
	\label{lsa3}
	\begin{split}
	& \rho_s \phi   \left(  \diff_t^{(0)}\bv^{(1)}  
	+  \bv^{(1)}  \ccdot  \nabla \bv^{(0)} \right)    
	= - \bN^{(0)}\nabla p^{(1)} \\
	&+ (  \dmu^{(0)}   -1 )  \frac{\mu^{(0)} p^{(0)}}{ \DDD}
	 (\bE^{(0)} \nabla )( \bE^{(0)}  \ccdot  \nabla \bv^{(1)} ) 
	 +   \frac{\mu^{(0)} p^{(0)}}{2 \DDD} \nabla^2 \bv^{(1)}   -  2 \chi \nabla^4 \bv^{(1)}  \\
	&+   \left[    \frac{ \mu^{(0)} (\ddmu^{(0)} - \dmu^{(0)} )}{4 p^{(0)}}  p^{(1)} 
	 +   \frac{\mu^{(0)} }{2 \DDD} (\dmu^{(0)}    -  \ddmu^{(0)} ) \bE^{(0)}   \ccdot \nabla \bv^{(1)}  \right]    \bE^{(0)}\nabla p^{(0)} , \\&\mbox{where}\:\:\bN^{(0)} = \bI - \alpha \bE^{(0)} \:\:\mbox{with}\:\:
	\alpha = \left[\frac{ (2 - \dmu )}{2} \mu\right]^{(0)}   ,
	\end{split}
\ee
and the equations of    Ref.\! 1 are  obtained by taking $\nabla p^{(0)}={\bf 0}$ and $\chi =0$. \\ 

With   $\DDD^{-1}$  as time scale and 
 $d$ as  length scale,
we henceforth adopt the following non-dimensional variables 
\be
	\label{NDV}
	\begin{split}
	&\bar{\bx} = \frac{\bx}{d}, \:\: \bar{t} = \surd{2}\DDD t,	\:\: 
	\bar{\bv} = \frac{\bv}{\surd{2}\DDD d}, \\&
%	\bar{\DD} = \frac{\DD}{\DDD }, \:\: \\
	\bar{p} = \frac{p}{2\rho_s d^2 \DDD^2}, \:\: 
		\bar{\chi} = \frac{\chi}{\surd{2}\rho_s d^4  \DDD }, \:\:
		\bar{g} = \frac{g}{2d \DDD^2}. 
		%\:\:\mbox{and}\:\
	\end{split}
\ee Various factors  of $\surd{2}$, included here to simplify  the results presented below  for simple shear,  can be eliminated by substituting  the  norm   $\Vert \bD \Vert=|\bD|/\surd{2}$ employed  in   Ref.\! 1 and mentioned above.  For later reference, we note that inspection of  the combination of terms 
$\nabla^2\bv$ and $\nabla^4\bv$ in (\ref{dg5a}) indicates that $\bar{\chi}$ is proportional to the  square  of a ratio of  microscopic to macroscopic length scales. \\

Employing the above non-dimensional variables and dropping the superimposed bars for simplicity, we can write
(\ref{lsa3})   in component form  as
\be
	\label{lsa4}
	\begin{split}
	& \phi   \left[ (\diff_t^{(0)} \bv^{(1)})_i  +    v_{i,j}^{(0)} v_j^{(1)}  \right] \\
	& = -   N_{ij}^{(0)} p^{(1)}_{,j}
	 -   2\beta \gamma  E_{ij}^{(0)} E_{kl}^{(0)}  v^{(1)}_{k, jl}
	 +   \gamma v^{(1)}_{i, jj}
	 -  2 \chi v^{(1)}_{i, jjll}, \:\:\mbox{and}\:\: v^{(1)}_{j,j}=0,\\
& \mbox{with}\:\: N_{ij}^{(0)}=\delta_{ij} - \alpha E_{ij}^{(0)}, \:\:
	\beta = 1 - \dmu^{(0)} \:\:\mbox{and}\:\: 
	\gamma =  \frac{\mu^{(0)} p^{(0)}}{\surd{2} },
\end{split}\ee
for the case of uniform  base pressure $ p^{(0)}$.
  
The  parameters 
$\dmu^{(0)}\!\!\!, \alpha, \beta$, and $\gamma$ of this study are related to the parameters $\nu, q, r$, and $\eta^{(0)}$ of  Ref.\! 1, respectively, by  
\[ \dmu^{(0)} = \nu,\: \alpha = q = \mu^{(0)},\:  
\beta = r = 1 - \nu,\:\:\mbox{and}\:\: \gamma = \eta^{(0)} \]  
%= \frac{\mu p^{(0)}}{2 ||\bD||}$
%and $ \DDD = \surd{2} ||\bD||$.
We employ $\bE$ for the normalized deformation rate denoted by $\bA$ in Ref.\! 1, with  different norms discussed above, and vector $\bk$ for wave number in {\it lieu} of $\boldsymbol{\xi}$ of Ref.\! 1,  which is defined below as $\imath \bk$.    
 \\

Note that the  quantities  $\gamma/\phi$ and $\chi/\phi$ represent ratios of inertial to dissipative forces, replacing the  inverse of the Reynolds
number for Newtonian fluids. Since these quantities are $O(1)$  in the 
following analysis,  inertia can be assumed to be relatively unimportant. Hence, we are led to characterize the instability found in Ref.\! 1 and analyzed  below as  {\it material instability}, of a type  which is already manifest   in the quasi-static equations  obtained by taking  $\phi \equiv 0$ on the left-hand side of (\ref{dg3}) or (\ref{lsa4}). As discussed below, this type of instability   
can be attributed to the loss of generalized ellipticity of the highest-order differential operators on the right-hand side of  
(\ref{lsa4}), as indicated e.g.~by the classical analysis of Browder \citep{Bro61,Brez98}.   The first of these papers  indicates clearly the connection to the   transient    instability found in Ref.~1. This type of  linear  instability,  
with initially large  transient   growth rates,  may of course be more relevant  to  the   numerical simulation of non-linear instability than the  asymptotic linear stability  established analytically by the following analysis.  In that respect, the transient instability  bears a certain resemblance to 
that arising from the non-normality of linear-stability operators \citep{Trefethen05} which can trigger non-linear effects.   \\

We now consider a  base state   $\bv^{(0)} = \bL^{(0)} \bx$ with  spatially uniform velocity gradient $\bL^{(0)}=(\nabla\bv^{(0)})^T$. Then, with the Fourier-space representation $\hat{f}(\bk)$ of  fields $f(\bx)$ and
the  duality $\bx \leftrightarrow \imath \hat{\nabla}$ and  $ \nabla \leftrightarrow \imath \bk $, 
 (\ref{lsa4}) can be written 
\be
	\label{lsa5}
	\begin{split}
	& \hat{\diff}_t^{(0)} \hat{\bv}^{(1)}   
	= \left[ 
	{\bf M}- \frac{1}{\phi} (\gamma k^2  +  2 \chi  k^4) \bI \right] 		
	\hat{\bv}^{(1)} + {\bf m} \hat{p}^{(1)}, 	\:\:\mbox{where} \:\: \\
	&  {\bf M}  =  \frac{2 \beta \gamma}{\phi} (\bE^{(0)} \bk)
	 \otimes  (\bE^{(0)}\bk)  -  \bL^{(0)} 
	\:\:\mbox{and} \:\:
	{\bf m}  =  -\frac{\imath}{\phi} \bN^{(0)} \bk ,\\
	& \mbox{with} \:\:
	k=|\bk|=\sqrt{k_ik_i} \:\: \mbox{and} \:\: 
	\hat{\diff}_t^{(0)}= \partial_t -(\bL^{(0)T}\bk)\ccdot\hat{\nabla},
	\end{split}
\ee
where $\bI$ denotes the unit tensor.\\

 It is further 
 worth noting that the term in $\bL^{(0)}$ appearing in the expression for ${\bf M}$ represents convective distortion 
 of the Fourier mode $\hat{\bv}$ by the base flow,  whereas that appearing in the convected derivative,  defined by the final member of (\ref{lsa5}),  represents  wave-vector stretching  \citep{Thomson1887,JG03}. We shall show presently that the latter term is crucial to the asymptotic stability for  $t\rightarrow \infty$. \\

Taking the scalar product  with $\bk$ of (\ref{lsa5}) and invoking  incompressibility,
$\bk \ccdot \hat{\bv}^{(1)} = 0$,  and its  consequence, \[\bk\ccdot(\hat{\diff}_t^{(0)} \hat{\bv}^{(1)} ) =  \bk \ccdot  \bL^{(0)} \hat{\bv}^{(1)},\] we obtain  $\hat{p}^{(1)}$ as one component 
of the oblique projection ${\bf m} \otimes \bk/{\bf m}\ccdot \bk$:
\be
	\label{p1}
	\begin{split}
	 \hat{p}^{(1)} 
	&=  
	\frac{1}{\bk\ccdot{\bf m} }\bk\ccdot(\bL^{(0)}  - {\bf M})\hat{\bv}^{(1)}. \\
	\end{split}
\ee
Substitution of (\ref{p1}) into (\ref{lsa5}) then gives 
\be
	\label{ODE}
	\begin{split} 
	&\hat{\diff}_t^{(0)} \hat{\bv}^{(1)} 
	 =  {\bf A}  \hat{\bv}^{(1)}, \:\:\mbox{where}\:\: \\
	&{\bf A} =
	 \left( \bI  -   \frac{(\bN^{(0)}\bk) \otimes \bk} {  \bk \ccdot \bN^{(0)}\bk}  \right)  ( {\bf M}  -  {\bL}^{(0)} )
	 + \bL^{(0)}  -  \frac{1}{\phi} (\gamma k^2  +  2 \chi  k^4) \bI.
	\end{split}
\ee 
  The   first term in round brackets  on the r.h.s. of  (\ref{ODE}) represents another oblique projection,  orthogonal to $\bk$,  and is independent of  $k=|\bk|$.  Note that   
$\bk \ccdot \bN^{(0)} \bk = k^2 - \alpha(\bk \ccdot \bE^{(0)} \bk) $  is   a positive-definite quadratic form in  $\:\bk$   since $\alpha <1 $ and $\vert \bE^{(0)} \vert = 1$. (cf.   Ref.\! 1.) \\

Transforming (\ref{ODE})  from coordinate ${\bf k}$ to the dual material coordinate $\bkap$ gives the linear-stability equation as canonical ODE:
\be\label{stability}\begin{split}
	& \frac{ \diff \hat{\boldsymbol{\upsilon}}  }{\diff t}= \bA(t, \bkap)					 \hat{\boldsymbol{\upsilon}}  , 
	\:\: \mbox{with}  \:\:
	 \hat{\boldsymbol{\upsilon}}  (t, \bkap)= \hat{\bf v}^{(1)}(t,{\bf k}), \:\: \frac{ \diff \hat{\boldsymbol{\upsilon}}  }{\diff t}=\left(\frac{\partial\bup}{\partial t}\right)_{  \bkap},\\ 
	& \mbox{with}\:\:
	\bk =({\bf F}^{(0)})^{-T} \bkap,\:\:\mbox{and} \:\:\diff_t {\bf F}^{(0)}= \bL^{(0)}{\bf F}^{(0)}, \:\: {\bf F}^{(0)}=\bI\:\:@\:\:t=0,
	\end{split} \ee
where $\bkap$ represents the initial wave vector ${\bf k}$ at $t=0$ and 
superscript $-T$ represents inverse transpose.
This provides the linear stability theory for general 
homogeneous shearing,   similar to  that 
treated elsewhere  \citep{JG03}, where it is  identified as  a time-dependent (or ``non-autonomous") stability problem.\\

For  time-dependent stability, we recall that the  eigenvalues  of $\bA(t, \bkap)$ in (\ref{stability})  serve mainly  to determine   local stability at 
 time $t$ and fixed $\bkap$, with logarithmic growth of spectral energy $| \hat{\boldsymbol{\upsilon}}  |^2(\boldsymbol{\kappa},t)= \hat{\boldsymbol{\upsilon}}  ^*\!\ccdot  \hat{\boldsymbol{\upsilon}}  $  given by the first equation of (\ref{stability}) as twice the Rayleigh quotient, namely
\be\label{energy} \frac{\diff } {\diff t}\ln | \hat{\boldsymbol{\upsilon}}  |^2=  2\frac{ \hat{\boldsymbol{\upsilon}}  ^*\!\ccdot\bA \hat{\boldsymbol{\upsilon}}  }{| \hat{\boldsymbol{\upsilon}}  |^2},\ee 
where $^*$ denotes complex conjugate.
Whenever $\bA$ has real eigenvalues we may take $ \hat{\boldsymbol{\upsilon}}  ^*= \hat{\boldsymbol{\upsilon}}  $,  and the maximum of the  r.h.s.  
over all $ \hat{\boldsymbol{\upsilon}}  $ equals $2\lambda(t)$, where $\lambda(t)$ is the largest eigenvalue. Otherwise, the complex eigenvalues of $\bA$ represent stationary 
points in the complex plane  \citep{Did93}.  In the former case, it follows
that the greatest eigenvalue  $\lambda(t)$ implies asymptotic  stability in the sense of energy if  $\lambda(t) < 0$ for $t\rightarrow \infty$.
\subsection{ Stability of steady simple shear}\label{subsect:shear}
Following  Ref.\! 1, we consider a homogeneous steady simple shearing as base state, with  non-dimensional version of the  the canonical  forms given by  $\bv^{(0)} = (x_2, 0, 0)$ and
\be\label{sshear} \bL^{(0)}=
\begin{bmatrix}
 0 &  1 & 0  \\
 0 &  0 & 0  \\
 0 & 0  &   0
\end{bmatrix} \:\:\mbox{and}\:\:
	{\bE}^{(0)} =
	\frac{1}{\sqrt{2}}
	\begin{bmatrix}  0 & 1 &0\\ 1 &0& 0\\0&0&0  \end{bmatrix}
	\:\:\mbox{with}\:\:  {\DDD}  = 1/\surd{2}.
\ee 
Hence,  $({\bf F}^{(0)})^{-T}= (\bI - {\bf L}^{(0)T}t)$, and the 
  wavevector $\bk$  is given 
in terms of the initial wave number $\bkap$  by (\ref{stability}) as\footnote{cf.  Eq.\! (40) in the remarkable paper of Thomson \citep[\S 32 - \S39] {Thomson1887}. His exact solution for the perturbation of a simple-shear flow governed by the 
Navier-Stokes equations illustrates the short-comings of  stability analyses based on initial growth rates  , as pointed out in several previous works, e.g. \citep{Alam97}  .  } 
\be
	\label{wavevector}
	 \bk (\bkap, t) = (k_1, k_2, k_3 ) 
	=  (\kappa_1, \kappa_2 - \kappa_1t, 	
	\kappa_3), 
\ee with wave-vector stretching restricted to  the component $k_2$. Appendix \ref{app:B}
presents a Squire's-type theorem showing that planar perturbations with $k_3=0$ are the least stable. \\

For  planar perturbations, the components of the tensors appearing in (\ref{ODE}) can be displayed in the 2D format:
\be
	\label{obliqueP}
	\begin{split}
	\bI -  \frac{\bN^{(0)} \bk \otimes \bk} {  \bk \ccdot {\bf N}^{(0)} \bk}  
	&= \frac{1}{\Phi_1}
	\begin{bmatrix} 
	\Phi_1  +  k_1 \left( \frac{\alpha}{\sqrt{2}} k_2  -  k_1 \right) &&        k_2 \left( \frac{\alpha}{\sqrt{2}} k_2  -  k_1 \right) \\
	 k_1 \left( \frac{\alpha}{\sqrt{2}} k_1  -  k_2 \right) 	&&       \Phi_1  +  k_2 \left( \frac{\alpha}{\sqrt{2}} k_1 -  k_2 \right)
	\end{bmatrix},\\
	{\bf M}  -  {\bf L}^{(0)} 
%	 =   \frac{1}{\phi} 
%        \left[  
%	2\beta \gamma  (\bE^{(0)} \bk)  \otimes  (\bE^{(0)} \bk ) 
%	 -  2 \phi \bL^{(0)}  \right] 
	 &=  \frac{1}{\phi}
	\begin{bmatrix} 
	\beta \gamma k_2^2 &  \beta \gamma k_1 k_2  -  2 \phi  \\
	\beta \gamma k_1 k_2 & \beta \gamma k_1^2 \end{bmatrix},
\end{split} \ee
  where $\Phi_1 = \bk \ccdot {\bf N}^{(0)} \bk = k^2 - \sqrt{2} \alpha k_1 k_2 $ and $\Phi_2 = \gamma k^2 + 2 \chi k^4$. Hence,
  \be 
	\label{Acomp}
	\begin{split} 
	&A_{11}  = 
	\frac{1}{\phi \Phi_1}  \left[
	- \beta \gamma k_2^2 (k_1^2 - k_2^2) - \Phi_1 \Phi_2
	\right],\\ 
	& A_{12}  = 
	\frac{1}{\phi \Phi_1}  \left[
	- \beta \gamma  k_1 k_2    +  \phi  \right] (k_1^2 - k_2^2),\\ 
	&A_{21}  = 
	\frac{1}{\phi \Phi_1}  \left[
	\beta \gamma k_1 k_2  (k_1^2 - k_2^2)   
	\right],\\
	&A_{22}  = 
	\frac{1}{\phi \Phi_1}  \left[
	 \beta \gamma k_1^2  (k_1^2 - k_2^2)  - \Phi_1  \Phi_2  
	 +  {2} \phi k_1  (k_2 -  \alpha k_1 /\surd{2} )  	\right].
	\end{split} 
\ee
  where the components of $\bA$ without convection  are obtained by setting 
$\phi=0$ in the numerator of the expressions for $A_{12}$ and $A_{22}$. \\

The determinantal condition $\det(\bA - \lambda \bI ) = 0$ gives   
the eigenvalue $\lambda$   of $\bA$ with largest real part  as 
\be
	\label{eigen01}\begin{split}
	\lambda& = \frac{A_{11}+A_{22}}{2}+\left[  \left(\frac{A_{11}-A_{22}} {2}\right)^2+A_{12}A_{21}\right]^{1/2}\\&= \frac{1}{2 \phi \Phi_1} 
	\left[ \beta \gamma  (k_1^2 - k_2^2 )^2 - 2 \Phi_1 \Phi_2 + 2\phi k_1  ( k_2 - \alpha k_1 /\surd{2})  + \Phi_3^{1/2} \right],
	\\\mbox{where}\\ 
	\Phi_3 & =  
	    \beta^2 \gamma^2   (k_1^2  -  k_2^2 )^4  
	 +  2\phi^2  k_1^2  (\alpha k_1  -   \surd{2} k_2 )^2  
	 \\&\quad-  2 \beta \gamma \phi  k_1^2 (k_1^2  -  k_2^2 ) 
	(\surd{2} \alpha k_1^2    - 4 k_1 k_2 + \surd{2} \alpha k_2^2 ).\end{split} 
\ee 
   Negative values of $\Phi_3$ represent oscillatory behavior, analogous to the 
``flutter" instabilities exhibited by certain elasto-plastic models, the subject of a comprehensive review by \cite{Big95}.\\
 
  When convection  is  neglected, we have 
  \be 
	\label{lambda01}
	\Phi_3 =   \beta^2 \gamma^2  (k_1^2  -  k_2^2 )^4,\:\:\mbox{and} \:\:\lambda = 	\frac{\beta \gamma  (k_1^2 - k_2^2 )^2 - \Phi_1 \Phi_2}{\phi \Phi_1}.
\ee
 Since $\Phi_3$ is non-negative    in this case, it follows that    there is no
oscillatory behavior without convection.   Moreover, since  $\Phi_3$ does not depend on $\chi$,  the  oscillatory frequencies arising from     convection   do not depend  on $\chi$.  However, the  amplitude  of oscillation does   depends on $\chi$ through the   real part of $\lambda$, as discussed further below.\\

%%%%%%%%%%%%%%%%%%%%%%%%%%%%%%%%%%%%%%%%%%%%%
\begin{figure} 
        \centering
        \begin{subfigure}[b]{0.475\textwidth}
            \centering
            \includegraphics[width=\textwidth]{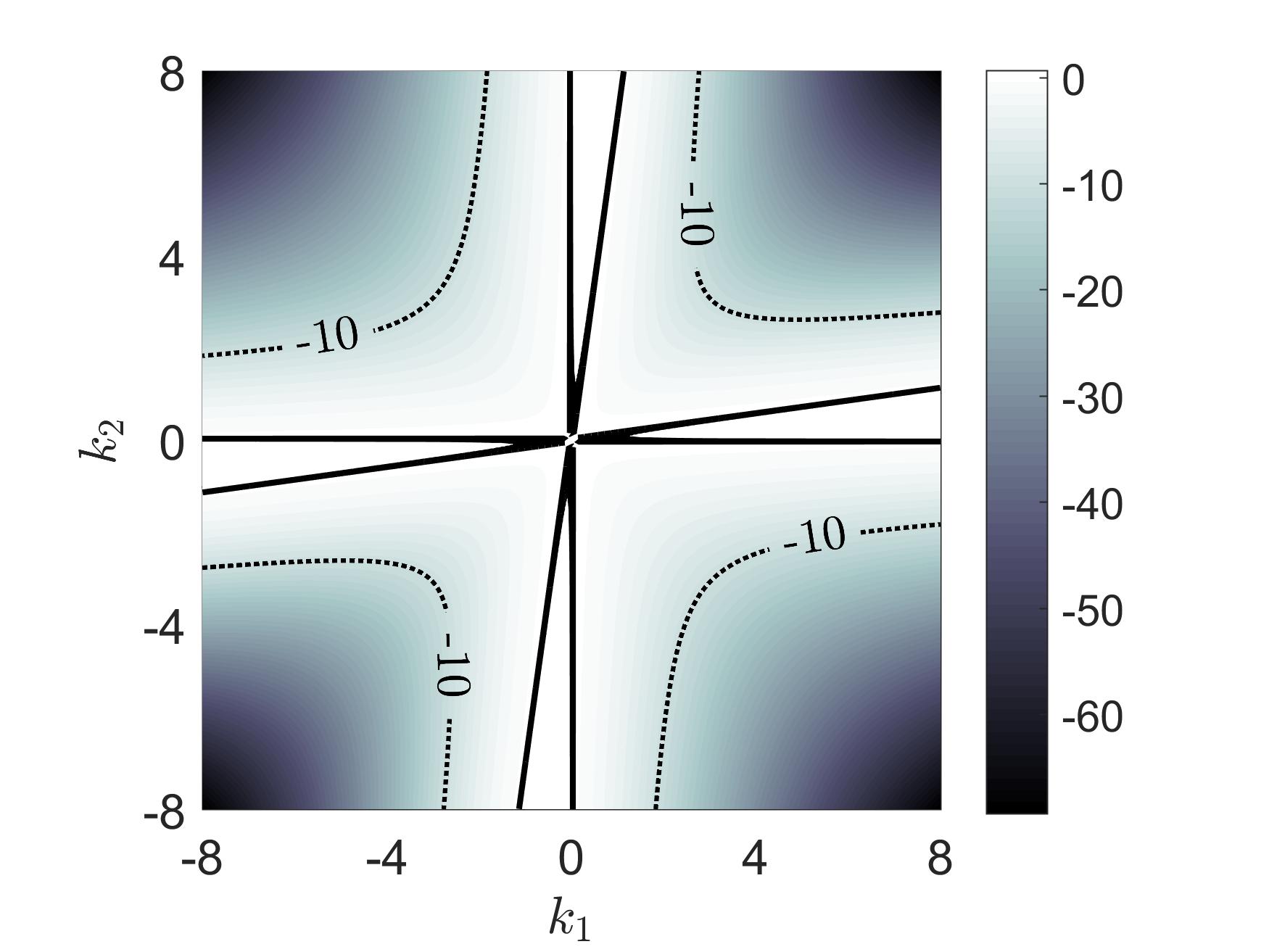}
            \caption[Fig1a]%
            {{\small }}    
            \label{fig: Fig1al}
        \end{subfigure}
        \hfill
        \begin{subfigure}[b]{0.475\textwidth}  
            \centering 
            \includegraphics[width=\textwidth]{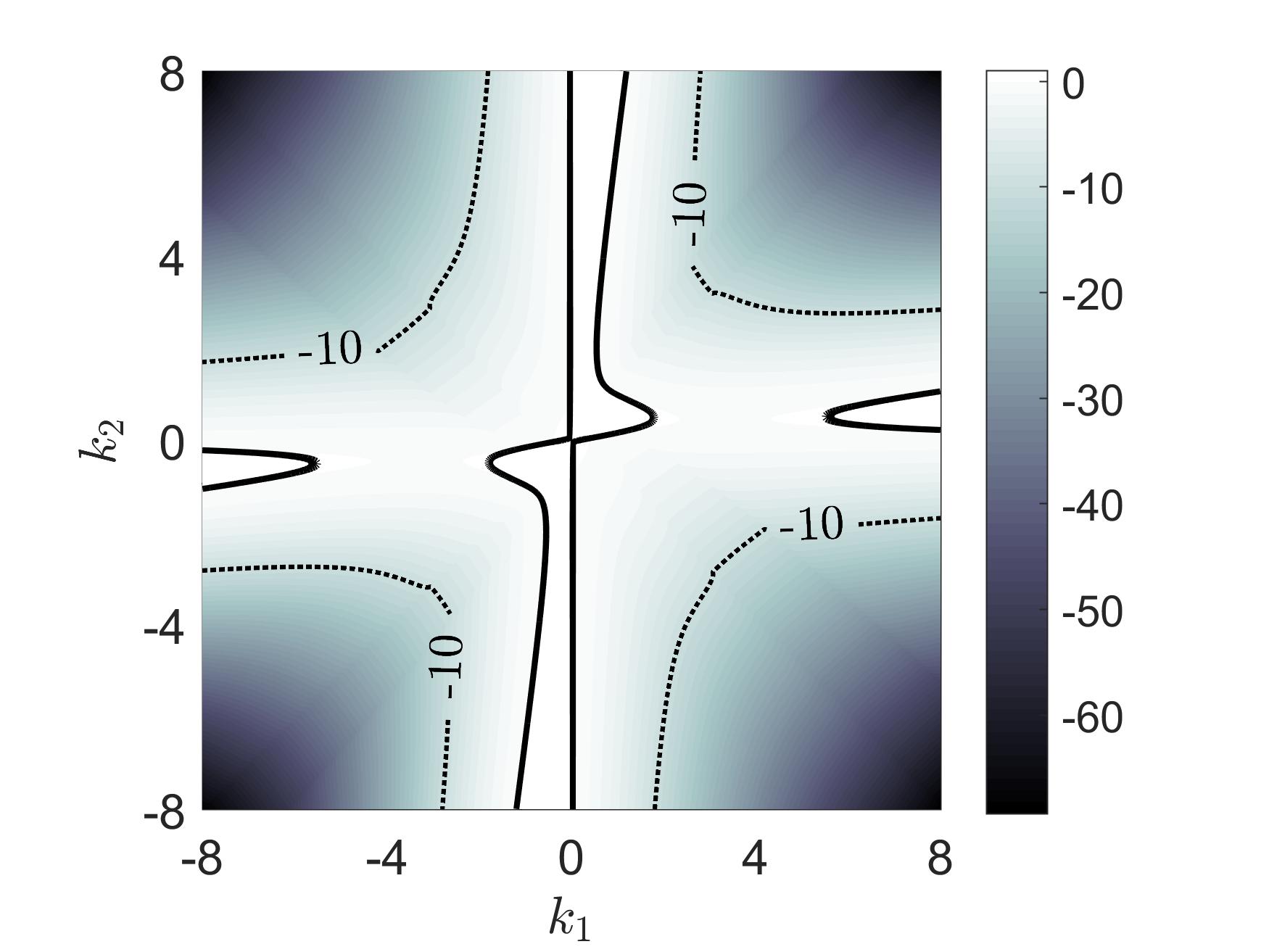}      
            \caption[Fig1b]%
            {{\small }}    
            \label{fig:Fig1bl}
        \end{subfigure}
        \caption[ Fig1b ]
        { Stability diagram for (a) without and (b) with  convection, for $\chi = 0$, with
 $I = 0.0001$,  $\phi = 0.5$, and $p^{(0)} = 1$. 
Solid curves represent neutral stability.}
        \label{fig:typical}
    \end{figure}
%%%%%%%%%%%%%%%%%%%%%%%%%%%%%%%%%%%%%%%%%%%%%%%%%%%%%%

These relations can be put in  polar form, based on the polar representation  
\be\label{axes}
	k = | \bk | = \sqrt{ k_1^2 + k_2^2 } \:\:\mbox{and}\:\: \vartheta = \mbox{tan}^{-1} (\kappa_2 / \kappa_1 ).
\ee
where $k$ is the magnitude of the in-plane wave vector and $\vartheta$ is its angle with the direction of flow.
With $c=\cos\vartheta$ and $s=\sin\vartheta$, the eigenvalue from (\ref{eigen01}) can be  expressed as
\be
	\label{eigen02}
	\begin{split}
	&\lambda = \frac{1}{2 \phi \tilde{\Phi}_1} 
	\left[ \beta \gamma k^2 (c^2 - s^2 )^2 - 2 \tilde{\Phi}_1 \Phi_2 + 2\phi c  (s - \alpha c /\surd{2}) + \tilde{\Phi}_3^{1/2} \right],
	\\\mbox{where}\\
	& \tilde{\Phi}_3 =  \Phi_3 /k^4=	
	\beta^2  \gamma^2 ( c^2  -  s^2 )^4 k^4
	 -    2  \beta \gamma \phi c^2 (c^2  -  s^2 ) 
	( \surd{2} \alpha  -  4cs )  k^2  \\&
	  +    2   \phi^2 c^2   (\alpha c  -  \surd{2} s)^2   
\:\mbox{and}\: \tilde{\Phi}_1 = \Phi_1 / k^2 =  1 - \surd{2} \alpha cs,
\end{split}\ee
in which the trigonometric relations $c^2-s^2 =\cos 2\vartheta$
and $2cs =\sin 2\vartheta$ also apply. \\ 
\subsection{  Transient    Instability}
Here, we consider the modifications of the stability analysis of  Ref.\! 1
by the inclusion of convection  and the vdW-CH gradient terms.  Since that analysis is strictly valid only for the    growth rates inferred from the   initial state, the wave number $\bk$  is to be interpreted here as its initial value $\bkap$.    We shall employ the term ``  transient    instability (or stability)" to denote positive (or negative) growth rates based on these initial values.    The parameters employed in this study are  the values proposed by \cite{Jop05}, namely $\mu_0 =  0.383$, $\mu_{\infty} = 0.643$, and $I_* = 0.279$,  unless otherwise specified.
 \\

Some contours of the initial growth rate, represented by the real part of the eigenvalue $\lambda$ from (\ref{eigen01}),  
are  shown in the stability diagram of Fig.  \ref{fig:typical},
where (a) and (b) represent, respectively, stability with and without  convection ($\bL^{(0)}\equiv {\bf 0}$   in all terms except $I$ ),  for $I = 0.0001$ and $\chi = 0$.
The solid curves represent neutral stability $\lambda=0$,  with lighter 
 zones   representing  unstable regions.
Fig.  \ref{fig:typical}(a) is identical with the result of  Ref.\! 1,  whereas  Fig.  \ref{fig:typical}(b) shows that convection causes strong distortion of the neutral stability curve, eliminating parts  of two unstable branches.
Note that convection  has a more pronounced effect  in the regions of small $k$, as it represents a contribution of order zero in $k$  to (\ref{eigen02}), compared to terms 
 of higher order in   $k$   arising from  the dissipative stresses. \\

\begin{figure} 
        \centering
        \begin{subfigure}[b]{0.45\textwidth}
            \centering
            \includegraphics[width=\textwidth]{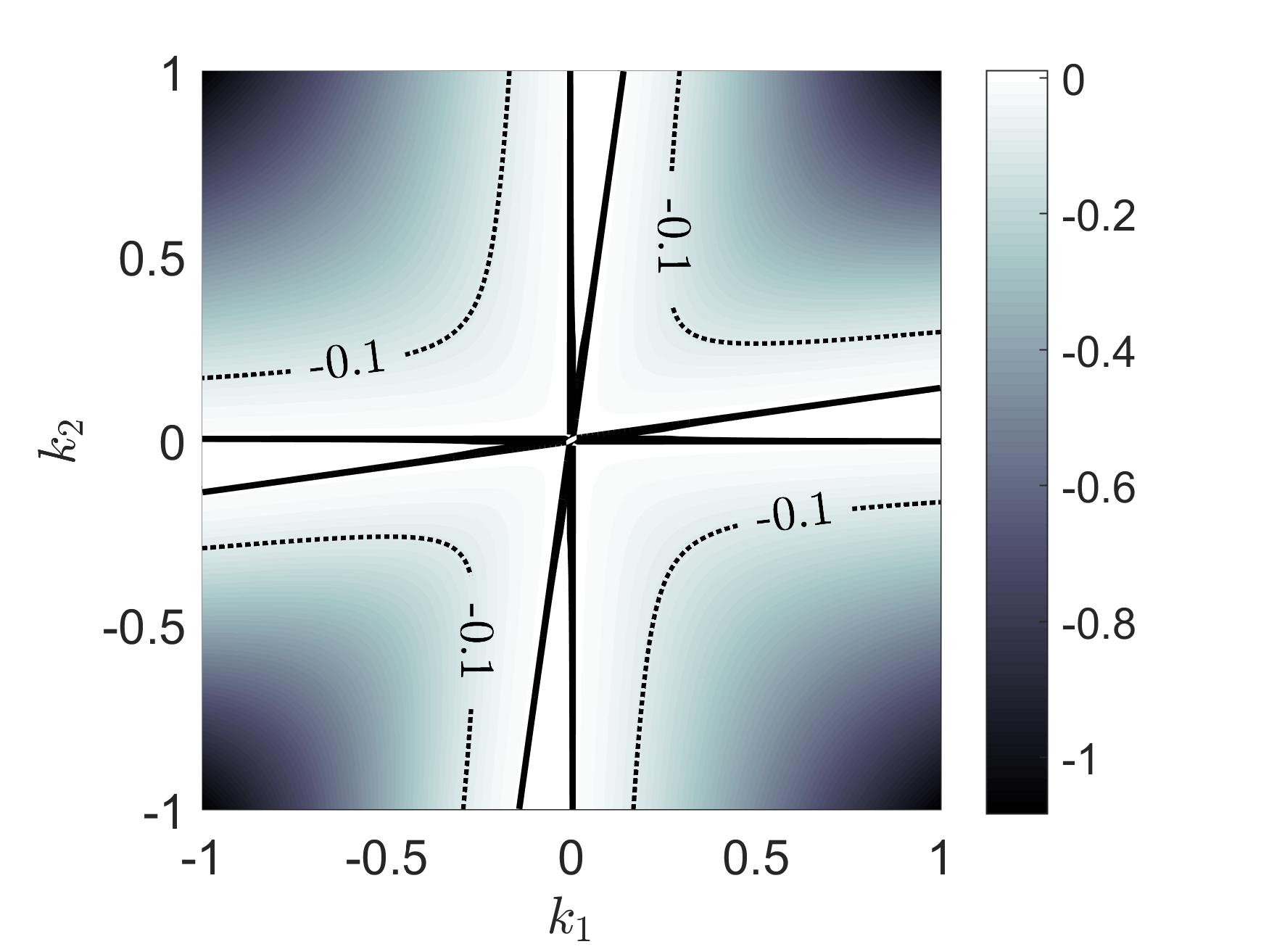}
            \caption[lambda03_OFF(a)]%
            {{\small }}    
            \label{fig:lambda03_OFF(a)}
        \end{subfigure}
        \hfill
        \begin{subfigure}[b]{0.45\textwidth}  
            \centering 
            \includegraphics[width=\textwidth]{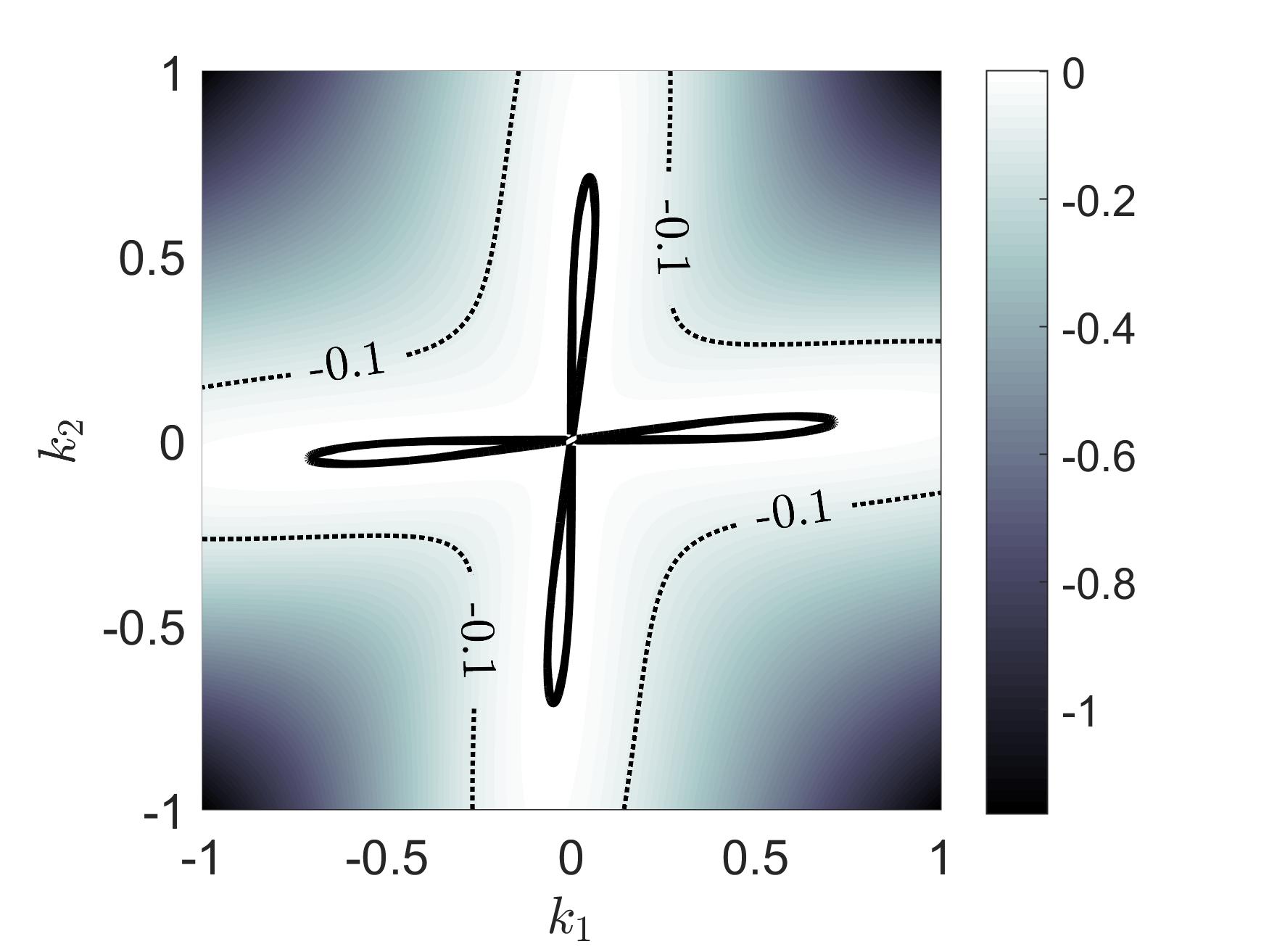}      
            \caption[lambda04_OFF(b)]%
            {{\small }}    
            \label{fig:lambda04_OFF(b)}
        \end{subfigure}
        \caption[ OFF_I_p0001 ]
        {Stability  without convection  for (a)  $\chi = 0$ and (b) $\chi= 0.005$, with  $I = 0.0001$, $\phi = 0.5$,  and $p^{(0)} = 1$. Solid curves represent neutral stability. } 
        \label{fig:OFF_I_p0001}
    \end{figure}

Fig. \ref{fig:OFF_I_p0001} illustrates the effect of $\chi$, where it is obvious
 that the vdW-CH model  provides a wave-number cut-off, shrinking unstable zones accordingly. By choosing very large $\chi$ one can, not surprisingly,  eliminate almost completely the unstable zones. \\
%%%%%%%%%%%%%%%%%%%%%%%%%%%%%%%%%%%%%%%%%%%%%
%%%%%%%%%%%%%%%%%%%%%%%%%%%%%%%%%%%%%%%%%%%%% 
\begin{figure} 
        \centering
        \begin{subfigure}[b]{0.45\textwidth}
            \centering
            \includegraphics[width=\textwidth]{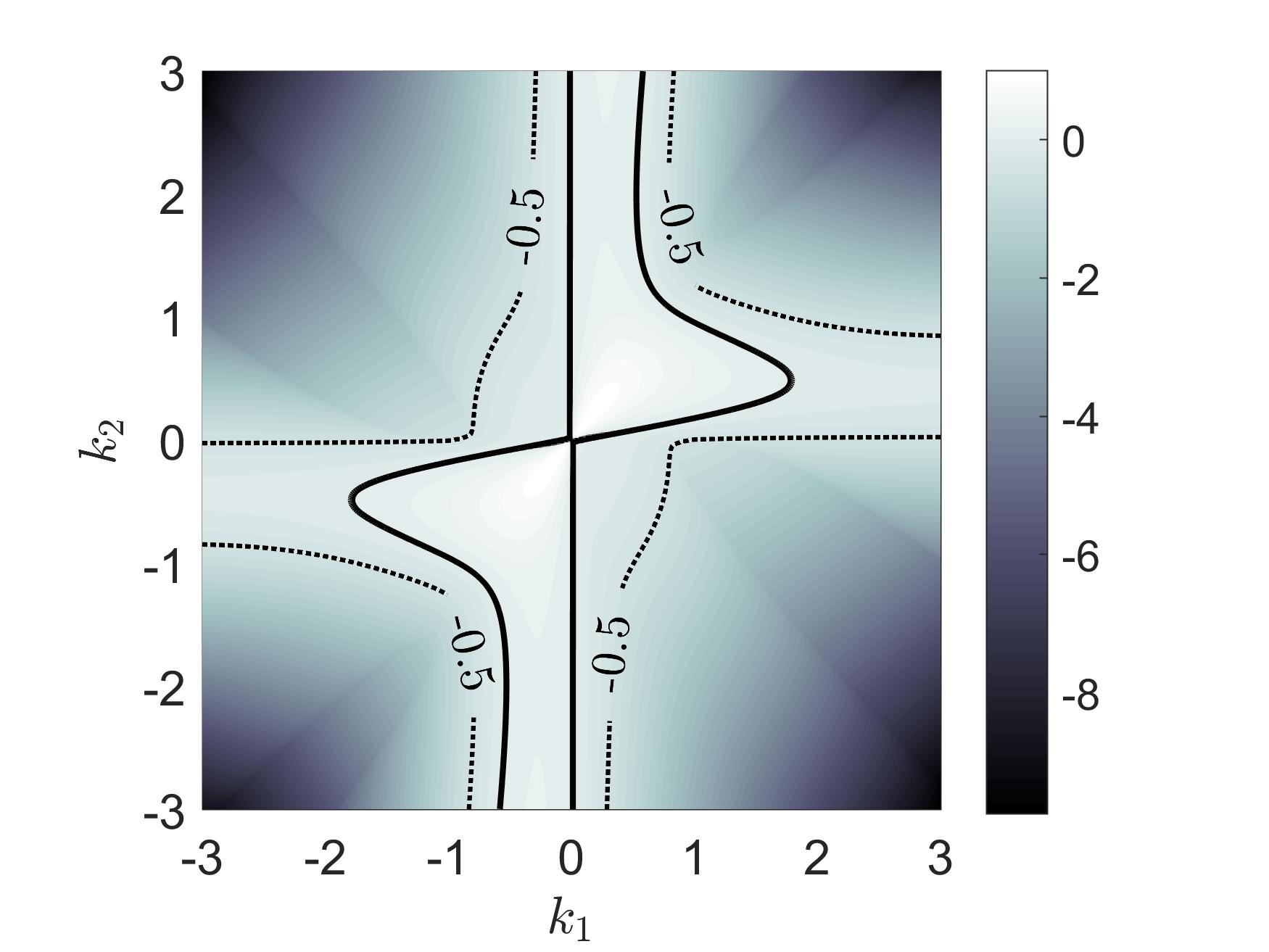}
            \caption[lambda03_ON(a)]%
            {{\small }}    
            \label{fig:lambda03_ON(a)}
        \end{subfigure}
        \hfill
        \begin{subfigure}[b]{0.45\textwidth}  
            \centering 
            \includegraphics[width=\textwidth]{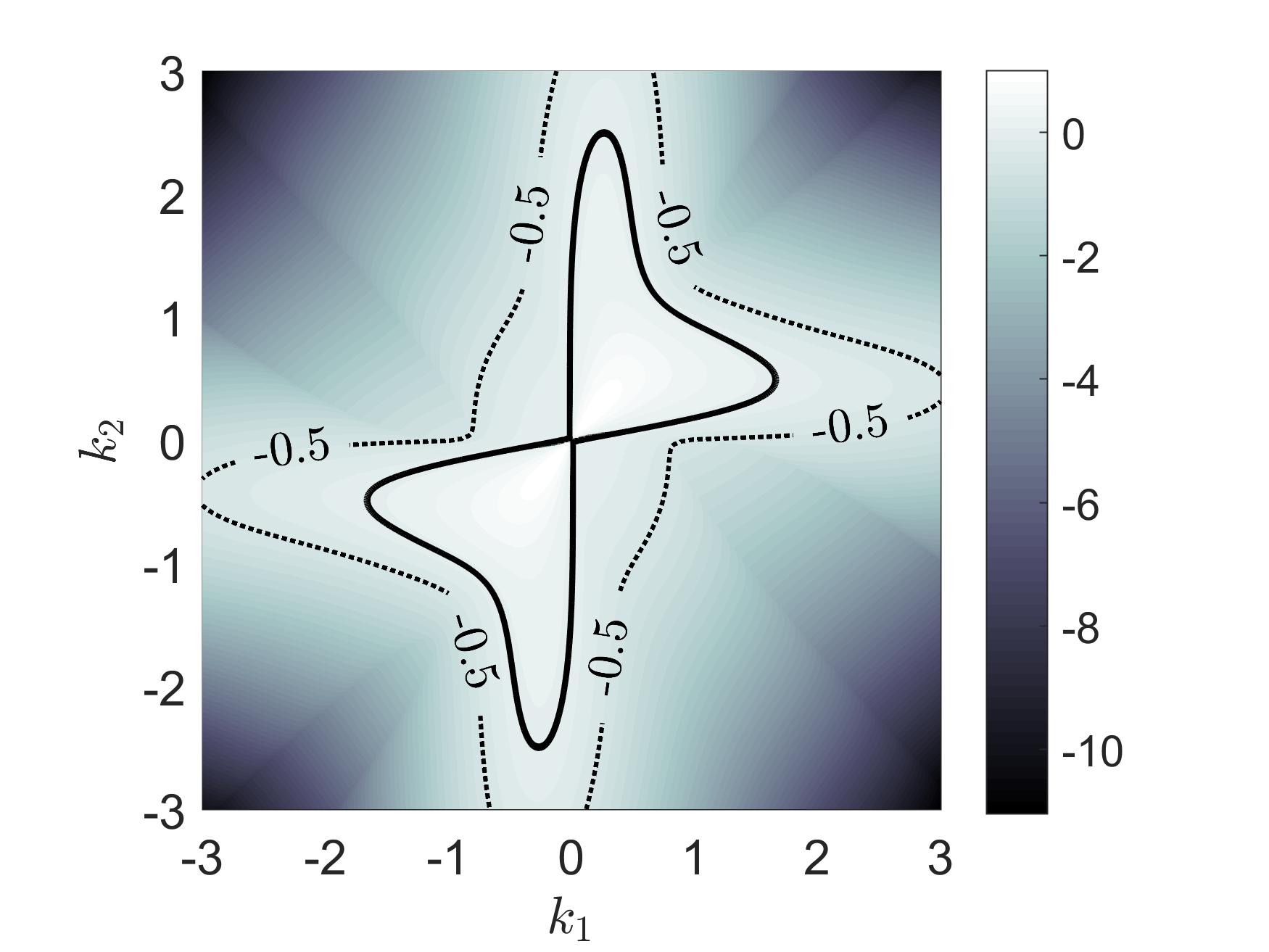}      
            \caption[lambda03_ON(b)]%
            {{\small }}    
            \label{fig:lambda03_ON(b)}
        \end{subfigure}
        \vskip\baselineskip
        \begin{subfigure}[b]{0.45\textwidth}
            \centering
            \includegraphics[width=\textwidth]{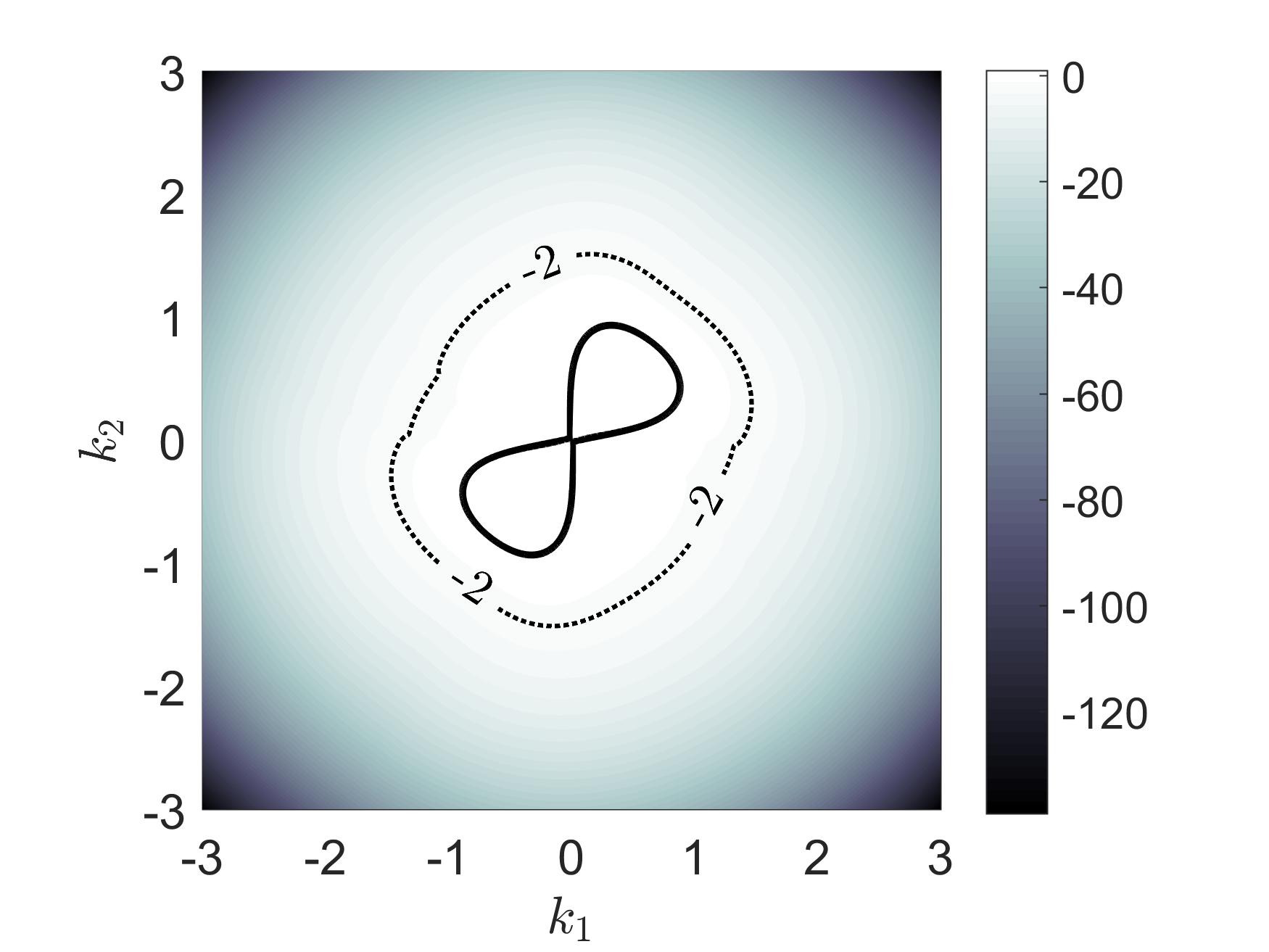}
            \caption[lambda03_ON(c)]%
            {{\small }}    
            \label{fig:lambda03_ON(c)}
        \end{subfigure}
        \hfill
        \begin{subfigure}[b]{0.45\textwidth}  
            \centering 
            \includegraphics[width=\textwidth]{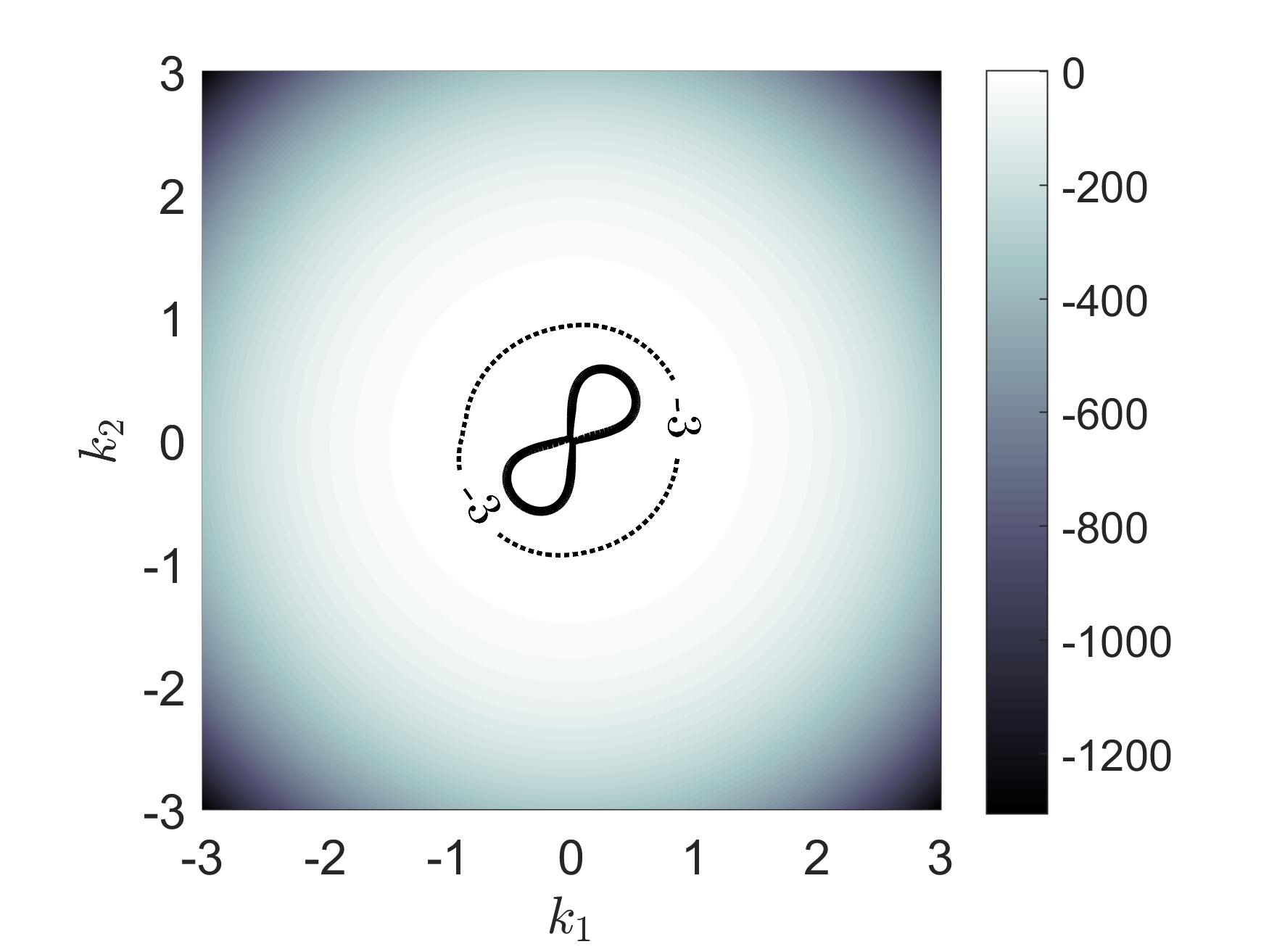}      
            \caption[lambda03_ON(d)]%
            {{\small }}    
            \label{fig:lambda03_ON(d)}
        \end{subfigure}
        \caption[ k-plane_ON ]
        { Stability  with convection  for increasing $\chi$ values: (a) $0$ (b) $0.001$ (c) $ 0.1$ and (d) $1$, with
  $I = 0.0001$,  $\phi = 0.5$, and $p^{(0)} = 1$.}
  \label{fig:k-plane_ON}
    \end{figure}
%%%%%%%%%%%%%%%%%%%%%%%%%%%%%%%%%%%%%%%%%%%%%  
\begin{figure} 
        \centering
        \begin{subfigure}[b]{0.45\textwidth}
            \centering
            \includegraphics[width=\textwidth]{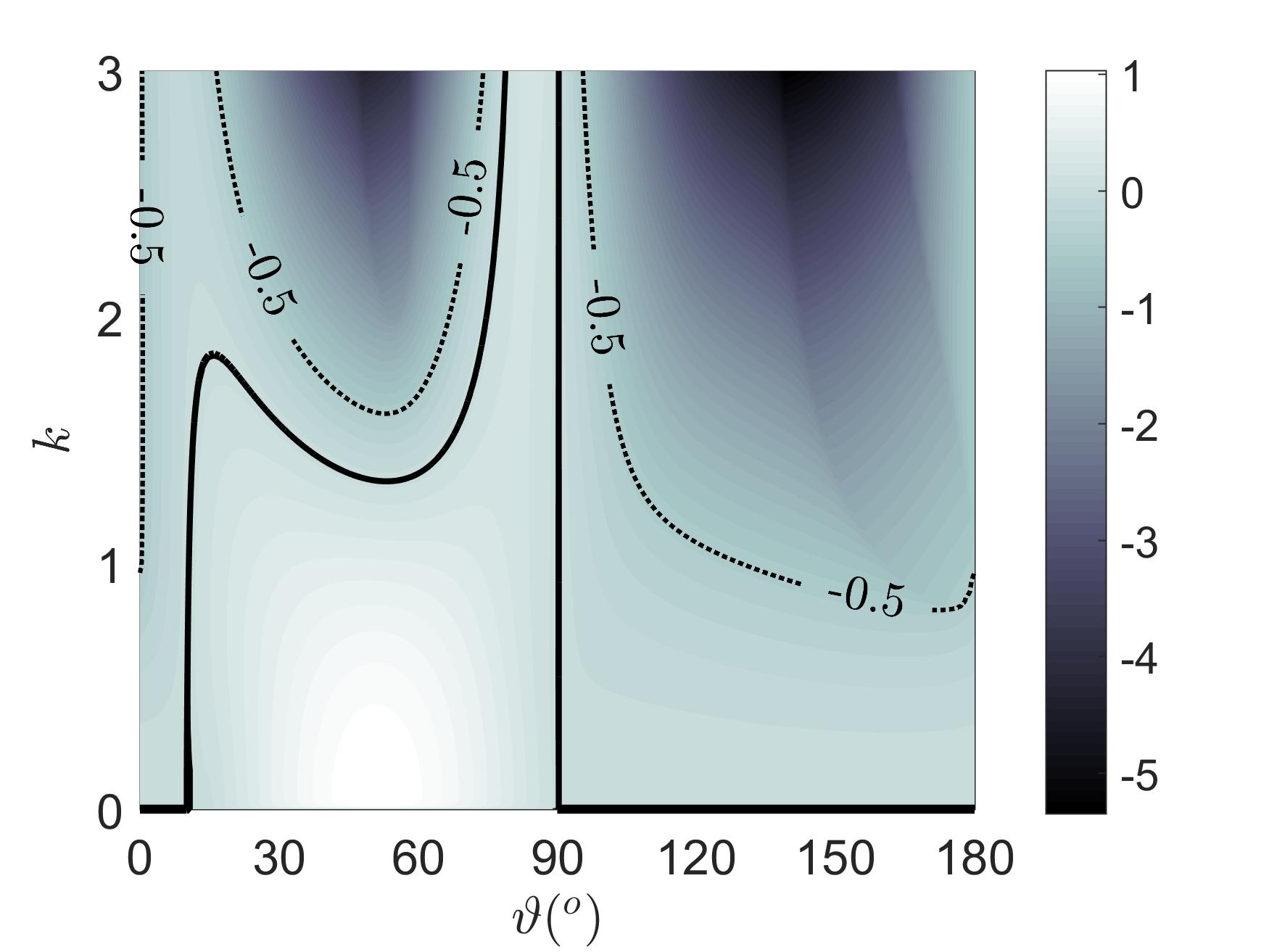}
            \caption[lambda04_ON_I_p0001(a)]%
            {{\small }}    
            \label{fig:lambda04_ON_I_p0001(a)}
        \end{subfigure}
        \hfill
        \begin{subfigure}[b]{0.45\textwidth}  
            \centering 
            \includegraphics[width=\textwidth]{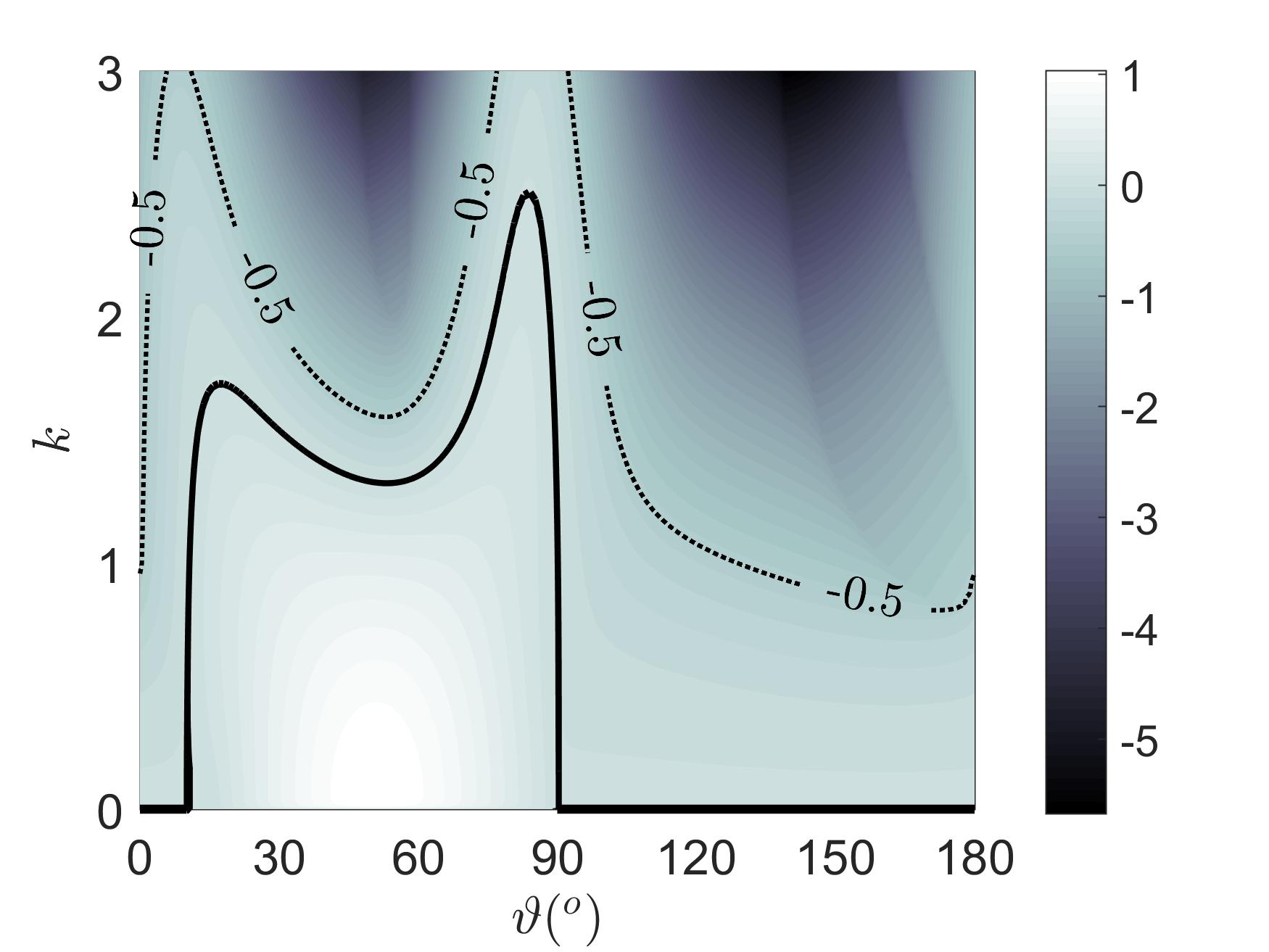}      
            \caption[lambda04_ON_I_p0001(b)]%
            {{\small }}    
            \label{fig:lambda04_ON_I_p0001(b)}
        \end{subfigure}
        \vskip\baselineskip
        \begin{subfigure}[b]{0.45\textwidth}
            \centering
            \includegraphics[width=\textwidth]{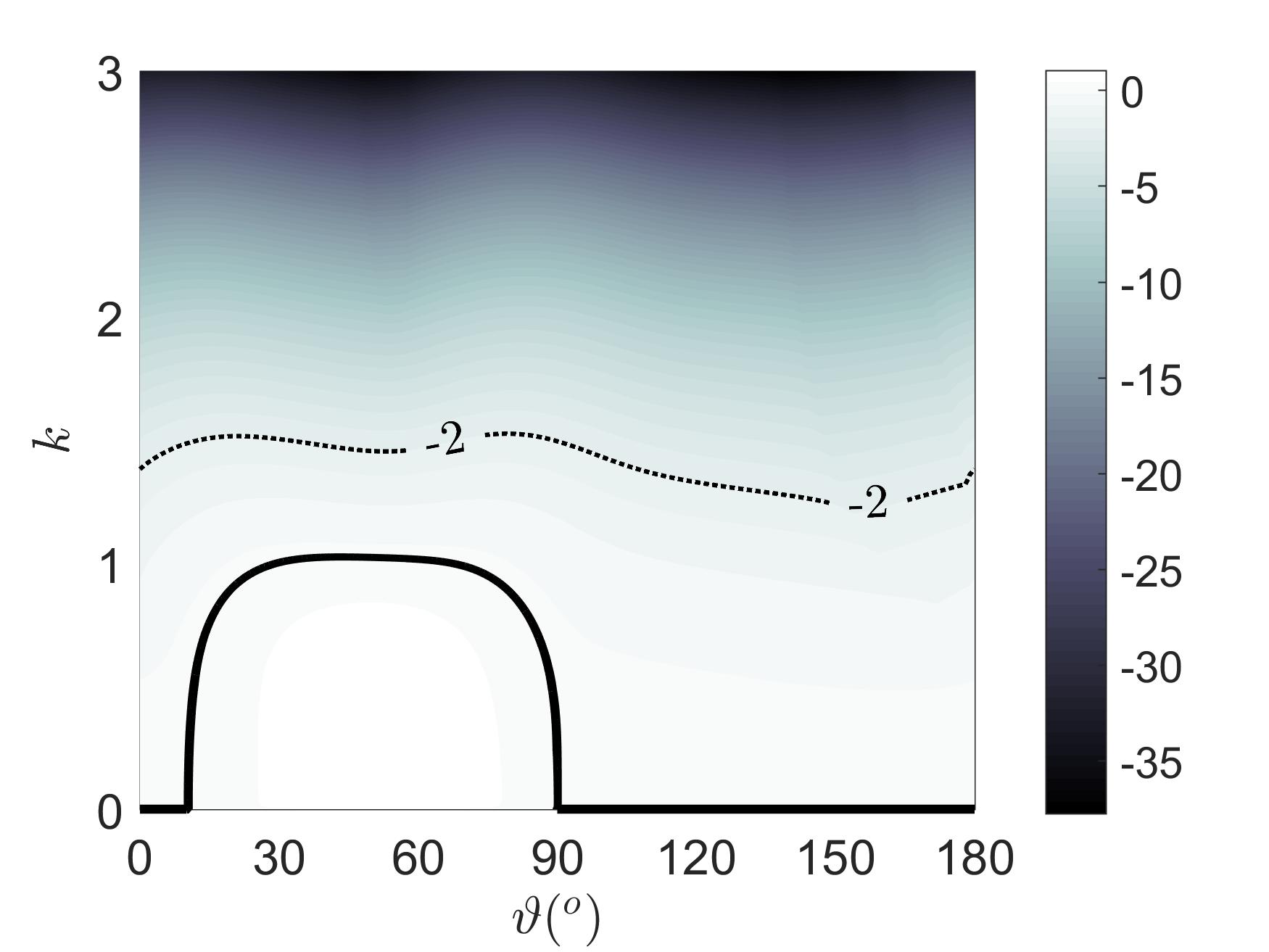}
            \caption[lambda04_ON_I_p0001(c)]%
            {{\small }}    
            \label{fig:lambda04_ON_I_p0001(c)}
        \end{subfigure}
        \hfill
        \begin{subfigure}[b]{0.45\textwidth}  
            \centering 
            \includegraphics[width=\textwidth]{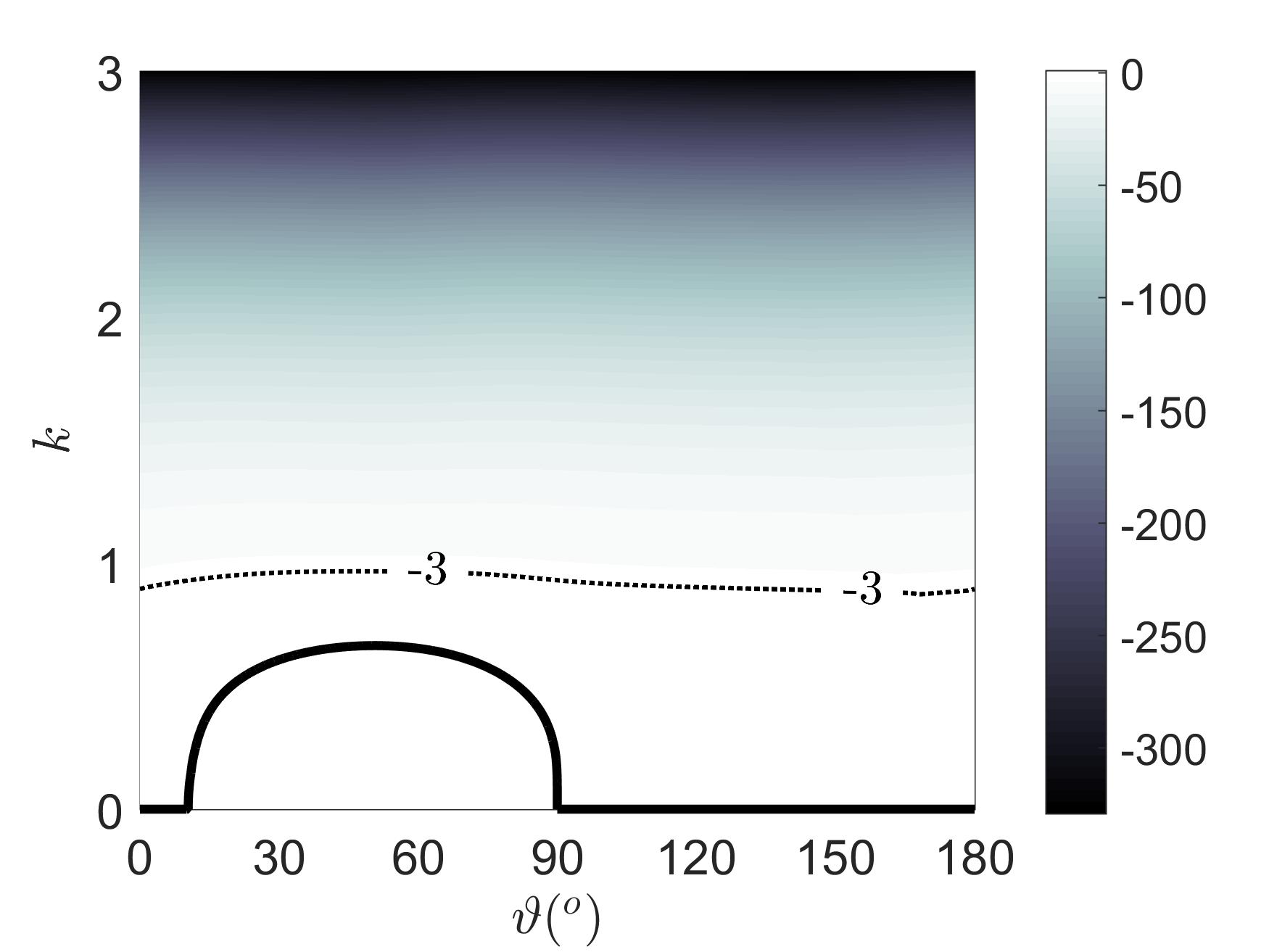}      
            \caption[lambda04_ON_I_p0001(d)]%
            {{\small }}    
            \label{fig:lambda04_ON_I_p0001(d)}
        \end{subfigure}
        \caption[ k-vartheta_ON_I_p0001 ]
        {Fig. \ref {fig:k-plane_ON} in polar variables.}
        \label{fig:k-vartheta_ON_I_p0001}
    \end{figure}
%%%%%%%%%%%%%%%%%%%%%%%%%%%%%%%%%%%%%%%%%%%%%

Fig. \ref{fig:k-plane_ON}
shows the combined effects of convection and wave-number cut-off  for various values of $\chi$, with $I= 0.0001$. It is once again clear that  the region of  instability is  greatly   reduced by increasing $\chi$. 
This is perhaps made clearer by the plots in terms of polar variables 
 in Fig. \ref{fig:k-vartheta_ON_I_p0001}. Note that the growth rates in $0 \le \vartheta < \pi $  are also repeated in $\pi \le \vartheta < 2\pi $ when convection is included. \\
 
As discussed above, oscillatory behavior with   $\mbox{Im}(\lambda)\ne 0$  arises for negative $\Phi_3$ and, writing  $X=k^2 > 0 $, we see from (\ref{eigen02}) that Re$(\lambda)$ and $\Phi_3$ are given by separate quadratic polynomials in $X$. Then, in order to have 
undamped oscillation, we must require the simultaneous conditions
\be \label{simult}  \Phi_3 < 0 \:\: \mbox{and}\:\:{\rm Re}(\lambda) \ge 0,\:\:\mbox{for} \:\: X>0, \ee
on the these two quadratics. While we have not been able to provide a straight-forward and rigorous algebraic proof of the impossibility of (\ref{simult}), a detailed numerical investigation for various values of $k$ at closely spaced increments of $\vartheta$ in $(0, 2\pi)$ and for various values of $\chi$ 
$I$ failed to achieve the condition. Hence, we are led to conjecture that oscillatory 
solutions are always damped. \\

\begin{figure} 
        \centering
              \includegraphics[width=0.70\textwidth]{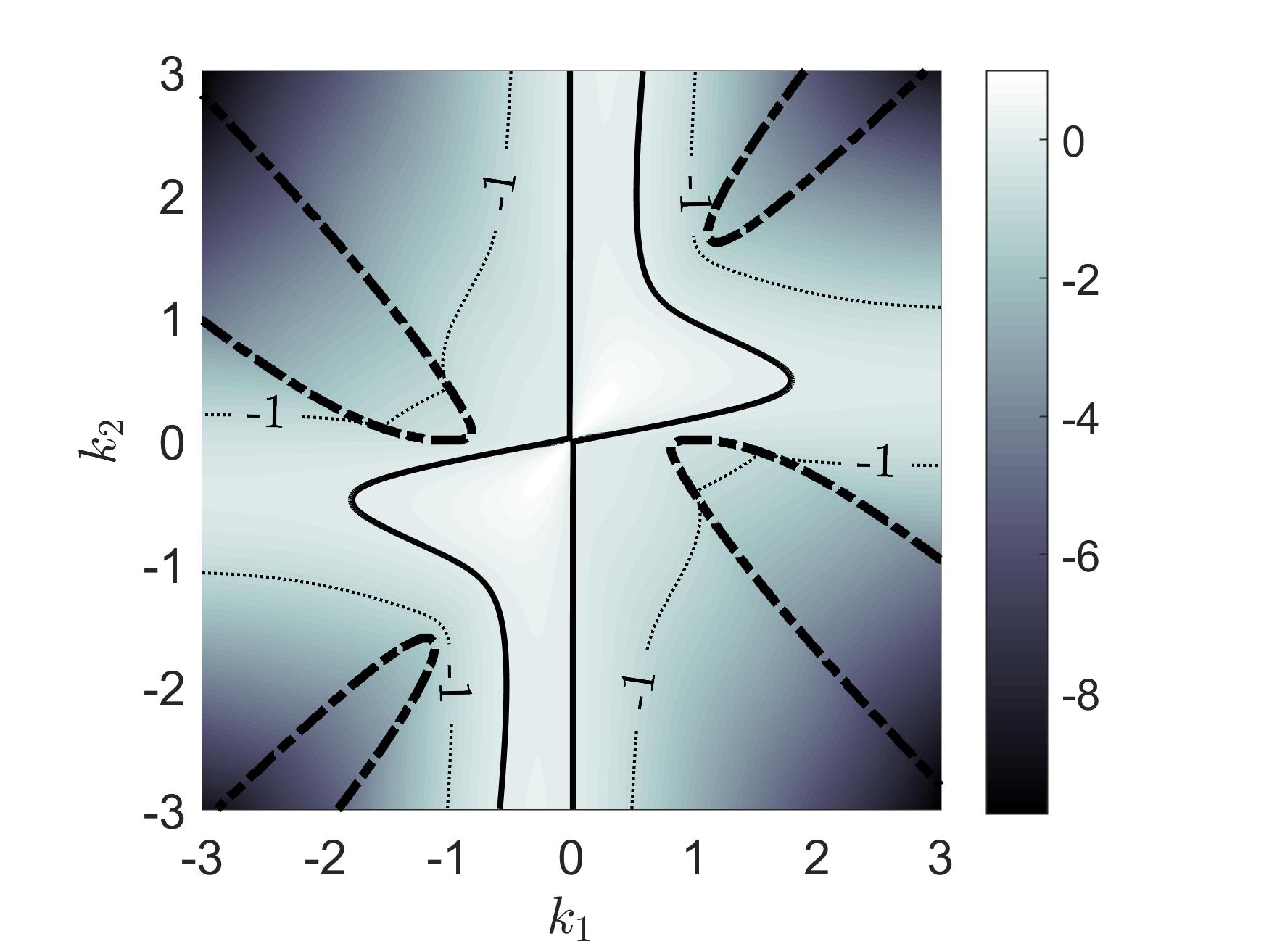}     
        \caption[ lambda_imag ]
        { 
Oscillatory behavior resulting from  convection,  for $\chi = 0$, $I = 0.0001$, $p^{(0)} = 1$,   and $\phi = 0.5$. 
The oscillation frequency is non-zero within finger-like regions bounded by dashed curves    and equal to zero upon and outside these curves. The solid and dotted curves are contours of the real part of $\lambda$, the former representing  neutral stability.}
        \label{fig:lambda05_imag_ON}
    \end{figure}

Fig.  \ref{fig:lambda05_imag_ON} shows the corresponding imaginary part of the eigenvalue $\lambda$ for a particular set of  parameter values,
where finger-like regions bounded by the dashed lines contain the non-zero imaginary values of $\lambda$.  Also shown are two contours of Re$(\lambda)$, the neutral stability 
curve and a curve representing damping. In line with the above conjecture, the oscillatory 
behavior is damped.   \\

\begin{figure} 
        \centering
        \begin{subfigure}[b]{.475\textwidth}
            \centering
            \includegraphics[width=\textwidth]{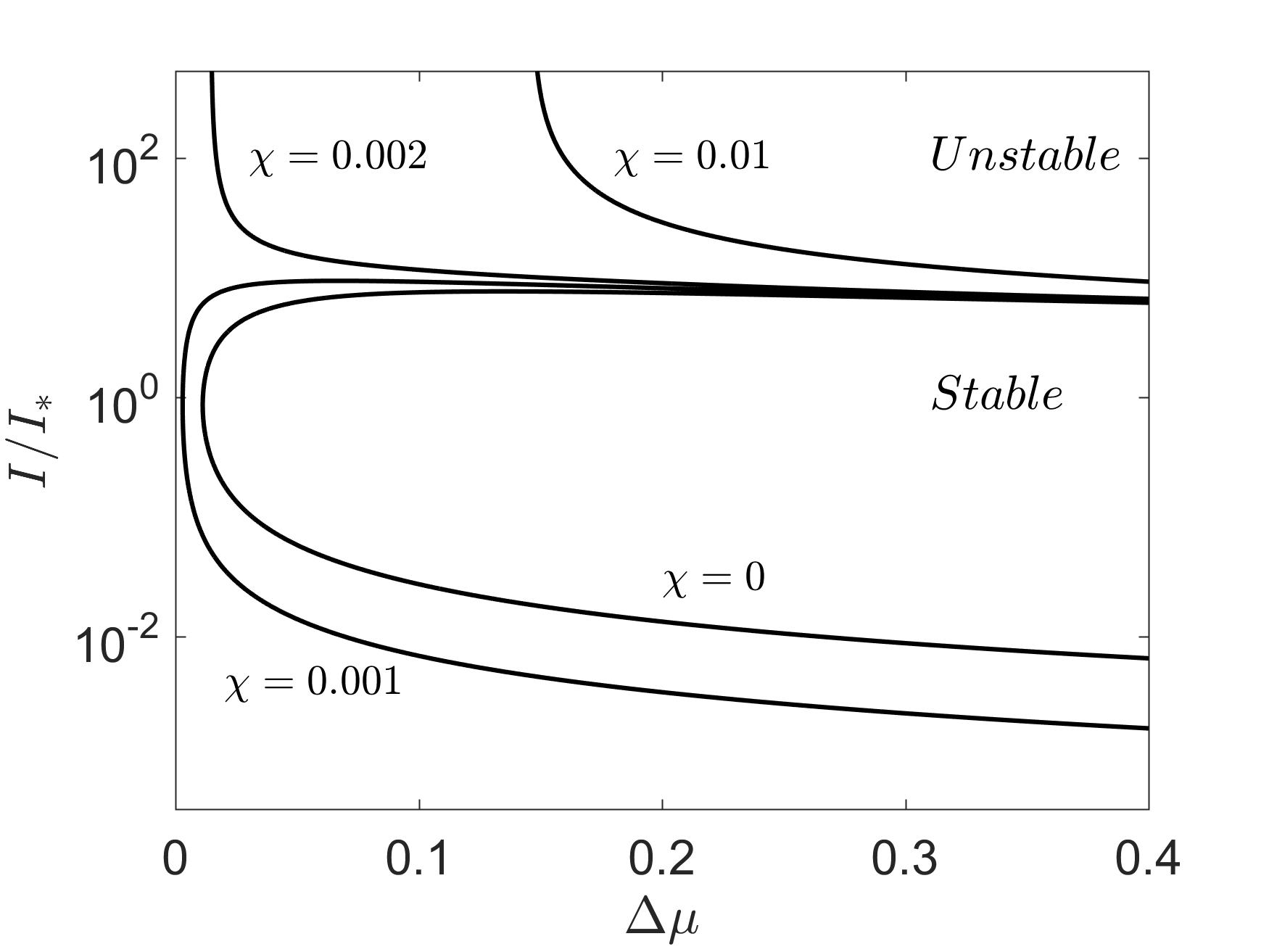}
            \caption[lambda06_OFF]%
            {{\small }}    
            \label{fig:lambda06_OFF}
        \end{subfigure}%
       % \vskip\baselineskip
        \begin{subfigure}[b]{0.475\textwidth}  
            \centering 
            \includegraphics[width=\textwidth]{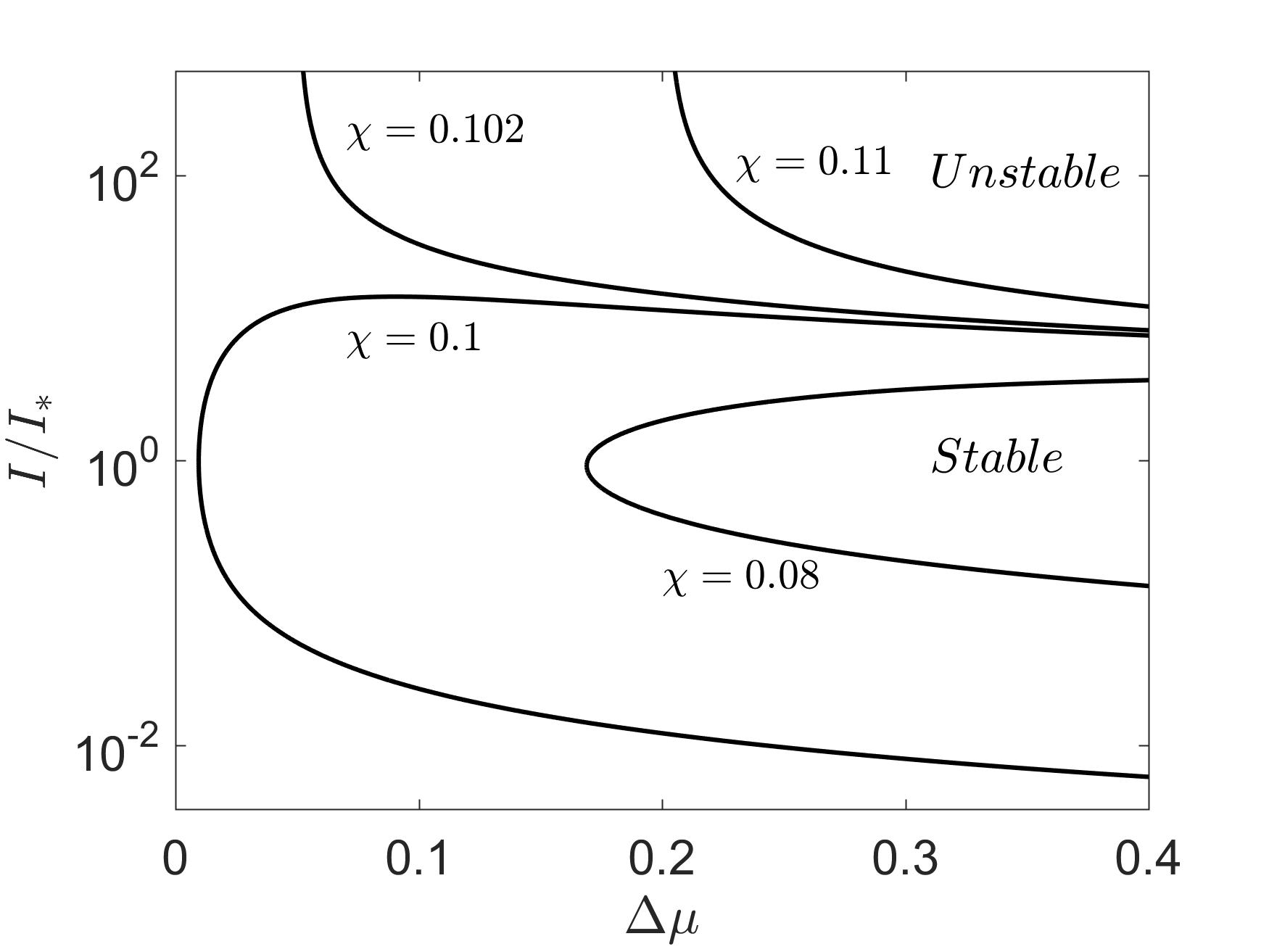}      
            \caption[lambda06_ON]%
            {{\small }}    
            \label{fig:lambda06_ON}
        \end{subfigure}
        \caption[ I-Deltamu ]%
        { Stability in the  $I$-$\Delta\mu$ plane,  (a) without and  (b) with convection,  
showing the effect of $\chi$,  where $\bk = (0.1, 0.8)$  ($k = 0.806, \:\:\vartheta = 82.9^\circ$ ),
 $ p^{(0)} = 1$,  and $\phi = 0.5$.} 
        \label{fig:I-Deltamu}
    \end{figure}

Following  Ref.\! 1, 
we present in Fig.  \ref{fig:I-Deltamu} a stability diagram in the $I$-$\Delta\mu$ plane  to show the effect of $\chi$ with and without  convection, for $\bk = (0.1,0.8)$ (or $ k = 0.806, \:\:\vartheta = 82.9^\circ )$. 
The curves shown there are  loci of neutral stability and  regions inside the closed curves or beneath the open curves are the stable
regions.  The curve $\chi=0$ in Fig. \ref{fig:lambda06_OFF} is a magnified version of that given in  Ref.\! 1, indicating 
that the perfectly-plastic case of constant $\mu$, with  $\Delta\mu=0$ is always unstable. However,  this instability is apparently removed at a critical $\chi$ value between 0.001 and 0.002, where the closed curve opens up to form an unbounded stable region. Interestingly, this critical $\chi$ value is nearly ten times larger when convection is included, as illustrated by Fig. \ref{fig:lambda06_ON}. \\

   In summary, the preceding results indicate that convection distorts the regions of   transient    instability, completely eliminating instability for some directions of $\bk$  while leading to short wavelength instability for others, in the absence of cut-off by the gradient terms in $\chi$.   
    %%%%%%%%%%%%%%%%%%%%%%
%%%%%%%%%%%%%%%%%%%
    
%%%%%%%%%%%%%%%%%%%%%%%%%

\subsection{   Asymptotic behavior}

According to (\ref{wavevector}), for large $t$ and $ \kappa_1 \ne 0$ we have
  $|k_2 | >> |k_1|$, resulting in the following asymptotic forms: 
\be 
	\label{Aasym}
	\begin{split} 
	&A_{11} \sim k_2^2 [ (\beta -1) \gamma- 2\chi k_2^2]/\phi,\:\: 
	A_{12} \sim - A_{12}\sim \beta\gamma k_1k_2/\phi, \\ 
	 &
	A_{22} \sim -k_2^2[\gamma + 2\chi k_2^2]/\phi , \:\:\mbox{and}\:\: \lambda\sim A_{11},
	\end{split} 
\ee
up to terms of relative magnitude   $O(|k_2|^{-1})$ .  
 Since   $\beta\!-\!1 = \!-\!\dmu \:\: \le  0$, for $\mu_\infty  \ge  \mu_0, $    it follows from (\ref{Aasym}) that  the local growth rate for   $\chi \ge 0$ is   never positive   for large $t$ whenever $\kappa_1\ne 0$. In the exceptional case $\kappa_1=0$, where $\bk(t) \equiv \bkap =(0, \kappa_2)$, it is easy to show by means of  (\ref{Acomp}) -(\ref{eigen01}) 
  that $\lambda$ is  never positive for $\chi\ge0$. Therefore, we conclude that any   transient    instability is eventually killed off by wave vector stretching, irrespective of vdW-CH regularization.  The resulting asymptotic stability is illustrated 
 by a numerical solution of (\ref{stability}). For the incompressible planar flows considered here, this is facilitated by the following  closed-form solution for  the stream function.   

\subsubsection{Scalar solution to (\ref{stability})}\label{sec:closed}
For planar flow, the condition of incompressibility can be expressed in the usual way in terms of a stream function $\uppsi(x_1, x_2)$ as
\be  v_1 = \partial_{x_2}\uppsi, \:\: v_2 = - \partial_{x_1}\uppsi \ee 
which in terms of Fourier transforms become 
\be\label{kbot}\begin{split}& \hat{v}_1= \imath k_2\hat{\uppsi}, \:\: \hat{v}_2 =-\imath k_1 \hat{\uppsi}, \:\: \mbox{or} \:\: \hat{\bv} = \imath \bk^\bot\hat{\uppsi}, \\ &\mbox{where} \:\:\bk^\bot =(k_2, -k_1)= {\bf Q}\bk, \:\: {\bf Q}= 
\left[
\begin{array}{rr}
0  & 1     \\
 -1 &  0    
\end{array}
\right], \end{split} \ee ${\bf Q}$ is orthogonal, and $\bk^\bot$ is identical with the flipped wave vector introduced in Ref.\! 1.
It is easy to see that the preceding linear relations also apply to perturbations. 
Hence, after  making use of $\diff \bk/\!\!\diff t= -{\bL}^{(0)T}\bk$  the ODE 
can be reduced to the form 
\be\label{stability1} \bk\frac{ \diff\hat{\uppsi} }{\diff t}= {\bf B}\bk \hat{\uppsi},
\:\:\mbox{where}\:\:{\bf B}= {\bf Q}^T \bA {\bf Q}+ \bL^{(0)T}\ee

Now, $\bk$ and $\bk^\bot$ serve as orthogonal basis for 2D vectors and it is easy to 
show that $\bk^\bot\ccdot{\bf B}\bk=0$, i.e. the right hand side of (\ref{stability1}) 
has  zero component in  direction $\bk^\bot$. By Fredholm's theorem,  
applied implicitly in Ref.\! 1, this establishes that $\bk$ is in the range of ${\bf B}$ and, hence, is an eigenvector of ${\bf B}$.  Thus, (\ref{stability1}) involves only 
 components in direction $\bk$, given by  orthogonal projection as:
\be\label{stability2}
	\frac{ \diff\hat{\uppsi} }{\diff t}= \Lambda(t, \bkap)\hat{\uppsi}, \:\: \mbox{where}\:\:
	\Lambda = \frac{\bk\ccdot{\bf B}\bk}{k^2},
\ee
where 
\be\label{bB}  \Lambda =\frac{1}{\phi k^2\Phi_1} \left\{ \phi\left[(k_1 k_2 (k^2+\Phi_1)-\surd{2}\:\alpha k_1^4\right]+\beta\gamma k^2 (k_1^2 - k_2^2)^2
-\Phi_1\Phi_2 k^2\right\}
\ee
is the eigenvalue for eigenvector $\bk$ of ${\bf B}$ .\\  

It follows that the solution to the linear stability problem for constant $\bL^{(0)}$ is given  by
\be\label{closed} \hat{\bv}^{(1)} = \hat{\uppsi}_0 {\bf Q}\exp\left\{ \int_0^t \Lambda(t', \bkap)\diff t' \right\}\bk=\hat{\uppsi}_0 {\bf Q}\exp\left\{\bI\! \int_0^t \Lambda(t', \bkap)\diff t' - \bL^{(0)T}t\right\}\bkap,\ee 
which gives
\be\label{absv}|\hat{\bv}^{(1)}|= |\hat{\uppsi}_0|k\exp\left\{ \int_0^t \Lambda(t', \bkap)\diff t' \right\}, \ee
where, in all the above formulae, 
 \be\label{kvec} \bk(t, \bkap)=  \exp\left\{-\bL^{(0)T}t\right\}\bkap, \:\:\mbox{with}\:\: k = \left[ \bkap\ccdot\exp\left\{-\bL^{(0)}t \right\} \exp\left\{-\bL^{(0)T}t\right\}\bkap\right]^{1/2} \ee
 It is easy to show that (\ref{absv}) gives the same result as that obtained by substitution of 
$\hat{\bv}^{(1)}=\uppsi\bk^\bot$ into (\ref{energy}). Moreover, when convection is neglected one finds that $\bk\ccdot{\bf B}\bk= \bk^\bot\ccdot \bA\bk^\bot$, and, hence, that $\bk^\bot$ is an eigenvector of $\bA$ with eigenvalue $\Lambda$, as already found in  Ref.\! 1. \\

    Making use of the asymptotic form (\ref{Aasym}), we find 
for large $k$ with $|k_2| >> |k_1|$ that 
\be \label{asymm} \Lambda \sim \frac{k_2^2}{\phi\tilde{\Phi}_1} [(\beta-\tilde{\Phi}_1)\gamma-2\tilde{\Phi}_1\chi k_2^2]\sim\lambda,\:\:\mbox{with} \:\:\tilde{\Phi}_1 = \Phi_1 / k^2\sim\Phi_1 / k_2^2, \:\:\mbox{for}\:\: t \rightarrow \infty,\:\: \ee
where $\lambda$ is the eigenvalue of $\bA$ with largest real part, which confirms the asymptotic behavior inferred previously from (\ref{Aasym}).\\

 \begin{figure} 
        \centering
        \begin{subfigure}[b]{0.45\textwidth}
            \centering
            \includegraphics[width=\textwidth]{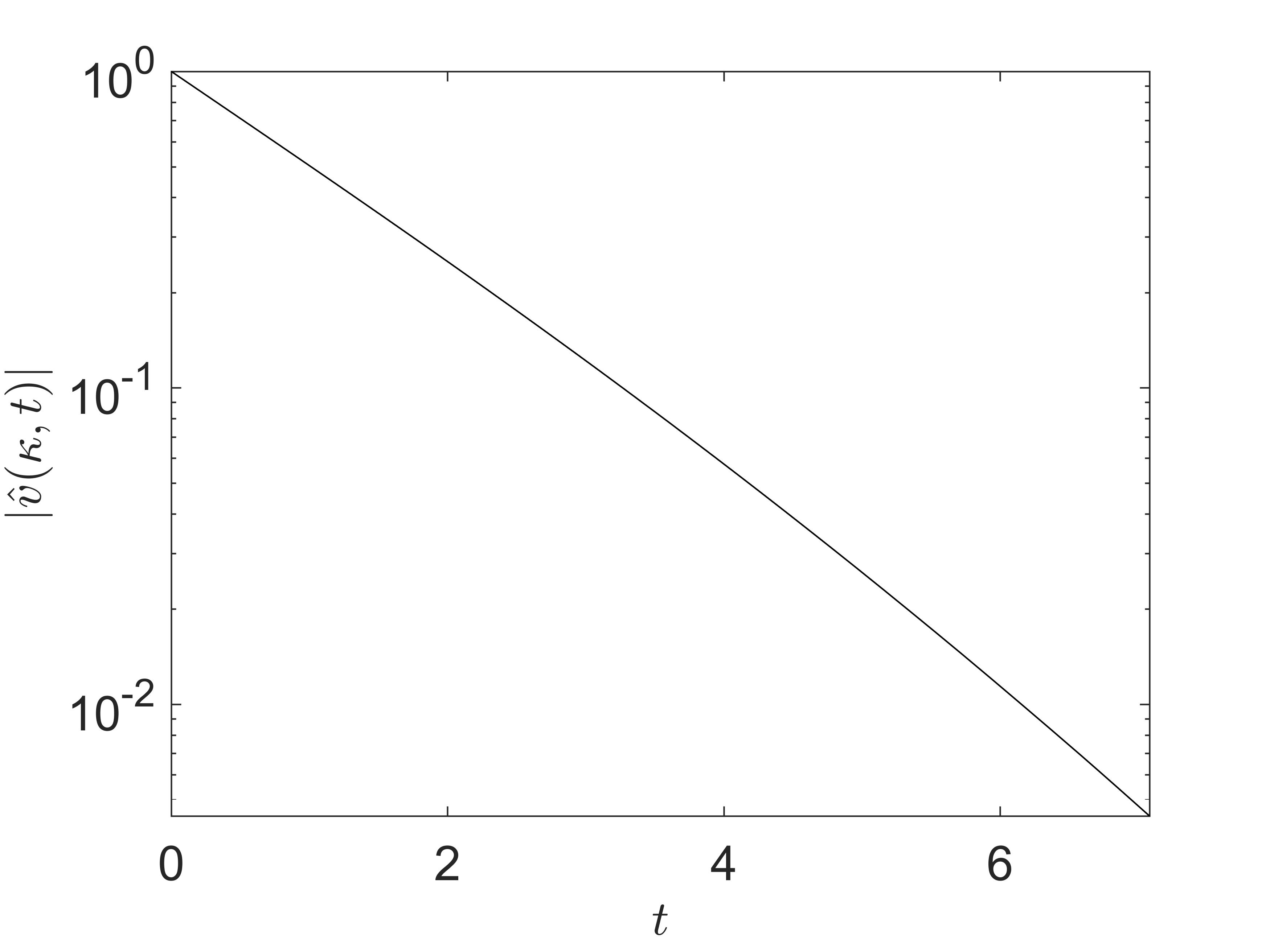}
            \caption[]%
            {{\small }}    
            \label{fig:kappas(a)}
        \end{subfigure}
 %       \hfill
	\quad 
       \begin{subfigure}[b]{0.45\textwidth}  
            \centering 
            \includegraphics[width=\textwidth]{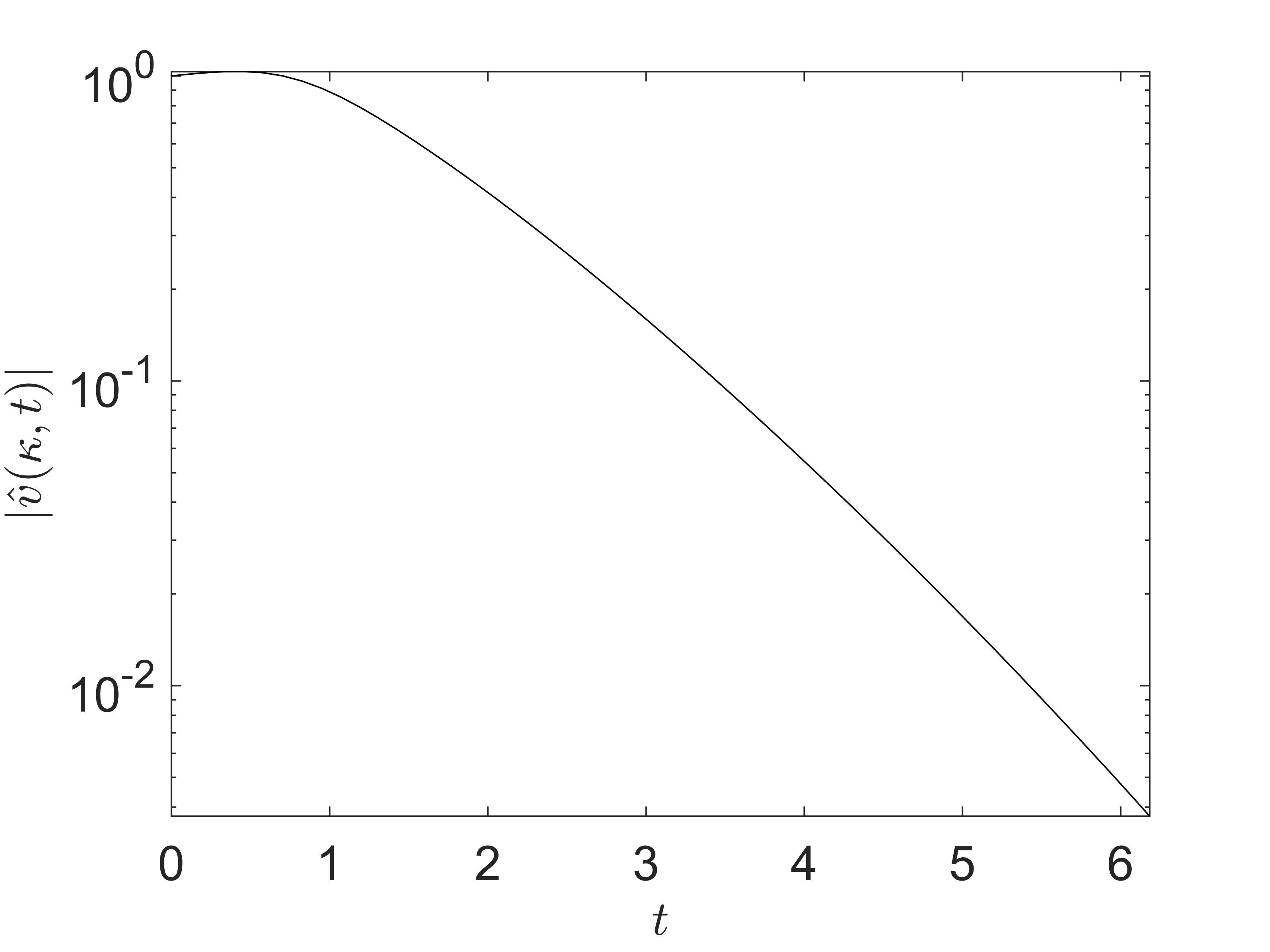}
            \caption[]%
            {{\small }}    
            \label{fig:kappas(b)}
        \end{subfigure}
        \vskip\baselineskip
        \begin{subfigure}[b]{0.45\textwidth}   
            \centering 
            \includegraphics[width=\textwidth]{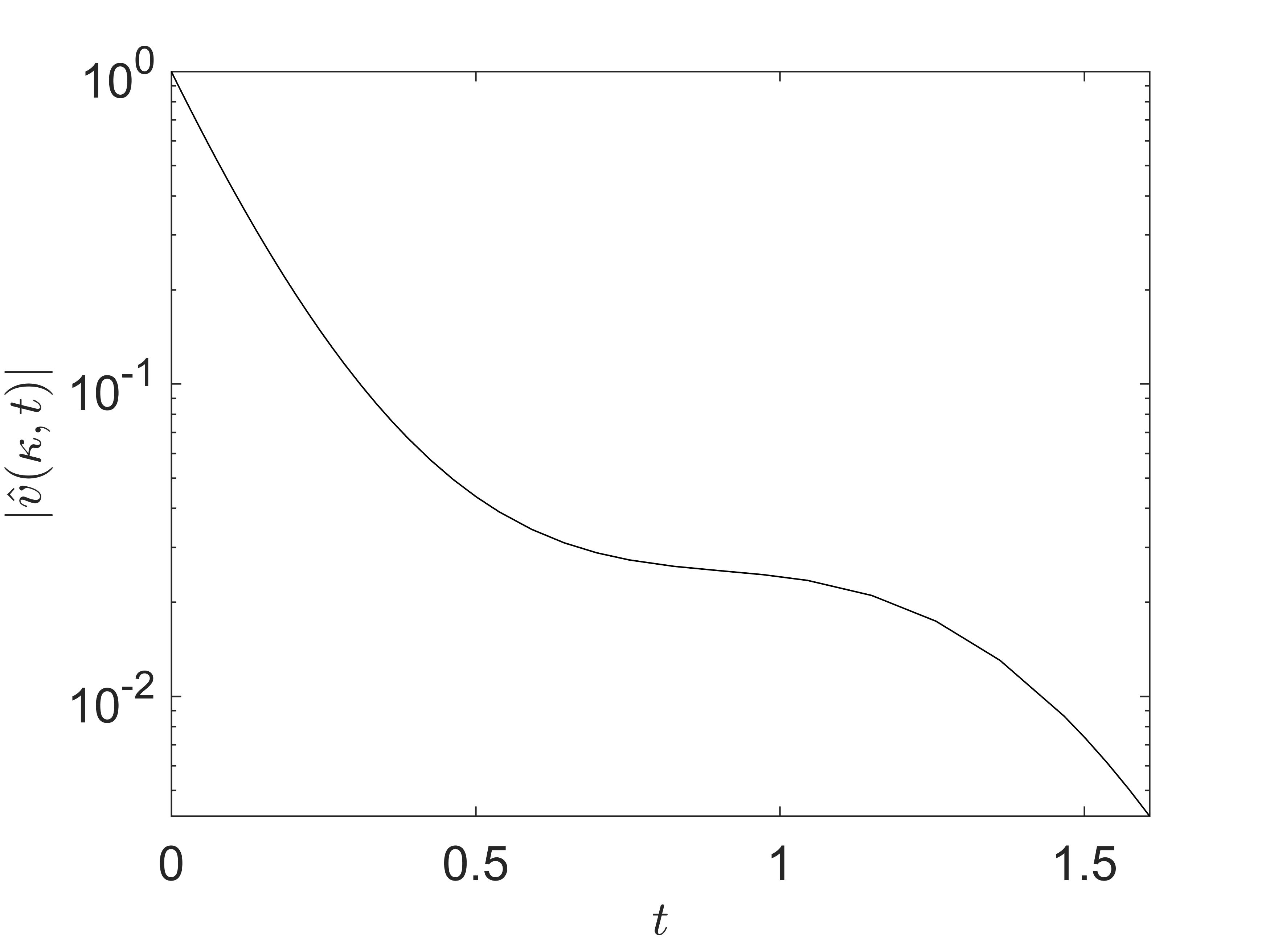}
            \caption[]%
            {{\small}}    
            \label{fig:kappas(c)}
        \end{subfigure}
        \quad
        \begin{subfigure}[b]{0.45\textwidth}   
            \centering 
            \includegraphics[width=\textwidth]{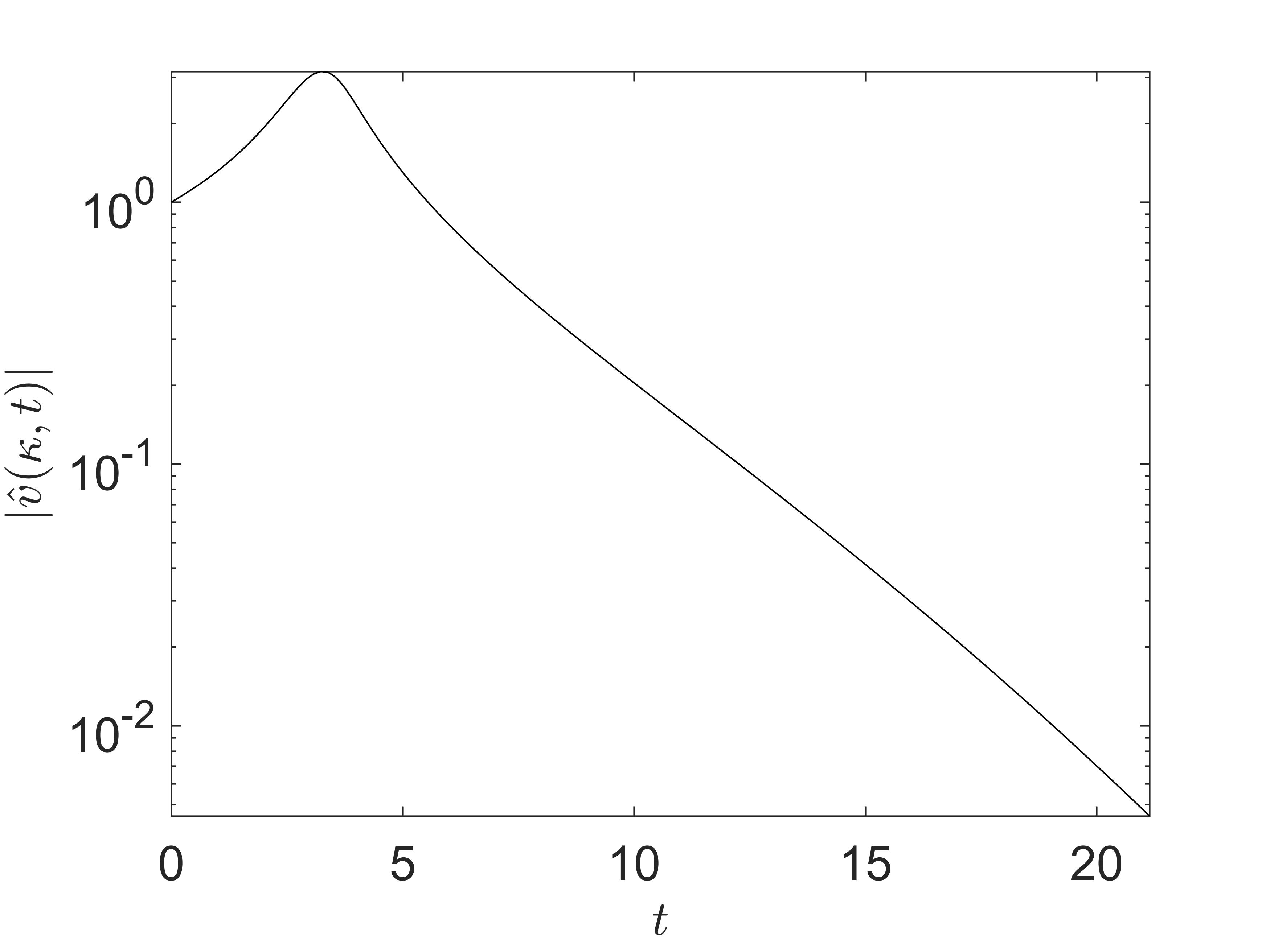}
            \caption[]%
            {{\small}}    
            \label{fig:kappas(d)}
        \end{subfigure}
        \caption[ Kappas ]
        {\small  Magnitude $| \hat{\boldsymbol{\upsilon}}  (\boldsymbol{\kappa}, t)|$ of various Fourier modes vs. $t$ for    $\chi = 10^{-6}$  and initial conditions (a) $\boldsymbol{\kappa}=(-.5, 0.866)$; (b) $\boldsymbol{\kappa}=(-0.707, -0.707)$; (c) $\boldsymbol{\kappa}=(-3.54, -3.54)$; (d) $\boldsymbol{\kappa}=(0.259, 0.966)$. } 
        \label{fig:kappasSS}
    \end{figure}

   Fig. \ref{fig:kappasSS} presents semi-log plots   of (\ref{absv}),  with $|\hat{\uppsi}_0|=1/\kappa$ and, hence, $|\hat{\bv}^{(1)}|=1$ at $t=0$, as obtained from the  integration of (\ref{stability2})    by 
means of the the finite difference approximation (FDA) and the \ML\: numerical integrator ``ode45" for   two values  $\kappa =1$ and $\kappa=5$    and parameter values \[ I^{(0)}= 0.001, \:\: I_*=0.2790,\:\:\mu_0=0.3830, \:\: \mu_\infty =0.6430, \phi =0.5,\:\:p^{(0)}=1,\]  with cessation of integration for values of $| \hat{\boldsymbol{\upsilon}}  |$   
less than 0.001.   
Figs.~\ref{fig:kappas(b)} and \ref{fig:kappas(c)} serve to illustrate the effect of $\kappa=|\bkap|$ for $\vartheta = 45^\circ$. These figures are not changed drastically by taking $\chi= 1$, whereas the transient instability in Fig.~\ref{fig:kappas(d)} is completely eliminated.\\ 

As pointed out  above  in the paragraph following (\ref{lsa4}),  material  instability can also be characterized as the 
loss of generalized ellipticity in the quasi-static (inertialess) field equations. The latter can be
obtained by omitting all inertia terms from  (\ref{stability}) and (\ref{stability1}), which gives the following expression for $\Lambda$ 
\be\label{newLam}\Lambda = \frac{\beta\gamma(k_1^2-k_2^2)^2}{\Phi_1(\bk)} -(\gamma k^2+ 2\chi k^4) = -\frac{\beta\gamma(\xi_1^2-\xi_2^2)^2}{\Phi_1(\boldsymbol{\xi})} +\gamma \xi^2- 2\chi \xi^4, \ee
where $\boldsymbol{\xi}= \imath\bk$,  the so-called {\it symbol}, is the algebraic representation of 
the differential operator $\nabla$ \citep{Renardy06}. Hence,  the positive
definite quadratic form $\Phi_1(\boldsymbol{ \xi})$ in real-valued $\boldsymbol{ \xi}$ represents an anisotropic Laplacian.  The latter can be reduced to the usual 
Laplacian on an appropriate coordinate system, with $\Phi_1^{-1}$ representing the
corresponding  Green's function. \\

Now, generalized ellipticity may  be defined as the requirement that the differential operator represented by  $-\Lambda$ be positive definite in real-valued $\xi$ \citep{Bro61,Brez98,Renardy06}, which in the present context requires   that $\chi > 0$. Therefore, the vdW-CH regularization is 
essential for material stability according to this criterion, and    
it seems worthwhile to illustrate how this  regularization might serve to impart
a diffuse length scale to shear bands.

 \section{ Model of a steady shear band   and its non-linear stability  } \label{sect:shearband}
Based on the preceding analysis, with large $k_2$ representing dominant gradients in
the direction $y = x_2$,  we assume a single  stabilized shear band localized at $y=0$ in an initially unperturbed shear flow   with ${\bf g}=0, \:\:\nabla p=0,$   and $\bv=(y, 0)$ will take the form of a stable,  fully-developed  flow in the $x = x_1$ direction with $\bv=(u(y), 0)$ and with the partial derivatives $(\partial_t, \partial_x)$ of all quantities vanishing. Hence, after one integration with respect to $y$   the momentum balance  (\ref{dg3})   reduces to the ODE 
\be \label{1dmom}\begin{split} u^{\prime\prime\prime} &= f(u') =\frac{\mu(I)-\mu(I^{(0)})}{2\surd{2}\:\chi\:I^{(0)2}} {\rm sgn}(u')\\&= \frac{(\mu_\infty-\mu_0)I_*(|u'|-1) {\rm sgn}(u')  }{2\surd{2}\:\chi\:I^{(0)}(I^{(0)}+I_*)(I^{(0)}|u'|+I_*)} , \\&\mbox{with}\:\:  I= I^{(0)}|u'| \:\:\mbox{and}\:\: I^{(0)}=1/\surd{p} , \end{split}\ee
 where  
  primes denote differentiation with respect to $y$  and we employ the non-dimensional variables (\ref{NDV}). We  further invoke the 
condition $u' \rightarrow 1$ for $|y|\rightarrow \infty$, corresponding to the unperturbed flow. Hence, assuming that $u(y)$ is an odd function of $y$ we may focus on the half space $y > 0$ with further  conditions $u(0)=0$.\\

As discussed below, the form of the ODE  (\ref{1dmom}) indicates that the limit $\chi\rightarrow 0$ leads to  a well-known {\it singular perturbation} for small $\chi$,  in which that one may generally neglect the vdW-CH terms in the  ``outer region" lying outside a thin shear band of thickness  $O(\chi^{1/2})$, which we recall represents the ratio of microscopic to macroscopic length scales.\\

To pursue an analytical treatment, we note that the function $f(u')$ in (\ref{1dmom}) can be written as  derivative $\diff \tilde{\psi}_0/\diff u'$  of a modified form of the
dissipation potential $\psi_0$: 
 \be \label{1diss} \tilde{\psi}_0(u') = 
\frac{(\mu_\infty-\mu_0)I_*}{2\surd{2}\chi I^{(0) 3 }}\left[ \frac{|u'|}{1 +I_*/I^{(0)}}-\ln \left(  \frac{I^{(0)} |u'| }{I_*}+1 \right)
\right].
 \ee

Then, by a standard method, (\ref{1dmom}) can be integrated twice  to yield $u'$ as implicit function of $y$: 
\be\label{1dint1} y =Y(u', u'_0)= \int_{u'}^{u'_0}\frac{\diff w}{\sqrt{2[  \tilde{\psi}_0(w)-\tilde{\psi}_0(1)]}}.
\ee Note that  (\ref{1dmom})-(\ref{1diss}) imply that $\tilde{\psi}_0(w)- \tilde{\psi}_0(1)\rightarrow0$ quadratically 
and $y\rightarrow \infty$    logarithmically   in $|w\! - \!1|$ as $w\rightarrow 1$.\\

It is an easy matter to convert the preceding relation to the second implicit form  
\be\begin{split}& \label{1dint2} u = y + \int_{u'}^{u'_0}\frac{(w-1)\diff w}{\sqrt{2[ \tilde{\psi}_0(w)- \tilde{\psi}_0(1)]}}  \sim y + \Delta_0, \:\:\mbox{for} \:\:|u'| \rightarrow 1, \\&   \mbox{where}\:\: \Delta_0 =\int_{1}^{u'_0}\frac{(w-1)\diff w}{\sqrt{2[ \tilde{\psi}_0(w)- \tilde{\psi}_0(1)]}},   
\end{split}\ee
which gives  $u'$ implicitly as function of $u$ and $y$ and shows that 
$u'_0= u'(0)$ is a free parameter that must be specified to complete the solution.
  The quantity $\Delta_0$, the intercept at $y=0$ of the asymptotic solution for $y\rightarrow\infty$,  represents  one-half     the       apparent slip on a shear band of zero thickness  at $y=0$  as seen in the far field. It can be calculated by numerical quadrature for given $u'_0$ and 
provides a quite useful criterion for convergence of the numerical solution considered next. The quadrature was performed in the present study by means of the\: \ML \: function 
"integral.m". \\

Note that under the rescaling  
\be \label{rescale}\bar{y} = y/I^{(0)}\surd{\chi}, \:\: \bar{u}= u/I^{(0)}\surd{\chi},\:\:\bar{\psi} =I^{(0)2}\chi \tilde{\psi}_0,\:\:\bar{\Delta}_0= \Delta_0/I^{(0)}\surd{\chi}, \ee   (where overbars are not to be confused with prior usage to denote the present non-dimensional variables)   $u'$ is invariant and    the relation (\ref{1dint2}) gives, upon division by   $I^{(0)}\surd{\chi}$ , a description of the inner layer discussed above
in terms of  $\bar{y}$,  $\bar{u}$, $\bar{\psi}$, and $\bar{\Delta}_0$. \\

Since numerical methods are generally  necessary to treat (\ref{1dint1}) or (\ref{1dint2}), it
is more efficacious to integrate (\ref{1dmom}) 
numerically.   Upon replacement   of    $u$ and $y$ by the scaled variables $\bar{u}$ and $\bar{y}$ defined in (\ref{rescale}), the ODE (\ref{1dmom}) maintains the same form, with  
\be \chi \rightarrow 1, \:\: I^{(0)} \rightarrow 1,\:\: I_*\rightarrow I_*/I^{(0)}\:\: \ee
Moreover, one can show that the replacement of $\chi$ by $\chi\mu_0$ in (\ref{rescale}), converts 
the parameter $\Delta\mu=\mu_\infty\! -\! \mu_0 $ to  $\Delta\mu/ \mu_0$, 
with $ \mu_0 \rightarrow 1$ elsewhere. Since the same transformation applies to the full 
momentum balance, it serves to justify the general validity of the stability plot of \cite{BSBG15} represented by  Fig.~\ref{fig:I-Deltamu} in the present paper. We do not
adopt this additional scaling in the present paper, since $\mu_0$ is not varied. \\

As pointed out above, the  quantity $u'_0= u'(0)$ is an unknown that must in principle be specified by the complete solution to the full momentum balance. We recall that such a numerical solution  was carried out in   Ref.\! 1 for a non-homogeneous shear flow   for $\chi=0$  .  While the authors found instability, it is not clear what role non-linear effects may play and, in line with comments in the Introduction, whether their numerics  serve to rule out asymptotically stable states with sharp  shear bands.\\    

 To pursue  numerical solutions of the full momentum balance would take us well beyond the scope of the present work . Instead, we provide below a one-dimensional analysis of the non-linear stability of steady shear bands against normal sinusoidal perturbations as a function of $u'(0)$.   For illustrative purposes, we consider here the assumption that,  up to factors of order unity,    $u'(0) \sim k_{2,\rm{max}}$, where $k_{2,\rm{max}}$ is the 
component $k_2$ of wave number in the vicinity of the maximum growth rate according to the linear stability analysis.
As indicated 
in the preceding paragraphs, a reasonable approximation for the transient growth rate is provided by the eigenvalue $\lambda$  given by  (\ref{eigen01}), whenever real. Hence, we have  made use
of the \ML \:program ``fminunc", with the quasi-Newton option,  to determine numerically the unconstrained minimum in $-\lambda(\bk)$. Thus, for parameter values  $\mu_0=0.383, \mu_\infty=0.643, I_*=0.279, I^{(0)}= 0.001,\chi = 10^{-6}$ , we find  $\bk_{\rm max} \approx \pm(2, 28)$. \\

\begin{figure}
\centering
\includegraphics[width=0.5\textwidth]{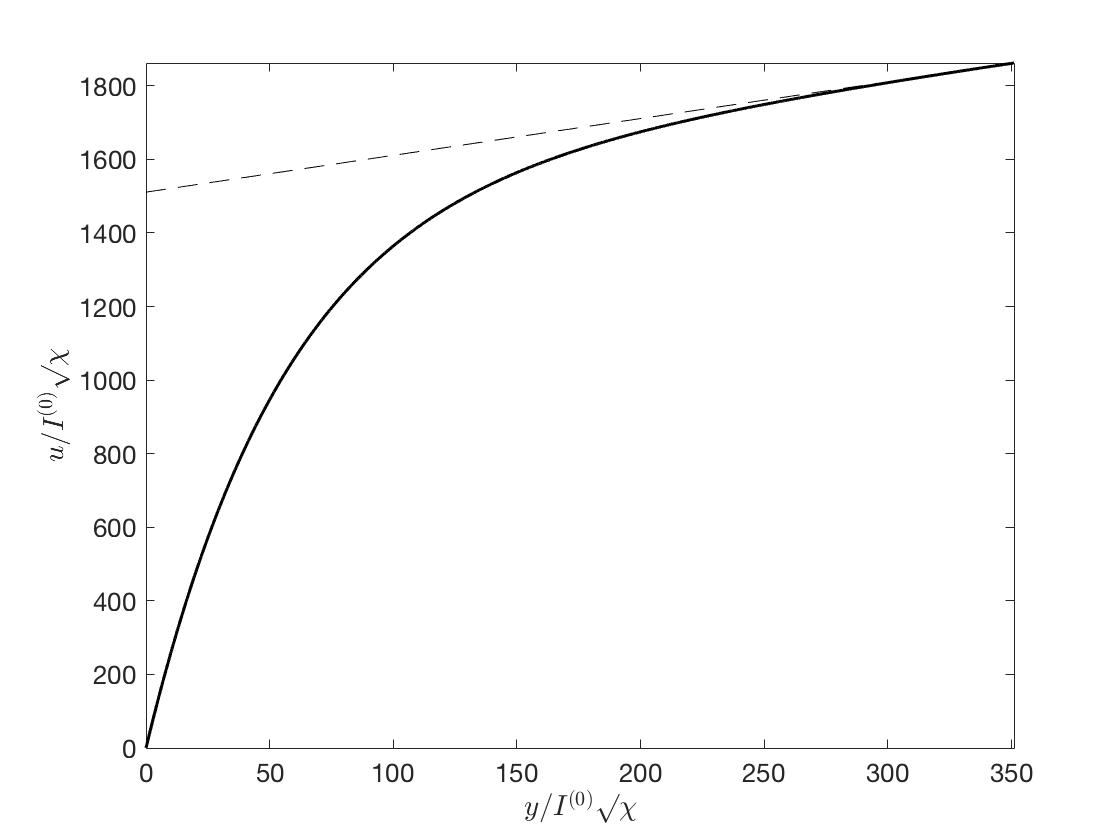}
\caption{Velocity profile in a model steady-state shear band with $u'_0=28, \:\:
I^{(0)}/I_*=0.036, \:\:\mu_0=0.383, and\:\:\mu_\infty =0.643$. }
\label{fig:band}
\end{figure}

Fig.  \ref{fig:band}  shows the corresponding curve of   $u/I^{(0)}\surd{\chi}$ vs. $y/I^{(0)}\surd{\chi}$      obtained for $u'(0)=28$ by
means of the \ML\: numerical integrator ``bvp4c.m",  a finite difference code applicable to two-point boundary-value problems. Also, shown     as dashed line    in Fig.~\ref{fig:band}        is the asymptote for large $y/  I^{(0)}  \surd{\chi}$, whose intercept at y=0 represents the 
quantity $\Delta_0$ defined in (\ref{1dint2}).  We note that the integration is sensitive   to the choice of the initial FDA mesh, as defined by the number of mesh points $N$ and the FDA step-size $\Delta y$, and the comparison of the intercept calculated from $u(y_f) -y_f  $, where $y_f = N\Delta y$,  against the exact value 
$\Delta_0$ calculated numerically from the last equation in (\ref{1dint2}).

%\newpage

       We note that in the case of constant friction coefficient $\mu_0=\mu_\infty$ that       the one-dimensional form of  (\ref{dg3}) becomes 
\be \label{1dmom1} 2\chi u''''   -    \frac {  \mu_0     p}{\surd{2}} (\mbox{sgn}(u'))'=0,  \ee
which formally involves a   Dirac    delta      at points of transition from  $u'=0$ to $u'\ne0$.
It is clear that the term in $\chi$ represents a singular surface traction that
balances the jump in frictional stress at such points.
% \footnote{  As an aside, we note that such effects could be operative at the nominal free surface of avalanching shallow granular layers.  }  
In any case,  (\ref{1dmom1}) can be integrated immediately to give    the odd    solution   in $y$ with representation    appropriate to the infinite half-space $y >0$:
\be \label{1dmom2} u  =
\left\{
\begin{array}{ll}
 c_0,\:\: \mbox{for} \:\: y > \delta, \\
  p\mu_0 y^3/12\surd{2}\chi +  u'_0y, \:\: \mbox{for} \:\:0\le y < \delta .
\end{array}
\right.
\ee
The continuity condition     $u=c_0$ at $y=\delta$  provides one condition on the three quantities $c_0,\: u'_0$, and $\delta$. For $u'_0>0$, it follows that there is a discontinuity $p\mu_0\delta^2/4\surd{2}\chi + u'_0$ in $u'$ at $y=\delta$, leading to the  Dirac delta anticipated above in (\ref{1dmom1}). 
 At any rate,  upon  specification of any two quantities,    the solution consists of a rigid, unyielded region $y > \delta$ riding on a shear band in $0\le y \le\delta$.   It appears that other solutions given by piecewise cubics in $y$  are  possible for  finite regions $0 \le y \le L$. \\

Based on the preceding analysis, one can envisage solutions with any number of diffuse shear bands interspersed with more gently sheared  regions, which not only reflects multiplicity of solution but is no doubt related to the mesh-size dependence in various numerical treatment of the underlying field equations   \citep{Bel94}.\\   

  While the above steady shear bands represents one possible solution of the steady field equations, 
it remains to show that they represent stable points in the space of all steady solutions.   

  \subsection{Non-linear stability of shear bands against parallel shearing}\label{subsect:1Dstab}
Without attempting a full two-dimensional analysis, we consider the nonlinear  stability 
governed by the  one-dimensional form of (\ref{dg3}) for $u(t, y)$:
\be\label{1dunsteady} \partial_{t}u = \partial_y\left[ h(\partial_y u) - 2\partial_y^3u\right],\:\: \mbox{with}\:\:h(\partial_y u)=\frac{1}{\surd{2}}\mu(I) {\rm sgn}(\partial_y u), \:\:\mbox{and}\:\: I=I^{(0)}|\partial_y u|,  \ee
where  $u, y$ denote the variables $\bar{u}, \bar{y}$ defined in (\ref{rescale}) and $t$ denotes the 
scaled time \be\label{tscale}\bar{t} = t/\phi I^{(0)4}\chi.\ee It is obvious that this scaling of variables restricts all parameter  dependence  to the function $\mu(I)$ and that the PDE (\ref{1dunsteady}) reduces to the rescaled ODE (\ref{1dmom}) at steady state.\\

 To investigate the stability of the steady-state velocity $u=u_s(y)$, we consider the special case of an initial condition on (\ref{1dunsteady}) given by a perturbation of $u_s$. After considering several
 possibilities, we have settled on what appeared to be a particularly unstable  perturbation represented by
 the  sinusoidal form:
\be\label{pert} u(0, y) = u_0(y) =  u_s(y) + A\sin k_2y,\ee with constant amplitude $A$ and spatial frequency $k_2$. We employ a standard finite-difference approximation with spatial discretization 
on nodal points $y_i, \:\: i=1,\ldots, N$, with $y_1=0$ and with representative spacing $\Delta y_i$.  We then  employ  the method of lines (MOL) \citep{Schiesser12} to
solve the ODEs resulting from the discretization of (\ref{1dunsteady}) numerically by means of the  \ML \:\: stiff integrator ode15s.m. The details  are summarized below in Appendix C, and,  as pointed out there, 
we employ a certain number $M$ of invariant ``ghost nodes" at each  end of the  $y$-interval at which $u(t, y_i)= u_s(y_i)$. \\
\begin{figure}
\centering\includegraphics[width=.8\textwidth]{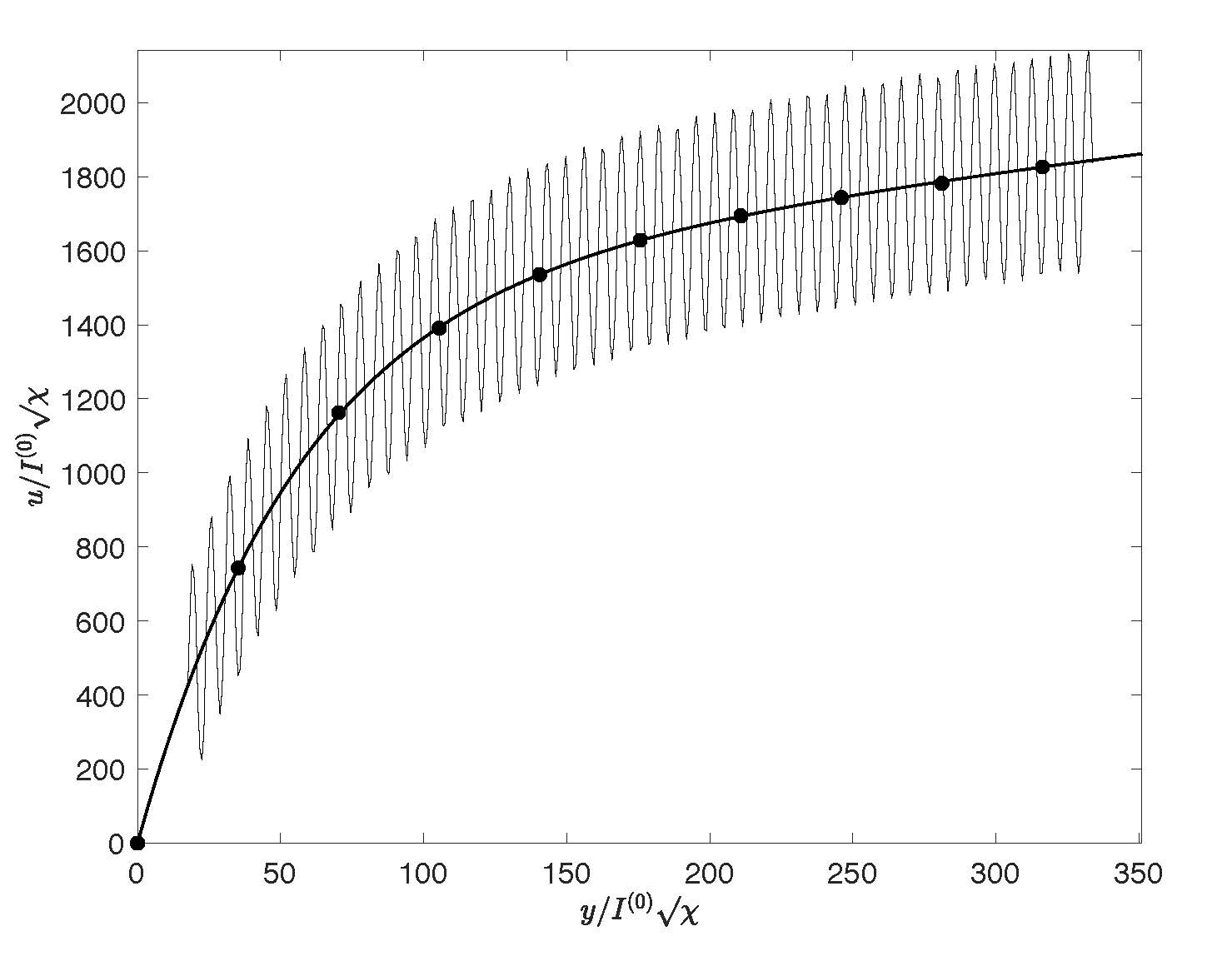}
\caption{Stable shear band with $k_2=27.8$ for $u'_0=28.0,\:\:  I^{(0)}/I_*=0.036, \:\:\mu_0=0.3830$, and
$\mu_\infty= 0.6430$. The solid curves represent the initial  steady-state and sinusoidally perturbed shear band with amplitude $A=300$, while  points $\bullet$ represent the final shear band.  
%The mean square error between initial and final shear bands is 0.000544.
 }
\label{fig:Kupopt}
\end{figure}

\begin{figure}
\centering\includegraphics[width=.6\textwidth]{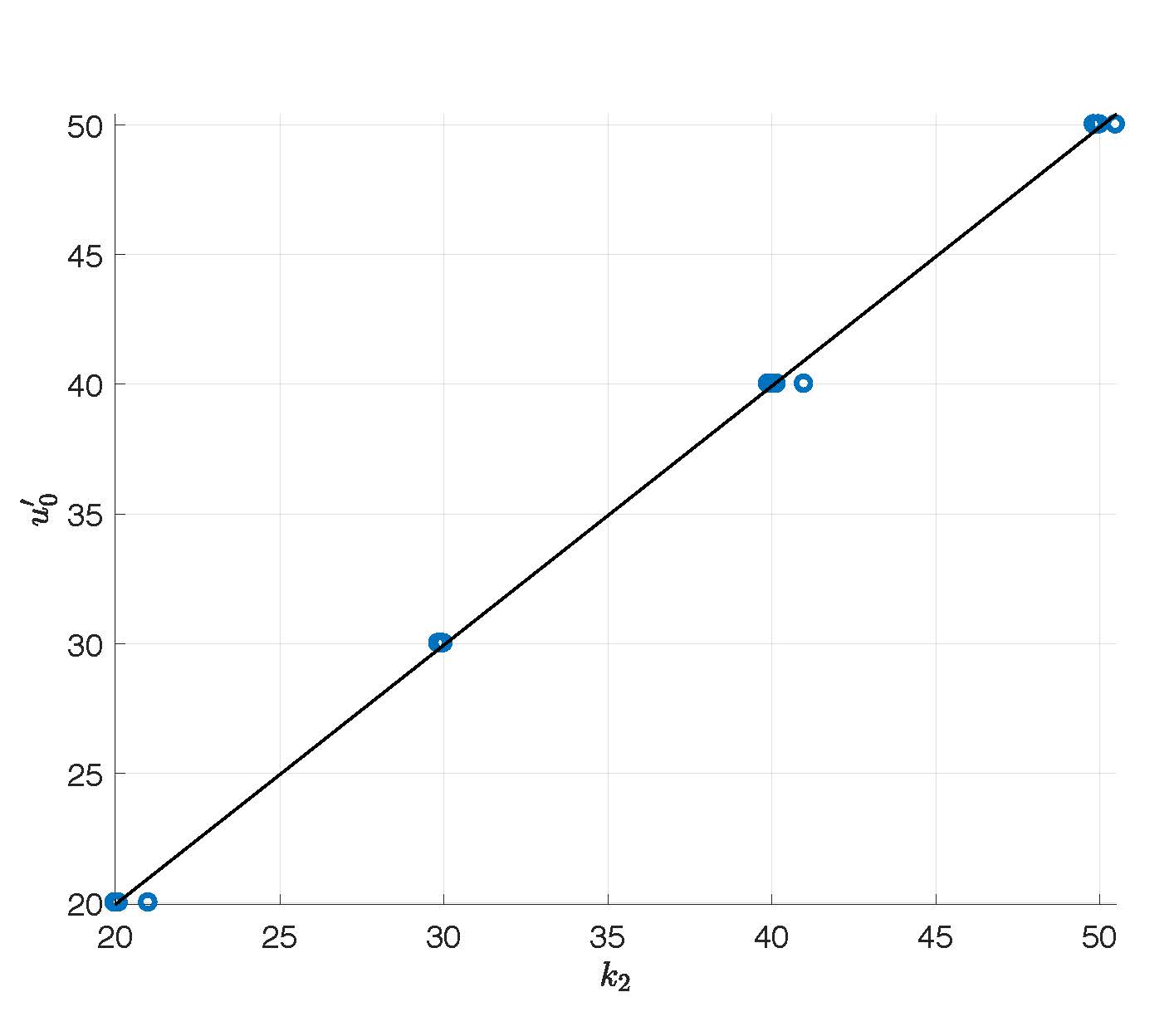}

\caption{Maximally stable $u'_0$ vs. $k_2$ for 72  different combinations $I^{(0)}/I*=
0.1,1,10, \:\: \Delta\mu = 0, 0.1, 0.2, 0.4$, and $M=10$  ghost nodes. The number of mesh points ranged from 700 to 1000, with FDA step-size $\Delta y=0.1$, and initial guesses of the form  $k_2=u'_0 =20, 30, 40,  50 $. 
 }
%\l
\label{fig:scatter}
\end{figure}
Stability of the initial steady state profile $u_s(y)$ is measured by the  mean square departure $\epsilon = \sum_i|u(t_{\rm max}, y_i)-u_s(y_i)|^2$ at the time $t=t_{\rm max}$ required for convergence of the unsteady solution. 
 The most stable shear band for various values of the parameters $\mu_0, \mu_\infty, I^{(0)}$ can be determined by  fixing one member  of the  pair $k_2, u'_0$,  where $u'_0= u'_s(0)$, and minimizing  $\epsilon$ with respect to the other.\\

However,  few initial  calculations revealed that $k_{2, {\rm opt}}\:\dot{=}\: u'_0$, so that it is  somewhat more efficient to minimize the quantity $\epsilon$  with respect to both members of the pair. This is accomplished by means of the \ML \:\: program fminunc.m,  with initial guess $k_2=u'_0$, for various values of $u'_0$.  Thus, with initial guess  $u'_0=k_2=  28$, one finds  $u'_0=28$ and $k_2=27.8$ as the optimal or maximally stable pair, for which the original, perturbed, and final shear bands are shown in Fig.~\ref{fig:Kupopt}. In this calculation the $y$-interval  is divided into $N$ nodes, with $M<<N$ ghost nodes at each end of the interval. As indicated in Fig.~\ref{fig:Kupopt}, the sinusoidal perturbation is suppressed at the ghost nodes in order to maintain the correct boundary conditions on the unsteady solution.\\  

Contrary to our initial hope of establishing a unique point $k_2, u'_0$,  we find instead  a fairly definite locus of optimal points in the $k_2$-$u'_0$ plane with $k_{2, {\rm opt}}\dot{=} u'_{0,{\rm opt}}$. This is summarized by the scatter plot in Fig.~\ref{fig:scatter} of 72 different runs with diverse values of the various parameters. The least-squares straight line shown there is given by $u'_0=0.998 k_2$  whose rms fractional error is approximately 1\%.  Remarkably the results appear to be almost 
independent of the sinusoidal-perturbation amplitude $A$, although $A$ does have some influence on the convergence of the unsteady solution. \\

For completeness, we have also investigated the stability of the homogenous shear field with  above sinusoidal  perturbation, i.e. the stability of the state with $u_s=y$ in (\ref{pert}). 
With $u=0$ imposed at $y=0$ as the only constraint, we find unbounded growth near $y=0$ without achieving a  profile that resembles our steady shear band. Although one might conclude that our steady shear band does not represent a general point of attraction in the space of steady-state solutions, and is perhaps attainable only through 2D effects, we are inclined to attribute the form of the instability to  our finite-difference implementation of the method of lines. While it would be interesting to employ more robust spectral methods, this would take us well beyond the scope of the present paper.

    \section{Pure Shear} 
\begin{figure} 
        \centering
        \begin{subfigure}[b]{0.45\textwidth}
            \centering
            \includegraphics[width=\textwidth]{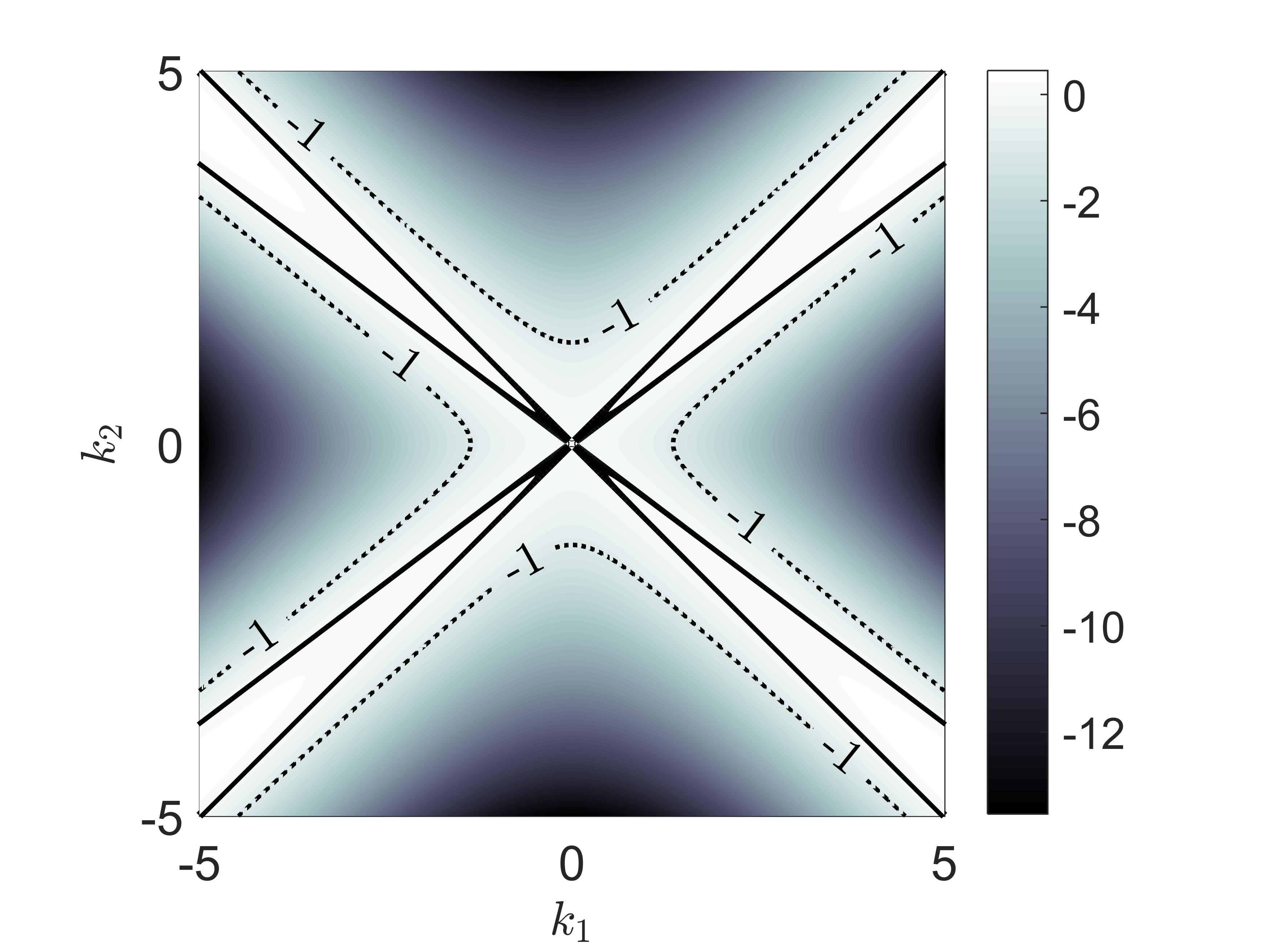}
            \caption[lambda01_OFF(a)]%
            {{\small }}    
            \label{fig:lambda01_OFF(a)}
        \end{subfigure}
        \hfill
        \hfill
        \begin{subfigure}[b]{0.45\textwidth}  
            \centering 
            \includegraphics[width=\textwidth]{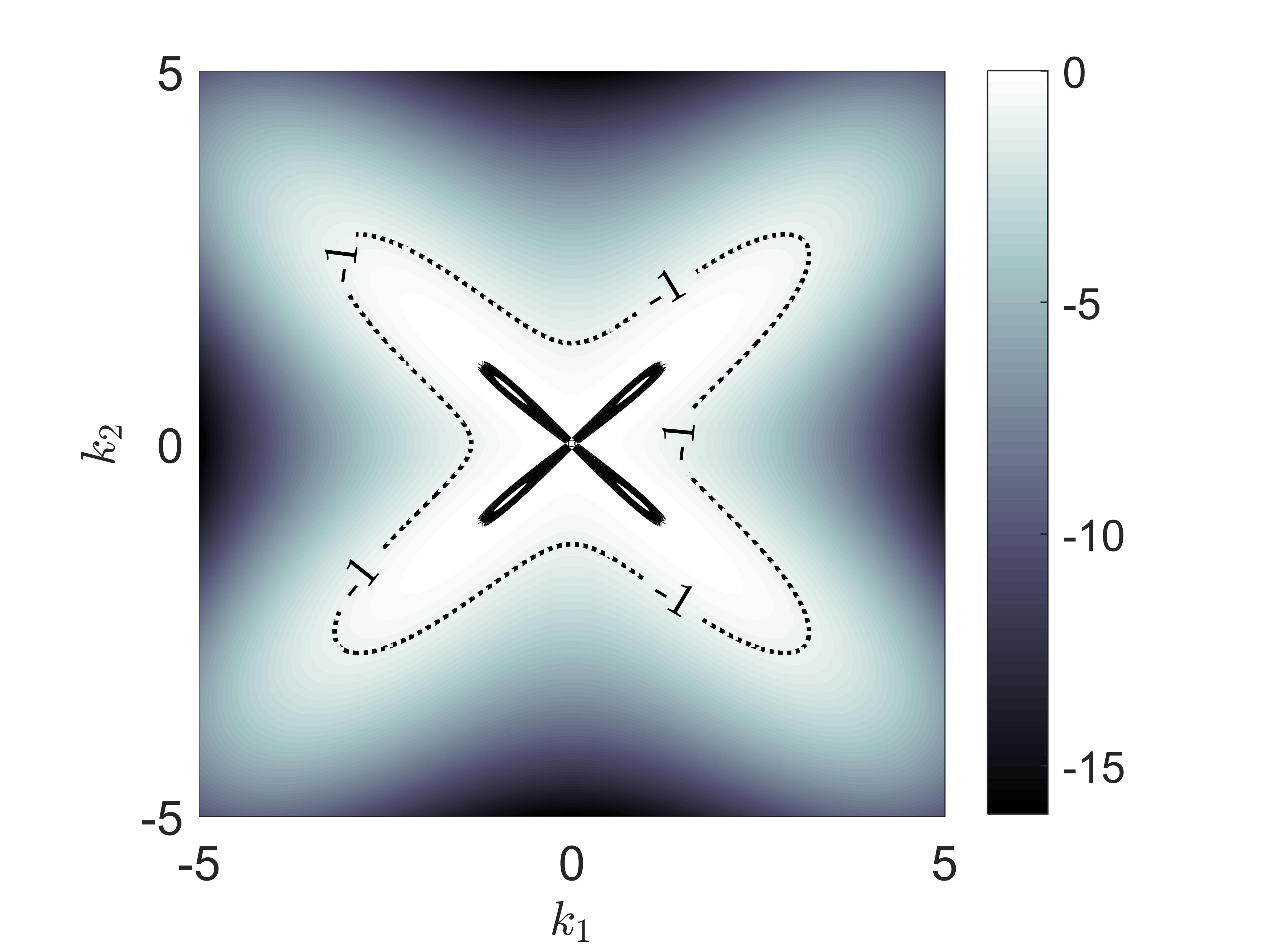}
            \caption[lambda01_OFF(d)]%
            {{\small }}    
            \label{fig:lambda01_OFF(d)}
        \end{subfigure}
        \begin{subfigure}[b]{0.45\textwidth}
            \centering
            \includegraphics[width=\textwidth]{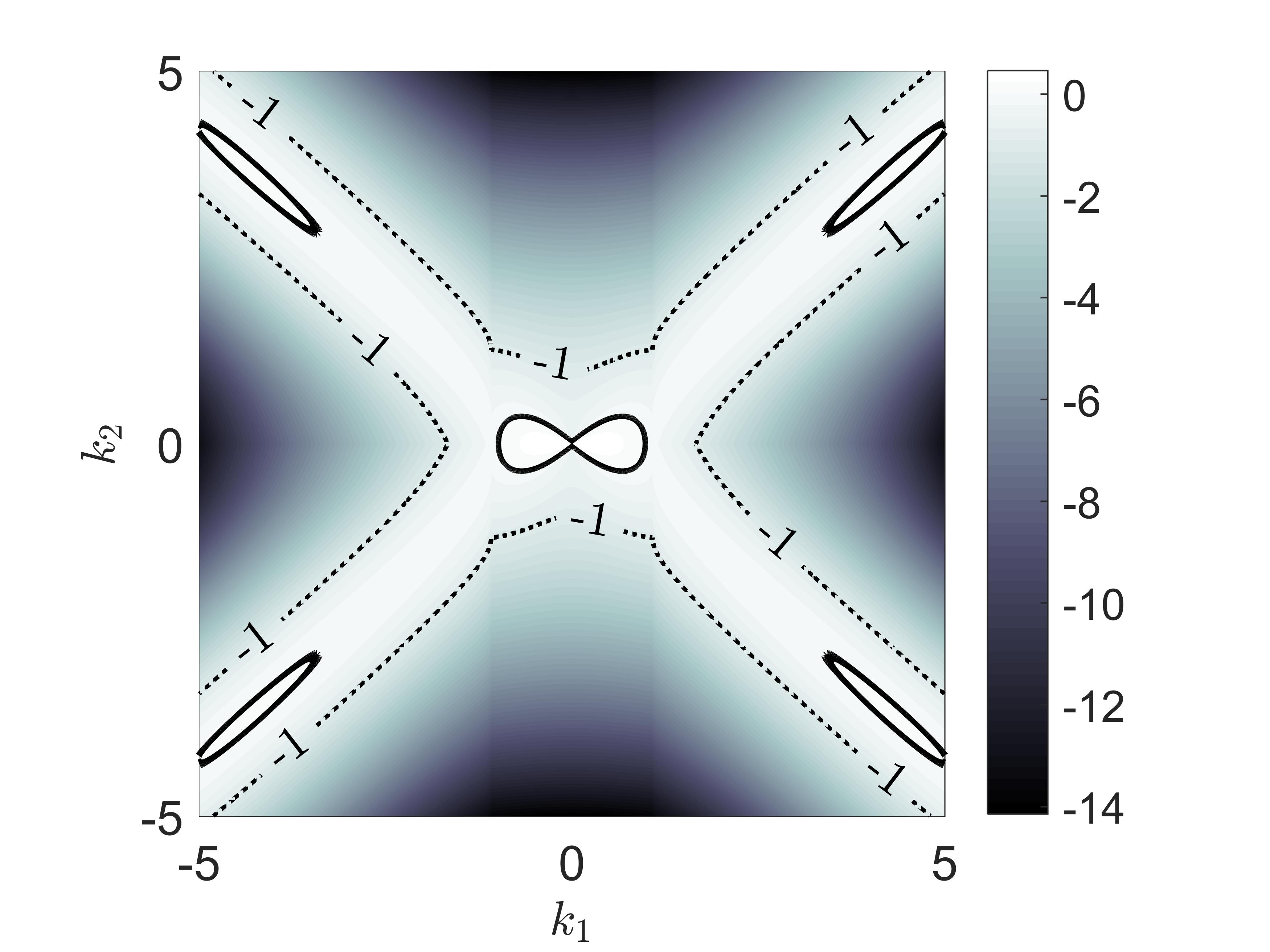}
            \caption[lambda01_ON(c)]%
            {{\small }}    
            \label{fig:lambda01_ON(c)}
        \end{subfigure}
        \hfill
        \begin{subfigure}[b]{0.45\textwidth}   
            \centering 
            \includegraphics[width=\textwidth]{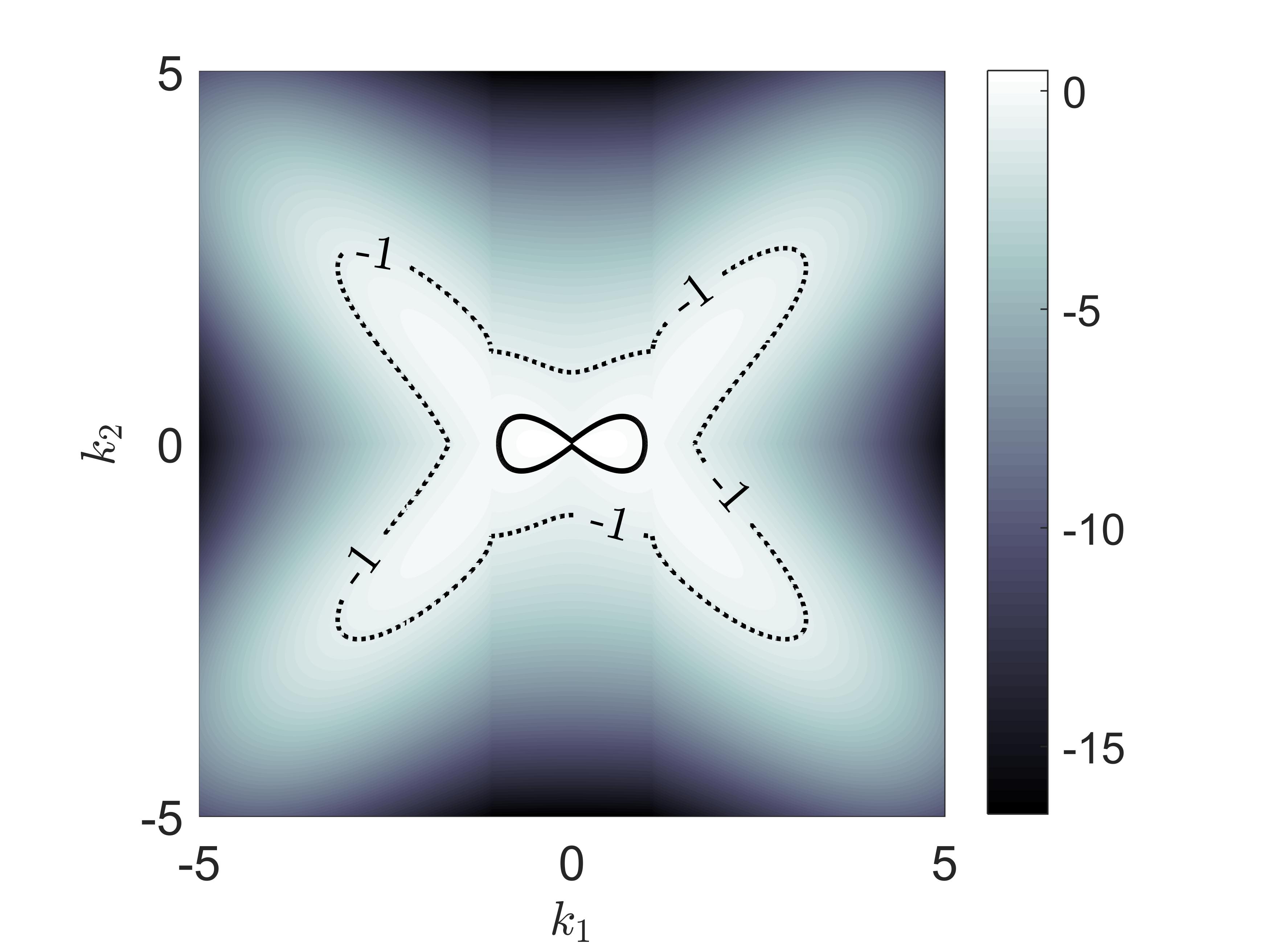}
            \caption[lambda01_ON(d)]%
            {{\small }}    
            \label{fig:lambda01_ON(d)}
        \end{subfigure}
        \caption[ k-plane_ON ]
        { Stability for pure shear without convection,  for $\chi$ values (a) $0$  (b) $0.001$,  and with 
        convection,  for $\chi$ values (c) $4 \times 10^{-5}$ and (d) $0.001$, for 
  $I = 0.0001$,  $\phi = 0.5$, and $p^{(0)} = 1$.}
  \label{fig:k-plane_OFF}
    \end{figure}

\begin{figure} 
            \centering
            \includegraphics[width=0.45\textwidth]{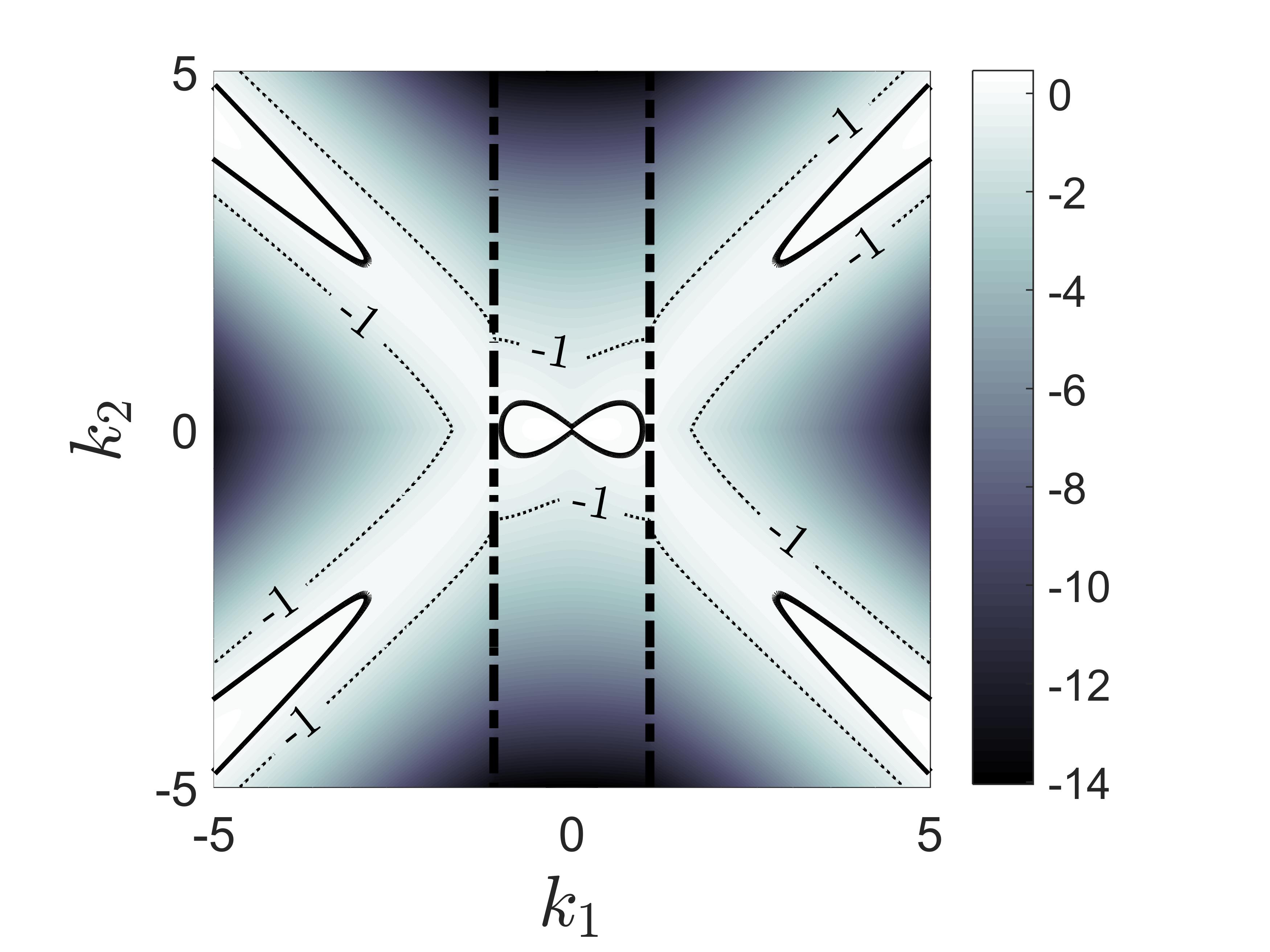}

      \caption{
Oscillatory behavior for pure shear resulting from  convection,  for $\chi = 0$, $\textit{I} = 0.0001$, $p^{(0)} = 1$,   and $\phi = 0.5$. 
The oscillation frequency is non-zero  within the central regions bounded by  dash-dotted  vertical lines. Solid and dotted curves are contours of the real part of $\lambda$, with the former representing  neutral stability.}
     \label{fig:lambda03_imag_ON}
    \end{figure}

To illustrate the effects of base-state shearing, we  consider
a base state  defined by  $\bv^{(0)}=( x_1, -x_2)/2$, with
\be \label{newL} \bD^{(0)} \equiv \bL^{(0)}= \frac{1}{\surd{2}}{\bf E^{(0)}}=
\frac{1}{2}\left[
\begin{array}{cc}
 1 &  0    \\
 0 &  -1   \    
\end{array}
\right],
\ee
Hence, for planar disturbances the relation for $\bk(t, \bkap)$ in (\ref{stability}) becomes
    \be\label{newK} \bk = \exp( - t\bL^{(0)T})\bkap,\:\: \mbox{with} \:\: k_1=\kappa_1\kappa_2/ k_2, \mbox{and} \:\:\: k_2 = \kappa_2 \exp(t/2)\ee   
 Then, by the methods employed above, we find the   
 eigenvalue $\lambda$  of $\bA$ with largest real part  to be
\be
	\label{eigen01PS}\begin{split}
	&\lambda = 
	\frac{1}{2 \phi \Phi_1} 
	\left[ 4 \beta \gamma  k_1^2 k_2^2  - 2 \Phi_1 \Phi_2 
	+ \phi ( k_1^2 - k_2^2 - \frac{\alpha}{\surd{2}} k^2 )  + \Phi_3^{1/2} \right],\\&
	\mbox{where}\:\: \Phi_1= (1-\frac{\alpha}{\surd2})k_1^2 +(1+\frac{\alpha}{\surd2})k_2^2,\\&\mbox{and}\:\:
	 \Phi_3  =  
	  2k_1^2 k_2^2\:[8 \beta^2 \gamma^2 k_1^2 k_2^2  
	 +  4 \beta \gamma \phi   
	 (  k_1^2 -  k_2^2  -  \alpha k^2 /\surd{2} ) 
	 +  \phi^2 ( \alpha^2 - 2 )].\end{split} 
\ee 
 When convection is neglected, we find that 
 \be\label{eigen03}
	\Phi_3 = 16 \beta^2 \gamma^2 k_1^4 k_2^4, \:\:\mbox{and}\:\:
	\lambda = 	\frac{ 4 \beta \gamma  k_1^2 k_2^2 - \Phi_1 \Phi_2}{\phi \Phi_1}.\ee
Fig.\! \ref{fig:k-plane_OFF} shows the effects of $\chi$ and convection on stability, and  Fig.\! \ref{fig:lambda03_imag_ON} illustrates oscillatory behavior. Note that the unstable regions 
in both figures are rotated at approximately $45^\circ$ relative to the corresponding regions for simple shear, which is relevant to the associated shear-band models.\\

Oscillatory  behavior occurs when $\Phi_3<0$ in (\ref{eigen01PS}), or in the region
\be
	\label{oscillation}
	8 \beta^2 \gamma^2 k_1^2 k_2^2  
	 +  4 \beta \gamma \phi   
	 (  k_1^2 -  k_2^2  -  \alpha k^2 /\surd{2} ) 
	 +  \phi^2 ( \alpha^2 - 2 )	< 0, 
\ee
whose boundaries are readily found to be    
 \be \label{bound} k_1 = \pm \left[ \frac{\phi}{2\beta\gamma} \left(1+ \frac{\alpha}{\surd{2}} \right) \right]^{1/2},\ee 
 upon solving for $k_2^2$ in terms of $k_1^2$. These are
 shown in the stability diagram of   
Fig.\! \ref{fig:lambda03_imag_ON}, where, in contrast to simple shear,  there now exist both stable and unstable oscillations. \\

Since $k_1 k_2 = \kappa_1 \kappa_2$, a constant, from (\ref{newK}),
the asymptotic behavior with or without  convection is stable, 
\be\label{asymlam} \lambda \sim-\frac{1}{\phi}\Phi_2\sim -\frac{1}{\phi}(\gamma k_2^2+2\chi k_2^4),\:\:\mbox{for}\:\:t\rightarrow \infty, \ee
with relative error $O(k_2^{-2})=O(e^{t})$.  Clearly,  we have stability even for $\chi=0$.\\

 Once again, one may derive a model for the shear band resulting from material instability. 
 Thus, making use once more of the 
of the \ML \:program ``fminunc"  to determine numerically the unconstrained minimum in $-\lambda(\bk)$, for parameter values  $\mu_0=0.383, \mu_\infty=0.643, I_*=0.279, I^{(0)}= 0.001, \chi = 10^{-6}$ we find  $\bk_{\rm max} \approx \pm( 26.0,  22.5 )$. This corresponds roughly  to the classical $\approx 45^\circ$ shear-band orientation found in previous experiments and numerical simulations.       Figure 13 shows  
 the growth of the Fourier mode, and it is seen from panel (b) that the transient growth can become 
 quite large before being quenched by a combination of wavevector stretching and damping by higher gradients.   
\\ 

The above result leads us to express (\ref{dg3}),  with ${\bf g}=0$ and $\nabla p$,  on a coordinate system oriented at $45^\circ$  to that
involved in  (\ref{newL}), reducing   it to the form (\ref{1dmom}) and  (\ref{newL})  to the form (\ref{sshear}). Therefore, the analysis of the shear band reduces essentially to that of Section \ref{sect:shearband}, with appropriate modification of the boundary condition on $u'(0)$.  On this  new coordinate system,  the above value of 
 $\bk_{\rm max}$ becomes $\approx \pm(2, 34)$ as opposed to the value $\pm( 2, 28)$
 found in Section \ref{sect:shearband}  for simple shear. Given the rough agreement of these values and in light of    the non-unique solution to    our shear-band model, it  hardly seems   worthwhile to repeat the calculation 
 leading to Fig. \ref{fig:band}  with an only  slightly modified boundary condition on $u'(0)$.

\begin{figure} 
        \centering
        \begin{subfigure}[b]{0.475\textwidth}
            \centering
            \includegraphics[width=\textwidth]{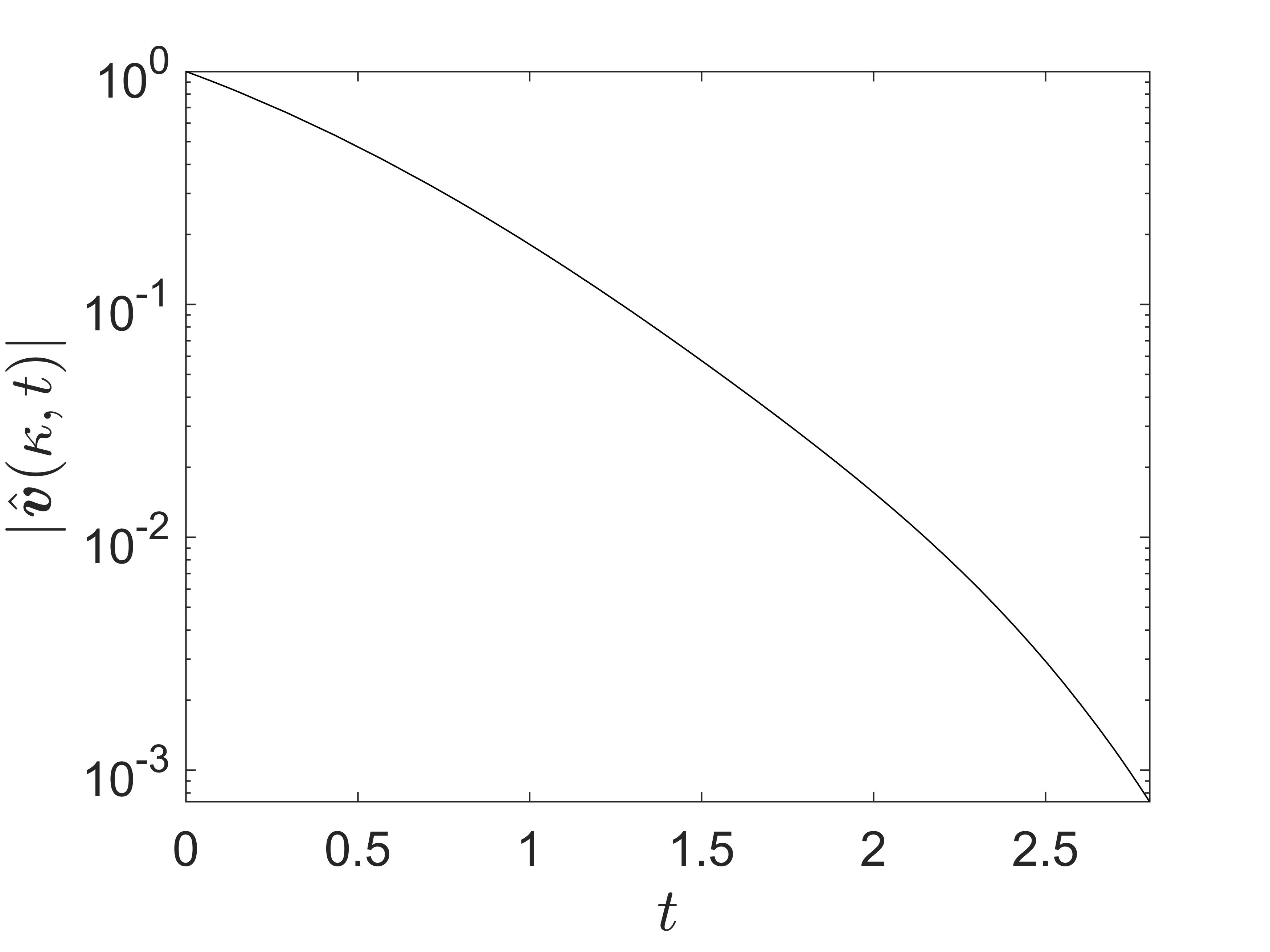}
            \caption[Figps1]%
            {{\small }}    
            \label{fig:PureShear1}
        \end{subfigure}
        \hfill
        \begin{subfigure}[b]{0.475\textwidth}  
            \centering 
            \includegraphics[width=\textwidth]{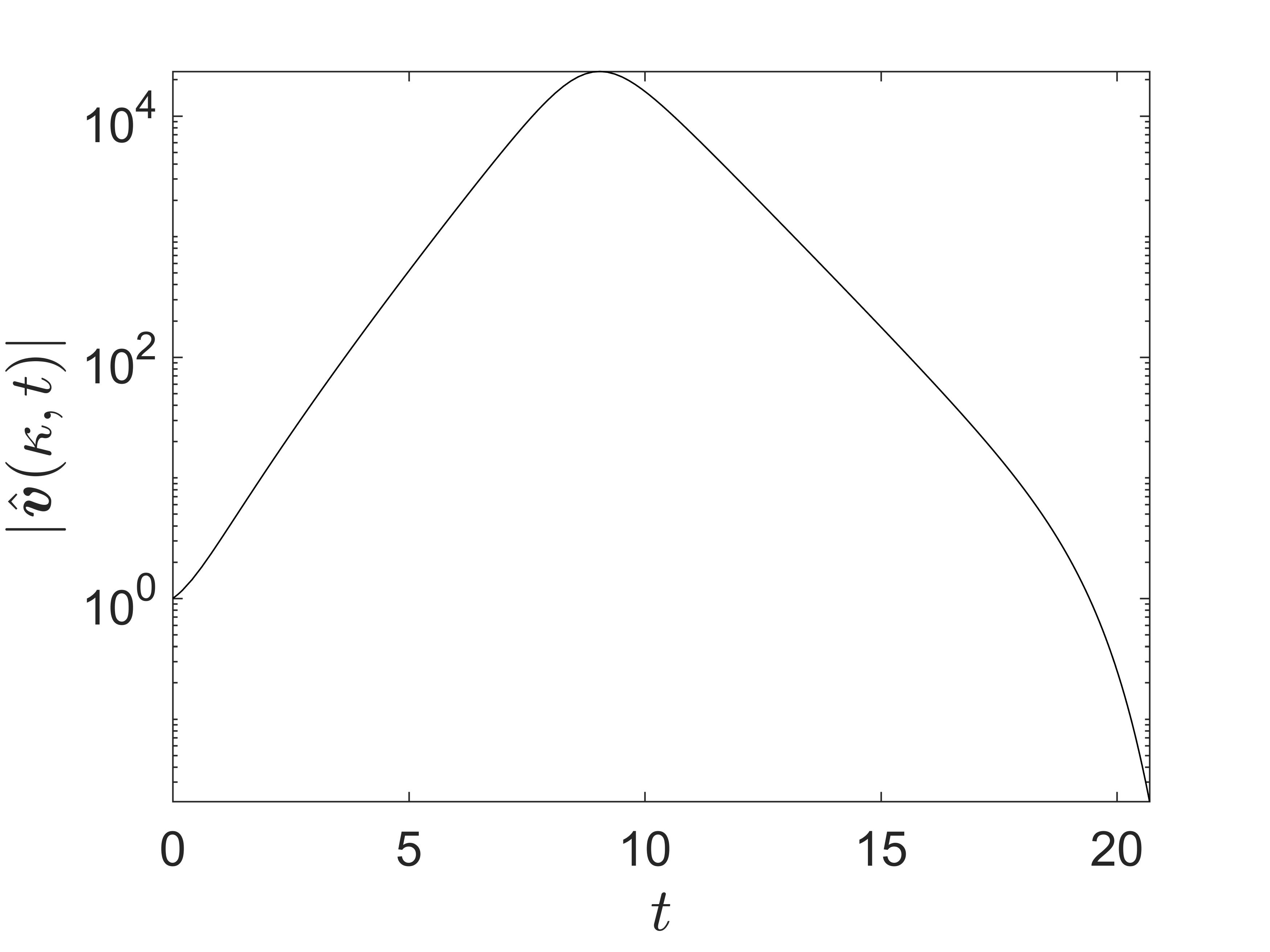}      
            \caption[Figps2]%
            {{\small }}    
            \label{fig:PureShear2}
        \end{subfigure}
        \caption[Figps2 ]
        {\small   Magnitude $|\hat{ \boldsymbol{\upsilon}}  (\boldsymbol{\kappa}, t)|$ of initially stable and unstable Fourier modes vs. $t$ for planar pure shear for   $\chi = 10^{-6}$  and initial conditions  (a) $\boldsymbol{\kappa}=(-0.5, -0.866)$,; (b) $\boldsymbol{\kappa}=(1, 0.0001)$.   }
   \label{fig:PureShear}
    \end{figure}
   
 % \newpage
\section{Concluding Remarks}\label{sect:conclude}
Our major findings are adequately summarized in the Abstract and Introduction. To briefly recapitulate the most important of these, we find for  both planar simple shear and pure shear    that the       linear instability of the $\mu(I)$ model  identified by \cite{BSBG15}  is modified through convection by the base flow, giving way to long-time stability induced by Kelvin wave-vector stretching. The addition of gradient effects via the vdW-CH model provides a wave-number cut-off  that serves to  stabilize the dynamical equations over the entire time domain,   to regularize  
the quasi-static field equations,  and to assign a diffuse length scale to eventual shear bands.\\ 

   We find that steady shear bands are stable against steady parallel sinusoidal 
shear fields, provided the normal velocity gradient of the shear band is very nearly equal to the wave number of the sinusoidal perturbation. To obtain a unique shear band, it is necessary to assume some preferred wave number, e.g. the  most unstable one according to linear theory. A challenge for future work is to elucidate shear band formation in the presence of more complex spatial perturbations to homogenous shearing.    
\\

 In summary, we conclude that the (Hadamard) short wave length instability found by  \cite{BSBG15} is  connected to the loss of ellipticity in the quasi-static field equations. Although transient and  eventually quenched by wave vector stretching according to linear theory,  the instability  is doubtless problematical for numerical simulation and it may also trigger non-linear instabilities. Addition of higher-gradient effects like those considered in the present work should go a long way towards alleviating such problems. \\

As a practical matter, the present work may suggest a stratagem for 
numerical simulation   of materially unstable visco-plasticity, in which a transient form of the vdW-CH or similar regularization is invoked 
to minimize transient instabilities, with the possibility of describing the structure of 
shear bands on longer time scales if so desired.  Our simple model of a shear band based on the vdW-CH model suggests that 
spatial boundary conditions,  required in principle by any higher-gradient model,   can be chosen somewhat arbitrarily, as the higher gradients tend to have a  spatially  localized domain of influence.  

As additional future work, it would be worthwhile to apply the current theory to  other homogeneous shear flows,  such as   axisymmetric straining of the kind that arises in the standard quasi-static tests of soil mechanics or  in more rapid converging-hopper flows. While somewhat distinct from the issue of material instability,  an investigation should be carried out of  the coupling of pressure gradients
in the base flow to the perturbed momentum balance, an effect neglected in Ref.\! 1 and the present study.\\

 As a deeper theoretical issue, it would  be interesting to investigate the possible relation between  the weakly non-local model of the present study and the fully non-local variants of the $\mu(I)$ model proposed by \cite{Poul09} and more recently by Kamrin and co-workers \citep[See e.g.][]{Hen14}. We recall that the latter model can be tied to a Ginzburg-Landau formalism, which shares a certain kinship to  the vdW-Cahn-Hilliard model of the present study \citep{G96}. \\

\bibliographystyle{jfm}
%\newpage
\section*{Acknowledgement}  A major part of this work  was carried out  during the tenure of the first author 
as Senior Fellow in the Institute of Advanced Study of Durham University, U.K., in 2016, and as a
  visiting professor in the \'Ecole Sup\'erieure de Physique et de Chimie Industrielles (ESPCI), Paris, in 2017.   He is grateful for the  support and  hospitality of the host institutions and for interactins with  Prof.~Jim  McElwaine in the Durham University Department of Earth Sciences, and Profs.~Eric Cl\'ement. Lev Truskinovsky  et al.  in the Laboratoire PMMH  in the ESPCI. Special thanks are due to Prof.~Truskinovsky for pointing out the seminal works of van der Waals. 
The second author wishes to thank   Prof.~Jan Talbot,  of the University of California, San Diego,  Department of Nanoengineering, for serving as host during his brief visit in 2016. Finally, we acknowledge a 2017-18 grant of HPC time by  the San Diego Supercomputer Center, at the  University of California, San Diego.

%\bibliography{CH}

\begin{thebibliography}{33}
\expandafter\ifx\csname natexlab\endcsname\relax\def\natexlab#1{#1}\fi
\def\au#1{#1} \def\ed#1{#1} \def\yr#1{#1}\def\at#1{#1}\def\jt#1{\textit{#1}}
  \def\bt#1{#1}\def\bvol#1{\textbf{#1}} \def\vol#1{#1} \def\pg#1{#1}
  \def\publ#1{#1}\def\arxiv#1{#1}\def\org#1{#1}\def\st#1{\textit{#1}}

\bibitem[Alam \& Nott(1997)]{Alam97}
{\sc \au{Alam, M} \& \au{Nott, P}} \yr{1997}  \at{The influence of friction on
  the stability of unbounded granular shear flow}.  \jt{J. Fluid Mech.}
  \bvol{343},  \pg{267--301}.

\bibitem[Anderson {\em et~al.\/}(1998)Anderson, McFadden \& Wheeler]{And98}
{\sc \au{Anderson, D.~M.}, \au{McFadden, G.~B.} \& \au{Wheeler, A.~A.}}
  \yr{1998}  \at{Diffuse-interface methods in fluid mechanics}.  \jt{Ann. Rev.
  Fluid Mech.}  \bvol{30}~(1),  \pg{139--165}.

\bibitem[Barker \& Gray(2017)]{Barker17b}
{\sc \au{Barker, T} \& \au{Gray, N}} \yr{2017}  \at{Partial regularisation of
  the incompressible $\mu(i)$-rheology for granular flow}.  \jt{J. Fluid Mech.}
   \bvol{828},  \pg{5--32}.

\bibitem[Barker {\em et~al.\/}(2015)Barker, Schaeffer, Bohorquez \&
  Gray]{BSBG15}
{\sc \au{Barker, T}, \au{Schaeffer, DG}, \au{Bohorquez, P} \& \au{Gray, JMNT}}
  \yr{2015}  \at{Well-posed and ill-posed behaviour of the-rheology for
  granular flow}.  \jt{J. Fluid Mech.}  \bvol{779},  \pg{794--818}.

\bibitem[Barker {\em et~al.\/}(2017)Barker, Schaeffer, Shearer \&
  Gray]{Barker17}
{\sc \au{Barker, T.}, \au{Schaeffer, D.}, \au{Shearer, M.} \& \au{Gray, N.}}
  \yr{2017}  \at{Well-posed continuum equations for granular flow with
  compressibility and $\mu${\rm(i)}-rheology}.  \jt{Proc. R. Soc. A}
  \bvol{473}~(2201),  \pg{20160846}.

\bibitem[Belytschko {\em et~al.\/}(1994)Belytschko, Chiang \& Plaskacz]{Bel94}
{\sc \au{Belytschko, T}, \au{Chiang, H-Y} \& \au{Plaskacz, E}} \yr{1994}
  \at{High resolution two-dimensional shear band computations: imperfections
  and mesh dependence}.  \jt{Comp. Meth. Appl. Mech. Eng.}  \bvol{119}~(1-2),
  \pg{1--15}.

\bibitem[Bigoni(1995)]{Big95}
{\sc \au{Bigoni, D}} \yr{1995}  \at{On flutter instability in elastoplastic
  constitutive models}.  \jt{Int. J. Solids Structs.}  \bvol{32}~(21),
  \pg{3167--3189}.

\bibitem[Brezis \& Browder(1998)]{Brez98}
{\sc \au{Brezis, H} \& \au{Browder, F}} \yr{1998}  \at{Partial differential
  equations in the 20th century}.  \jt{Adv. Math.}  \bvol{135}~(1),
  \pg{76--144}.

\bibitem[Browder(1961)]{Bro61}
{\sc \au{Browder, F}} \yr{1961}  \at{On the spectral theory of elliptic
  differential operators. \rm{I}}.  \jt{Math. Ann.}  \bvol{142}~(1),
  \pg{22--130}.

\bibitem[Cahn \& Hilliard(1958)]{CH58}
{\sc \au{Cahn, J~W} \& \au{Hilliard, J~E}} \yr{1958}  \at{Free energy of a
  nonuniform system. \rm{I}. interfacial free energy}.  \jt{J. Chem. Phys.}
  \bvol{28}~(2),  \pg{258--267}.

\bibitem[Didwania \& Goddard(1993)]{Did93}
{\sc \au{Didwania, A~K} \& \au{Goddard, J~D}} \yr{1993}  \at{A note on the
  generalized \rm{R}ayleigh quotient for non-self-adjoint linear stability
  operators}.  \jt{Phys. Fluids A}  \bvol{5}~(5),  \pg{1269--1271}.

\bibitem[Edelen(1972)]{Edelen72}
{\sc \au{Edelen, D. G.~B.}} \yr{1972}  \at{A nonlinear \rm{O}nsager theory of
  irreversibility}.  \jt{Int. J. Eng. Sci.}  \bvol{10}~(6),  \pg{481--90}.

\bibitem[Edelen(2005)]{Edelen05}
{\sc \au{Edelen, D G~B}} \yr{2005} {\em Applied exterior calculus\/}.
  \publ{Mineola, NY: Dover Publications Inc.}

\bibitem[Forest \& Aifantis(2010)]{Forest10}
{\sc \au{Forest, S} \& \au{Aifantis, E}} \yr{2010}  \at{Some links between
  recent gradient thermo-elasto-plasticity theories and the thermomechanics of
  generalized continua}.  \jt{Int. J. Solids Structs.}  \bvol{47}~(25),
  \pg{3367--3376}.

\bibitem[Fornberg(1988)]{Fornberg88}
{\sc \au{Fornberg, B.}} \yr{1988}  \at{Generation of finite difference formulas
  on arbitrarily spaced grids}.  \jt{Mathematics of computation}
  \bvol{51}~(184),  \pg{699--706}.

\bibitem[Goddard(2003)]{JG03}
{\sc \au{Goddard, J~D}} \yr{2003}  \at{Material instability in complex fluids}.
   \jt{Ann. Rev. Fluid Mech.}  \bvol{35}~(1),  \pg{113--133}.

\bibitem[Goddard(2014)]{JG14}
{\sc \au{Goddard, J~D}} \yr{2014}  \at{Edelen's dissipation potentials and the
  visco-plasticity of particulate media}.  \jt{Acta Mech.}  \bvol{225}~(8),
  \pg{2239--2259}.

\bibitem[Gurtin(1996)]{G96}
{\sc \au{Gurtin, M~E}} \yr{1996}  \at{\rm{G}eneralized
  \rm{G}inzburg-\rm{L}andau and \rm{C}ahn-\rm{H}illiard equations based on a
  microforce balance}.  \jt{Physica D}  \bvol{92}~(3),  \pg{178--192}.

\bibitem[Henann \& Kamrin(2014)]{Hen14}
{\sc \au{Henann, D~L} \& \au{Kamrin, K}} \yr{2014}  \at{Continuum
  thermomechanics of the nonlocal granular rheology}.  \jt{Int. J. Plasticity}
  \bvol{60},  \pg{145--162}.
  
   
  \bibitem[Heyman {\em et~al.\/}(2017)Heyman, Delannay, Tabuteau \&
  Valance]{Hey16}
{\sc \au{Heyman, J}, \au{Delannay, R}, \au{Tabuteau, H} \& \au{Valance, A}}
  \yr{2017}  \at{Compressibility regularizes the $\mu$(i)-rheology for dense
  granular flows}.  \jt{J. Fluid Mech.}  \bvol{830},  \pg{553--568}.  


\bibitem[Hill(1956)]{Hill56}
{\sc \au{Hill, R.}} \yr{1956}  \at{New horizons in the mechanics of solids}.
  \jt{J. Mech. Phys. Solids}  \bvol{5}~(1),  \pg{66--74}.

\bibitem[Jop {\em et~al.\/}(2005)Jop, Forterre \& Pouliquen]{Jop05}
{\sc \au{Jop, P}, \au{Forterre, Y} \& \au{Pouliquen, O}} \yr{2005}  \at{Crucial
  role of sidewalls in granular surface flows: consequences for the rheology}.
  \jt{J. Fluid Mech.}  \bvol{541},  \pg{167--192}.

\bibitem[Jop {\em et~al.\/}(2006)Jop, Forterre \& Pouliquen]{JFP06}
{\sc \au{Jop, P}, \au{Forterre, Y} \& \au{Pouliquen, O}} \yr{2006}  \at{A
  constitutive law for dense granular flows}.  \jt{Nature}  \bvol{441}~(7094),
  \pg{727--730}.

\bibitem[Leonov(1988)]{Leonov88}
{\sc \au{Leonov, A.~I.}} \yr{1988}  \at{Extremum principles and exact two-side
  bounds of potential: Functional and dissipation for slow motions of nonlinear
  viscoplastic media}.  \jt{J. Non-Newton. Fluid Mech.}  \bvol{28}~(1),
  \pg{1--28}.

\bibitem[MiDi(2004)]{GDR04}
{\sc \au{MiDi, GDR}} \yr{2004}  \at{On dense granular flows}.  \jt{Europ. Phys.
  J. E}  \bvol{14}~(4),  \pg{341--365}.

\bibitem[Mindlin(1964)]{M64}
{\sc \au{Mindlin, R~D}} \yr{1964}  \at{Micro-structure in linear elasticity}.
  \jt{Arch. Ration. Mech. Anal.}  \bvol{16}~(1),  \pg{51--78}.

\bibitem[Pouliquen \& Forterre(2009)]{Poul09}
{\sc \au{Pouliquen, O} \& \au{Forterre, Y}} \yr{2009}  \at{A non-local rheology
  for dense granular flows}.  \jt{Phil. Trans. Roy. Soc. Lond. A}
  \bvol{367}~(1909),  \pg{5091--5107}.

\bibitem[Renardy \& Rogers(2006)]{Renardy06}
{\sc \au{Renardy, M} \& \au{Rogers, R}} \yr{2006} {\em An introduction to
  partial differential equations\/}.  \publ{Springer Science}.

\bibitem[Rowlinson(1979)]{R79}
{\sc \au{Rowlinson, J~S}} \yr{1979}  \at{\rm{Translation of J D van der Waals'
  ``The thermodynamik theory of capillarity under the hypothesis of a
  continuous variation of density"}}.  \jt{J. Stat. Phys.}  \bvol{20}~(2),
  \pg{197--200}.

\bibitem[Saramito(2016)]{Saramito16}
{\sc \au{Saramito, P.}} \yr{2016} {\em \rm{Complex Fluids - Modeling and
  Algorithms}\/}.  \publ{SMAI/Springer Int. Publishing}.

\bibitem[Schiesser(2012)]{Schiesser12}
{\sc \au{Schiesser, W.~E.}} \yr{2012} {\em The numerical method of lines:
  integration of partial differential equations\/}.  \publ{Elsevier}.

\bibitem[Thomson(1887)]{Thomson1887}
{\sc \au{Thomson, W}} \yr{1887}  \at{\rm{XXI. Stability of fluid motions
  (continued from the May and June numbers). - Rectilineal motion of viscous
  fluid between two parallel planes}}.  \jt{Phil. Mag.}  \bvol{24}~(147),
  \pg{188--196}.

\bibitem[Trefethen \& Embree(2005)]{Trefethen05}
{\sc \au{Trefethen, L.} \& \au{Embree, M.}} \yr{2005} {\em Spectra and
  pseudospectra: the behavior of nonnormal matrices and operators\/}.
  \publ{Princeton University Press}.

\end{thebibliography}

\appendix
%%%%%%%%%%%%%%%%%%%
\vspace{-.5cm}
\section{Derivation of the perturbed equations (\ref{lsa3}) }\label{app:A}
Substituting (\ref{lsa1}) and (\ref{lsa2}) into (\ref{dg5a}),
the perturbed equations of motion become 
\be
	\label{alsa1}
	\begin{split}
	&\rho_s \phi   \left(  \partial _t\bv^{(1)} +  \bv^{(0)}  \ccdot  \nabla \bv^{(1)}  +  \bv^{(1)}  \ccdot  \nabla \bv^{(0)} \right)  = -   \nabla p^{(1)}
	\\&+  \frac{ (2  -  \dmu^{(0)}) \mu^{(0)}}{2} \bE^{(0)} \nabla p^{(1)}
	 +   
	\left[  \left( \frac{\partial \mu}{\partial p}  \right)^{(0)}   p^{(1)}
	 +  \left(  \frac{\partial \mu}{\partial \bD}  \right)^{(0)}   \ccdot   \nabla  \bv^{(1)}
	  \right]  
	 \bE^{(0)} \nabla p^{(0)} \\	
	 &- \frac{1}{2} 
	\left\{ \mu^{(0)}   \left[
	  {\left(  \frac{\partial \dmu}{\partial p}  \right)}^{(0)}    p^{(1)}
	 +  {\left(  \frac{\partial \dmu}{\partial \bD}  \right)}^{(0)} 
	  \ccdot  \nabla \bv^{(1)}   \right]\right.
	\\&\left.+  \dmu^{(0)}
	   \left[ 
	\left( \frac{\partial \mu}{\partial p}  \right)^{(0)}    p^{(1)}
	 +  \left(  \frac{\partial \mu}{\partial \bD}  \right)^{(0)}   \ccdot    
	\nabla  \bv^{(1)}	  \right] \right\}  
	\bE^{(0)} \nabla p^{(0)}  \\	
	& +   \frac{ ( \dmu^{(0)} -1) \mu^{(0)} p^{(0)} }{ \DDD} 
	 (\bE^{(0)} \nabla )( \bE^{(0)} \ccdot \nabla \bv^{(1)} )    
	 +   \frac{\mu^{(0)} p^{(0)} }{2 \DDD}   \nabla^2 \bv^{(1)} 
	 -  2 \chi \nabla^4 \bv^{(1)},
	\end{split}
\ee
where
\be
	\label{alsa2}
	\begin{split}
	\left( \frac{\partial \mu}{\partial p}  \right)^{(0)} 
	&= \left( \frac{\diff \mu }{\diff I} \frac{\partial I}{\partial p}  \right)^{(0)}
	= \left(  - \frac{I}{2p}  \frac{\diff \mu }{\diff I} \right)^{(0)} 
	= - \frac{\mu^{(0)} \dmu^{(0)}}{2p^{(0)}}, \:\: \\
	\left( \frac{\partial \dmu}{\partial p}  \right)^{(0)} 
	&= \left[  
	\frac{\partial I}{\partial p} \left(  \frac{1}{\mu}\frac{\diff \mu }{\diff I} 
	 	 -  \frac{I}{\mu^2} \left(\frac{\diff \mu }{\diff I} \right)^2
	 	  +   \frac{I}{\mu} \frac{\diff^2 \mu}{\diff I^2} \right)    \right]^{(0)}   
	\\&=  - \frac{1}{2p^{(0)}}  \left[  \dmu^{(0)}    -  ( \dmu^{(0)} )^2   +  \ddmu^{(0)}  \right], \\
	\left( \frac{\partial \mu}{\partial \bD}  \right)^{(0)} 
	&= 	\left( \frac{\diff \mu }{\diff I} \frac{\partial I}{\partial \bD} \right)^{(0)}
	=    \left( I \frac{\diff \mu }{\diff I} \frac{\bE }{\DD}   \right)^{(0)}
	= \mu^{(0)} \dmu^{(0)}   \frac{\bE^{(0)}}{ \DDD }, \\
	\left( \frac{\partial \dmu}{\partial \bD}  \right)^{(0)} 
	&=  \left[  
	\frac{\partial I}{\partial \bD} \left(  \frac{1}{\mu}\frac{\diff \mu }{\diff I} 
	 	 -  \frac{I}{\mu^2} \left(\frac{\diff \mu }{\diff I} \right)^2
	 	  +   \frac{I}{\mu} \frac{\diff^2 \mu}{\diff I^2} \right)    \right]^{(0)}  
	\\&=  \left[ \dmu^{(0)}    -  (\dmu^{(0)}  )^2  +  \ddmu^{(0)}    \right]   \frac{\bE^{(0)}}{\DDD}.
	\end{split}
\ee
Rearranging  (\ref{alsa1}) and making use of (\ref{alsa2}) gives
\be
	\label{alsa3}
	\begin{split}
	&\rho_s \phi   \left(  \partial _t\bv^{(1)}  +   \bv^{(0)}  \ccdot  \nabla \bv^{(1)}  
	 +   \bv^{(1)}  \ccdot  \nabla \bv^{(0)}  \right)  
	 =   -  \nabla p^{(1)} 
	 +   \frac{ (2  -  \dmu^{(0)}) \mu^{(0)}}{2}  \bE^{(0)} \nabla p^{(1)}  \\
	& +    \frac{ (\dmu^{(0)}   -1) \mu^{(0)} p^{(0)} }{ \DDD}  
	(\bE^{(0)} \nabla )( \bE^{(0)}   \ccdot \nabla \bv^{(1)} )   
	+  \frac{\mu^{(0)} p^{(0)} }{2  \DDD }  \nabla^2 \bv^{(1)}  -  2 \chi \nabla^4 \bv^{(1)}  \\
	& +   \left[    \frac{ \mu^{(0)} (\ddmu^{(0)} - \dmu^{(0)} )}{4 p^{(0)}}  p^{(1)} 
	 +   \frac{\mu^{(0)} }{2 \DDD} (\dmu^{(0)}    -  \ddmu^{(0)} ) \bE^{(0)}   \ccdot \nabla \bv^{(1)}  \right]    \bE^{(0)}\nabla p^{(0)}. 
	\end{split}
\ee

%%%%%%%%%%%%%%%%%%%
\section{ A Squire's-type  theorem for $\mu (I)$-rheology}\label{app:B} 

%  [ It would be better to use ``$k$" instead of  ``$k_{3D}$". This is an appendix and we do not have to worry about conflict with other notation. Also, why do you wish to show
%the components of $\bA$, since we never use them for calculation and it suffices to
%show the $\lambda_i$? Once again, the idea is to avoid complicated algebraic expressions which take up valuable journal space and which can be derived by computer algebra!]\\  
%
%[The above comments still apply. We can keep a record of the A's in case we need 
%them for future work, but they should not be included as in the appendix of the 
%current paper. ] \\ 

Allowing for out-of-plane perturbations, the components of the tensors  in (\ref{obliqueP})  are given by 
\be
	\begin{split}
	\bI \!-\!  \frac{\bN^{(0)} \bk \otimes \bk} {  \bk \ccdot {\bf N}^{(0)} \bk}  
	&\!=\! \frac{1}{\Phi_1} \!
	\begin{bmatrix} 
	\Phi_1 \!+\! k_1 \! \left( \!\frac{\alpha}{\sqrt{2}} k_2 \!-\! k_1 \! \right) 
	&& \!\!\!\!\!\! k_2 \left( \frac{\alpha}{\sqrt{2}} k_2 \!-\! k_1 \! \right) 
	&& \!\!\!\!\!\! k_3 \left( \frac{\alpha}{\sqrt{2}} k_2 \!-\! k_1 \! \right) \\
	 k_1 \left( \! \frac{\alpha}{\sqrt{2}} k_1 \!-\! k_2  \!\right) 	
	&& \!\!\!\!\!\! \Phi_1 \!+\! k_2 \left( \! \frac{\alpha}{\sqrt{2}} k_1\!- \!k_2 \! \right)
	&&  \!\!\!\!\!\! k_3 \left( \! \frac{\alpha}{\sqrt{2}} k_1\!- \!k_2 \! \right) \\
	-k_1 k_3 && -k_2 k_3  && \Phi_1 - k_3^2
	\end{bmatrix},\\
	{\bf M}\! -\! {\bf L}^{(0)} 
	\!&=\! \frac{1}{\phi}
	\begin{bmatrix} 
	\beta \gamma k_2^2 &  \beta \gamma k_1 k_2 \!-\! 2 \phi  & 0 \\
	\beta \gamma k_1 k_2 & \beta \gamma k_1^2  & 0 \\
	0 & 0 & 0 \end{bmatrix}
\end{split} \ee
where $\Phi_1 = \bk \ccdot {\bf N}^{(0)} \bk$,   
and
 $ k^2 = k_{2D}^2 + k_3^2$, $k_{2D}^2 = k_1^2 + k_2^2$, and $\Phi_2 = \gamma k^2 + 2 \chi k^4$.   

Making use of  the relation $\det(\bA - \lambda \bI ) = 0$, 
the eigenvalues $\lambda$ of $\bA$ are found  to be
\be
	\label{eigenvalue3D}
	\begin{split}  
	&\lambda_{  {\textrm{\scriptsize 1}} \atop {\textrm{\scriptsize 2}}   }=
	\frac{ \beta \gamma [ (k_1^2 - k_2^2 )^2 + k_3^2 k_{2D}^2  ] -  2 \Phi_1 \Phi_2  \pm {\Phi_4}^{1/2}}{2 \phi \Phi_1},\\ 
	&\lambda_3 = - \frac{ \Phi_2}{ \phi} =
	- \frac{1}{ \phi} \left(  \gamma k^2 + 2 \chi k^4 \right),  
 \end{split}\ee
where  
\[ \label{Phi4}
	\begin{split}
	\Phi_4 =&
	\beta^2  \gamma^2  (k_1^2 \!-\! k_2^2)^4  \!+\! 2 \phi^2  k_1^2  (\! \alpha k_1 \!-\! \sqrt{2} k_2 \!)^2  
	\!-\! 2 \beta \gamma \phi  k_1^2 (k_1^2 \!-\! k_2^2)   (\! \sqrt{2} \alpha k_1^2 \!-\! 4 k_1 k_2 \!+\! \sqrt{2} \alpha k_2^2 )
	\\ &+k_3^2 \left[ \beta^2 \gamma^2  k_{2D}^2 ( k_3^2 k_{2D}^2 \!+\! 2  
	(k_1^2 - k_2^2)^2  ) 
	\!+\! 2 \beta \gamma \phi k_1 [ (k_1^2 - k_2^2) (2 k_2 \!-\! \sqrt{2} \alpha k_1 ) \!-\! 2 k_2 k_3^2  ] \right].
	\end{split}
\]
Comparison of   the three eigenvalues indicates that the in-plane eigenvalue
 $\lambda_1 $ has  largest real part, so that planar disturbances are the least stable,{\it Q.E.D.}. 
 
\section{Numerical solution of (\ref{1dunsteady})}\label{app:C}
To solve (\ref{1dunsteady}) numerically we employ the well-known method of lines (MOL) \citep{Schiesser12}, with  spatial discretization at $N$ nodal points in
$y$, converting (\ref{1dunsteady}) to a set of  ODEs that can be solved by standard solvers. Specifically, we choose a standard finite-difference approximation with 
\be \label{fda}\begin{split} &y\rightarrow [ y_i], \:u(t, y)\rightarrow \ul{u}(t)\!=\! [u_i(t)],\: \partial^m_yu \rightarrow \ul{D}^{(m)}\ul{u}\!=\![D^{(m)}_{ij}u_j],\:\mbox{with}
\: u_i(t) \!=\! u(t, y_i),\\&\mbox{for}\: i, j\!=\!1, \ldots, N, \: \mbox{and}\:\frac{\diff \ul{u}}{\diff t} \!=\!\ul{D}^{(1)}\left( \ul{h}-2\ul{D}^{(3)}\ul{u}\right), 
\:\mbox{with}\:\: h_i\!=\!h(D^{(1)}_{ij}u_j),
\end{split}\ee
where  the rows of the matrices $\ul{D}^{(m)}$ are given by the classic interpolation coefficients of \cite{Fornberg88}, generalized to an arbitrary number  of interpolation points $ P > m$. To generate
these coefficients for $m=1, 3$  we have made use of the third-party \ML\:\:   program ``fdcoeffF.m".

We employ the \ML \:\: stiff integrator ``ode15s.m" to  integrate the N-dimensional  ODE in subject to the initial condition $\ul{u}(0)= \ul{u}_0= [u_0(y_i)]$ given by (\ref{pert}). We note that for the above ODE, it is rather easy to derive the analytical Jacobian, which is essentially the same as the matrix defining linear stability. \\ 

To satisfy 
the spatial boundary at the ends of the interval $(y_1, y_N)$ we employ $M$ ``ghost nodes" $y_1,\ldots, y_M$ and $y_{N-M+1}, \ldots,  y_N$ at either end,  where $u_i(t)= u_s(y_i)$ for all $t>0$. To enforce
this condition on the ODE solver, we
 we set the top and bottom $M$ rows of $\ul{D}^{(1)}$ equal to zero, i.e.  $D^{(1)}_{i, j}=0$, for $\:\:i=1, \ldots, M$ and $i=N-M+1, \ldots, N$, and we adopt the modified initial condition
 \be \label{IC}    u_0(y_i) =  u_s(y_i) + R(y_{M+1}, y_{N-M})A\sin k_2y_i,\ee 
where $R$ denotes the rectangular pulse vanishing on the ghost nodes and otherwise equal to unity. This leaves $N-2M$ ``active" nodes, and we choose $P=2M$ as the 
number of Fornberg interpolation points, so that interpolation on the $N-P$ active nodes in the center of the interval (0,1) may include up to
$M$ ghost node at either end.   \\

The ODE solver is run for a preset time $t=t_{\rm max}$,  long enough to ensure convergence of   $\ul{u}(t)$.  As indicated above in Subsection \ref{subsect:1Dstab}, we employ as measure of  stability of $\ul{u}_s$ the mean of $\epsilon=|\ul{u}(t_{\rm max})-\ul{u}_s|^2$,  as given by the \ML \:\: function ``std.m". 
For total number of nodes $\approx 1000$ on the $y$-interval (0, 1) we find a nearly neglible effect of the
number $M$  of ghost nodes, with $M$ ranging from $10$ to $50$. 		 
\end{document}